\documentclass[PAPER, cernpreprint, USenglish, texmf, orcidlogo]{atlasdoc}
\usepackage{atlaspackage}

\usepackage{atlasbiblatex}

\usepackage{atlasphysics}
\graphicspath{{logos/}{figures/}}

\usepackage{ANA-HIGP-2024-14-PAPER-defs}

\usepackage{multirow}
\usepackage{makecell}
\usepackage{feynmp-auto}
\usepackage{ifpdf}
\ifpdf
\fi

\newif\ifshowappendix
\showappendixtrue

\AtlasTitle{Measurement of off-shell Higgs boson production in the $H^*\rightarrow ZZ\rightarrow 4\ell$ decay channel using a neural simulation-based inference technique in 13 TeV $pp$ collisions with the ATLAS detector}

\AtlasAbstract{%
A measurement of off-shell Higgs boson production in the $H^*\to ZZ\to 4\ell$ decay channel is presented. The measurement uses 140~fb$^{-1}$ of proton--proton collisions at $\sqrt{s}=13$~TeV collected by the ATLAS detector at the Large Hadron Collider and supersedes the previous result in this decay channel using the same dataset. The data analysis is performed using a neural simulation-based inference method, which builds per-event likelihood ratios using neural networks. The observed (expected) off-shell Higgs boson production signal strength in the $ZZ\to 4\ell$ decay channel at 68\% CL is $0.87^{+0.75}_{-0.54}$ ($1.00^{+1.04}_{-0.95}$). The evidence for off-shell Higgs boson production using the $ZZ\to 4\ell$ decay channel has an observed (expected) significance of $2.5\sigma$ ($1.3\sigma$). The expected result represents a significant improvement relative to that of the previous analysis of the same dataset, which obtained an expected significance of $0.5\sigma$. When combined with the most recent ATLAS measurement in the $ZZ\to 2\ell 2\nu$ decay channel, the evidence for off-shell Higgs boson production has an observed (expected) significance of $3.7\sigma$ ($2.4\sigma$). The off-shell measurements are combined with the measurement of on-shell Higgs boson production to obtain constraints on the Higgs boson total width. The observed (expected) value of the Higgs boson width at 68\% CL is $4.3^{+2.7}_{-1.9}$ ($4.1^{+3.5}_{-3.4}$)~MeV.
}

\AtlasRefCode{HIGP-2024-14}

\PreprintIdNumber{CERN-EP-2024-298}

\arXivId{2412.01548}

\HepDataRecord{156805}

\AtlasJournal{Rep. Prog. Phys.}
\AtlasJournalRef{Rep. Prog. Phys. 88 (2025) 057803}
\AtlasDOI{DOI:10.1088/1361-6633/adcd9a}

\hypersetup{pdftitle={ATLAS document},pdfauthor={The ATLAS Collaboration}}

\begin{document}

%

%
\fmfcmd{%
style_def gluino expr p =
shrink(0.5);
cdraw (curly p);
endshrink;
cdraw p;
enddef;
style_def ewkino expr p =
cdraw (wiggly p);
cdraw p;
enddef;
style_def gravitino expr p =
cdraw (wiggly p);
cdraw p;
enddef;
style_def gluon expr p =
shrink(0.5);
cdraw (curly p);
endshrink;
enddef;}


%
\maketitle
\tableofcontents

\section{Introduction}
%

The Higgs boson was observed in 2012 by the ATLAS~\cite{HIGG-2012-27} and CMS~\cite{CMS-HIG-12-028} collaborations at the Large Hadron Collider (LHC). Since then, a series of measurements were performed to establish the fundamental nature of this new particle. Measurements of the spin~\cite{HIGG-2013-17,CMS-HIG-14-018}, mass~\cite{HIGG-2022-20,CMS-HIG-19-004,CERN-EP-2024-210}, and couplings~\cite{HIGG-2021-23,CMS-HIG-22-001} of the Higgs boson have been performed in several decay channels.  The total width of the Higgs boson ($\Gamma_H$) in the Standard Model (SM) is calculated to be 4.10~\MeV~\cite{deFlorian:2016spz}, and is a particularly challenging parameter to be measured. Direct measurements of the Higgs boson lineshape are unable to reach the precision required to measure the Higgs boson width due to limited detector resolution, $\mathcal{O}$(1.5--3~\GeV)~\cite{CERN-EP-2024-210,CMS-HIG-14-036}.

Due to the small total width, the differential cross-section with respect to the four-lepton invariant mass $d\sigma/dm_{4\ell}$ in the $H\to ZZ\to 4\ell$ decay channel falls steeply for invariant masses larger than the Higgs boson mass of $125\,\GeV$. Several authors have pointed out that this differential cross-section increases again at the kinematic threshold at twice the $Z$ boson pole mass $m_{4\ell}=2m_Z$ to a level that could be probed by the LHC experiments~\cite{Kauer:2012hd,Caola:2013yja,Campbell:2013una}. Similar kinematic thresholds can be exploited at twice the pole mass of the $W$ boson~\cite{HIGG-2014-10} and of the top quark~\cite{TOPQ-2023-22} in different production and decay channels. Both the ATLAS and CMS collaborations have recently reported evidence for the production of off-shell Higgs bosons using the \ohtollll and \ohtollnunu decay channels~\cite{HIGG-2018-32,CMS-HIG-21-013,CERN-EP-2024-210}.

Measurements of the Higgs boson production rate in $H\to VV$ decay channels, where $V=Z$ or $W$, are used to probe the Higgs boson couplings to vector bosons and fermions. These measurements depend on the Higgs boson total width if performed with on-shell Higgs bosons, but are independent of the width if the Higgs boson is off-shell, as can be seen from the Breit--Wigner model of the Higgs boson's resonance:

\begin{equation}
\begin{split}
\frac{d\sigma^{H\to VV}}{dm_{VV}^2}&\propto \frac{g_{\text{prod}}^2(\widehat{s})g_{\text{decay}}^2(\widehat{s})}{(m_{VV}^2-m_H^2)^2+m_H^2\Gamma^2_H}\\\Rightarrow \frac{d\sigma^{H^*\to VV}_{\text{off-shell}}}{dm_{VV}^2} \propto g_{\text{prod}}^2(\widehat{s})&g_{\text{decay}}^2(\widehat{s}),\quad \sigma^{H\to VV}_{\text{on-shell}} \propto \frac{g_{\text{prod}}^2(m_H)g_{\text{decay}}^2(m_H)}{\Gamma_H},
\end{split}
\end{equation}
where $g_{\text{prod}}(\widehat{s})$ and $g_{\text{decay}}(\widehat{s})$ are, respectively, the effective couplings of Higgs boson production and decay, as a function of the Higgs boson virtuality $\widehat{s}$. A comparison of the off-shell and on-shell Higgs boson production rates allows an indirect measurement of the Higgs boson total width assuming that there is no beyond SM (BSM) physics that alters the on-shell and off-shell couplings differently~\cite{Campbell:2013wga,Caola:2013yja,Campbell:2013una}.

In the presence of light BSM states, off-shell Higgs boson production can probe new kinematic thresholds produced by one-loop electroweak (\ew) corrections~\cite{Goncalves:2017iub}. Off-shell Higgs boson production can also be used to constrain BSM physics at high energy. Using the formalism of SM Effective Field Theory~\cite{Weinberg:1980wa, Coleman:1969sm, Callan:1969sn}, off-shell Higgs boson production breaks the degeneracy between the top-quark Yukawa and Higgs-gluon interaction which exists for on-shell Higgs boson production~\cite{Azatov:2014jga}. In these cases, using the SM prediction for on-shell and off-shell Higgs boson production would no longer provide an indirect measurement of the Higgs boson total width, since the new states can have a non-negligible contribution to the Higgs boson propagator and decay. The ATLAS Collaboration has recently reported limits on dimension-6 operators that modify the production of off-shell Higgs bosons~\cite{ATL-PHYS-PUB-2023-012}, while the CMS Collaboration reported limits on dimension-6 operators that modify both production and decay~\cite{CMS-HIG-21-013,CMS-HIG-21-019}.

This paper reports a new measurement of off-shell Higgs boson production in the \ohtollll\ decay channel using a novel method to interpret the data. The method, known as neural simulation-based inference (NSBI)~\cite{Brehmer:2018hga,Brehmer:2018eca,Brehmer:2019xox,BrehmerPNAS}, uses neural networks (NN) to estimate the per-event likelihood ratio of different hypotheses instead of using histograms of kinematic observables as an approximation to probability density ratios. The specific implementation of NSBI used in this paper is described in \Sect{\ref{sec:neural}} and in more detail in Ref.~\cite{atlasML}.

The NSBI method improves several aspects of a histogram-based analysis by providing a better approximation of the exact likelihood ratio between different hypotheses~\cite{Brehmer:2018hga}. The likelihood ratio obtained by the NSBI approach is optimally sensitive to any value of the off-shell Higgs boson production signal strength. This cannot be achieved by analyses that interpret a ratio of likelihoods constructed from a single observable when the signal model is non-linear, even if this observable is optimal for a given parameter value~\cite{atlasML}. Multidimensional optimal observables~\cite{CMS-HIG-18-002} and parameterized optimal observables~\cite{CMS-PAS-HIG-23-016} can alternatively be used with the same goal. In addition, the NSBI construction allows for unbinned measurements of parameters. Binned histograms treat all events inside a given bin as equivalent leading to an unavoidable loss in the power to separate different hypotheses. These losses can be minimized by choosing bin widths compatible with the observable resolution. This optimization may be challenging in high-resolution final states and may require large simulated samples, while an unbinned NSBI analysis can make better use of limited-size simulation samples.

The NSBI construction can be understood as a machine learning approach to the matrix-element method~\cite{D0:2004rvt,CDF:2006nne} where NNs learn both the theory dependency of the likelihood ratio relative to a high-dimensional final-state phase space and the complex detector response using simulated samples. The use of large NN models and the unbinned character of the NSBI analysis require a more advanced computational infrastructure than that used in histogram-based analyses~\cite{Megino:2024htf}.

The results of this paper supersede those of Ref.~\cite{HIGG-2018-32} for the $4\ell$ channel, using the same dataset, object selection, and event selection. The measurement is combined with the \llvv channel, which is not re-analyzed, applying the same method for its interpretation as in Ref.~\cite{HIGG-2018-32}.


%

%
\section{ATLAS detector}

The ATLAS detector~\cite{PERF-2007-01} at the LHC covers nearly the entire solid angle around the collision point.\footnote{ATLAS uses a right-handed coordinate system with its origin at the nominal interaction point (IP) in the center of the detector and the \(z\)-axis along the beam pipe. The \(x\)-axis points from the IP to the center of the LHC ring, and the \(y\)-axis points upwards. Polar coordinates \((r,\phi)\) are used in the transverse plane,  \(\phi\) being the azimuthal angle around the \(z\)-axis. The pseudorapidity is defined in terms of the polar angle \(\theta\) as \(\eta = -\ln \tan(\theta/2)\) and is equal to the rapidity $ y = \frac{1}{2} \ln \left( \frac{E + p_z}{E - p_z} \right) $ in the relativistic limit. Angular distance is measured in units of \(\Delta R \equiv \sqrt{(\Delta y)^{2} + (\Delta\phi)^{2}}\).} It consists of an inner tracking detector surrounded by a thin superconducting solenoid, electromagnetic and hadronic calorimeters, and a muon spectrometer incorporating three large superconducting air-core toroidal magnets.

The inner-detector system (ID) is immersed in a \qty{2}{\tesla} axial magnetic field  and provides charged-particle tracking in the range \(|\eta| < 2.5\). The high-granularity silicon pixel detector covers the vertex region and typically provides four measurements per track, the first hit generally being in the insertable B-layer (IBL) installed before Run~2~\cite{ATLAS-TDR-19,PIX-2018-001}. It is followed by the SemiConductor Tracker (SCT), which usually provides eight measurements per track. These silicon detectors are complemented by the transition radiation tracker (TRT), which enables radially extended track reconstruction up to \(|\eta| = 2.0\).  The TRT also provides electron identification information based on the fraction of hits (typically 30 in total) above a higher energy-deposit threshold corresponding to transition radiation.

The calorimeter system covers the pseudorapidity range \(|\eta| < 4.9\). Within the region \(|\eta|< 3.2\), electromagnetic calorimetry is provided by barrel and endcap high-granularity lead/liquid-argon (LAr) calorimeters, with an additional thin LAr presampler covering \(|\eta| < 1.8\) to correct for energy loss in material upstream of the calorimeters. Hadronic calorimetry is provided by the steel/scintillator-tile calorimeter, segmented into three barrel structures within \(|\eta| < 1.7\), and two copper/LAr hadronic endcap calorimeters. The solid angle coverage is completed with forward copper/LAr and tungsten/LAr calorimeter modules optimized for electromagnetic and hadronic energy measurements respectively.

The muon spectrometer (MS) comprises separate trigger and high-precision tracking chambers measuring the deflection of muons in a magnetic field generated by the superconducting air-core toroidal magnets. The field integral of the toroids ranges between \num{2.0} and \qty{6.0}{\tesla\metre} across most of the detector. Three layers of precision chambers, each consisting of layers of monitored drift tubes, cover the region \(|\eta| < 2.7\), complemented by cathode-strip chambers in the forward region, where the background is highest. The muon trigger system covers the range \(|\eta| < 2.4\) with resistive-plate chambers in the barrel, and thin-gap chambers in the endcap regions.

The luminosity is measured mainly by the LUCID--2~\cite{LUCID2} detector that records Cherenkov light produced in the quartz windows of photomultipliers located close to the beampipe.

Events are selected by the first-level trigger system implemented in custom hardware, followed by selections made by algorithms implemented in software in the high-level trigger~\cite{TRIG-2016-01}. The first-level trigger accepts events from the \qty{40}{\MHz} bunch crossings at a rate below \qty{100}{\kHz}, which the high-level trigger further reduces in order to record complete events to disk at about \qty{1}{\kHz}.

A software suite~\cite{SOFT-2022-02} is used in data simulation, in the reconstruction and analysis of real and simulated data, in detector operations, and in the trigger and data acquisition systems of the experiment.


%

%
\section{Modeling of off-shell Higgs boson production}

The measurement of off-shell Higgs boson production presented in this paper is interpreted using a model in which the effective couplings of the Higgs boson to gluons and to \ew\ vector bosons in the SM can have anomalous scalar modifications. This analysis framework, known as the $\kappa$-framework~\cite{LHCHiggsCrossSectionWorkingGroup:2013rie}, allows changes in the overall cross-section without changing the process kinematics, and can be understood as a sector of a larger Higgs Effective Field Theory (HEFT)~\cite{Ghezzi:2015vva}.

\begin{figure}[tbp]
\centering
\subfloat[]{\label{fig:gg_diagrams_s}
\includegraphics[width=0.35\textwidth,trim={0 1cm 0 1cm},clip]{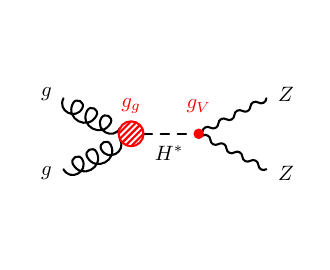} }\hspace{2cm}
\subfloat[]{\label{fig:gg_diagrams_b}
\includegraphics[width=0.35\textwidth,trim={0 1cm 0 1cm},clip]{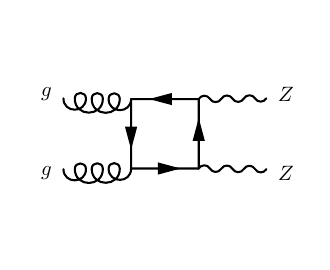}}
\caption{Illustrative leading-order Feynman diagrams for $gg\to ZZ$ production. The diagrams indicate the effective couplings $g_g$ and $g_V$. Diagram \protect\subref{fig:gg_diagrams_s} corresponds to the signal component and diagram \protect\subref{fig:gg_diagrams_b} corresponds to the background component. A large destructive interference between the two components is present in the off-shell regime.}
\end{figure}

The gluon--gluon fusion (\ggf) off-shell Higgs boson production and subsequent decay into a $ZZ$ pair, $gg\to ZZ$, can be described by using the Higgs boson effective coupling constants $g_g(\widehat{s})$ and $g_V(\widehat{s})$ to gluons and vector bosons, shown in \Fig{\ref{fig:gg_diagrams_s}}. The effective couplings $g_g$ and $g_V$ depend on the Higgs boson virtuality $\widehat{s}$, but the notation is suppressed hereafter for simplicity. The Higgs boson is represented by $H^*$ to denote explicitly that its virtuality is well above the pole mass $m_H=125\,\GeV$~\cite{HIGG-2020-07}.

In $gg\to ZZ$ production, the \textit{signal} ($\text{S}$) component is defined at leading order (LO) in perturbation theory by the absolute value squared of the amplitude of the diagram in \Fig{\ref{fig:gg_diagrams_s}}. The signal contribution scales as $g_g^2g_V^2$. The \textit{background} ($\text{B}$) component is defined at LO in perturbation theory by the absolute value squared of the amplitude of the diagram in \Fig{\ref{fig:gg_diagrams_b}}. The background component is independent of $g_g$ and $g_V$. The \textit{interference} ($\text{I}$) between the two diagrams scales as $g_gg_V$. The integrated interference between the two diagrams is negative, as required by perturbative unitarity conservation~\cite{Passarino:2012ri}.

The effective coupling between gluons and the Higgs boson $g_g$ is represented by a blob in \Fig{\ref{fig:gg_diagrams_s}} since in the SM, at LO in perturbation theory, it can be resolved to a fermion triangle loop dominated by the top-quark contribution. In several BSM scenarios, the contribution of new heavy particles cannot be resolved at the scales probed by this measurement, but can modify the Higgs boson effective couplings as well as the background process.

\begin{figure}[tbp]
\subfloat[]{\label{fig:ew_diagrams_vbf}
\includegraphics[width=0.24\textwidth,trim={0.5cm 0.6cm 0.5cm 0.2cm},clip]{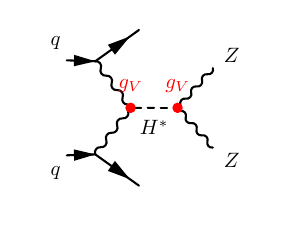}
}
\subfloat[]{\label{fig:ew_diagrams_t}
\includegraphics[width=0.24\textwidth,trim={0.5cm 0.6cm 0.5cm 0.2cm},clip]{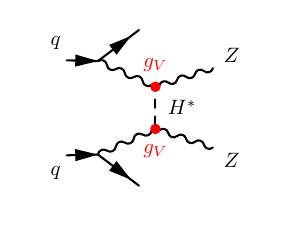}
}
\subfloat[]{\label{fig:ew_diagrams_vh}
\includegraphics[width=0.22\textwidth,trim={0.5cm 0.3cm 0.2cm 0.2cm},clip]{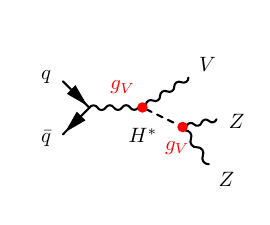}
}
\subfloat[]{\label{fig:ew_diagrams_vbs}
\includegraphics[width=0.24\textwidth,trim={0.5cm 0.6cm 0.5cm 0.2cm},clip]{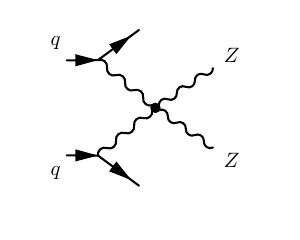}
}
\caption{Illustrative leading-order Feynman diagrams for the electroweak $qq \to ZZ+2j$ processes. The diagrams indicate the effective couplings $g_V$ used to define the off-shell Higgs boson production signal strengths. Diagrams \protect\subref{fig:ew_diagrams_vbf}, \protect\subref{fig:ew_diagrams_t}, and \protect\subref{fig:ew_diagrams_vh} correspond to the vector boson fusion, $t$-channel and $VH$ signal components, respectively. In diagram \protect\subref{fig:ew_diagrams_vh}, one of the vector bosons decays into a $q\bar{q}$ pair. Diagram \protect\subref{fig:ew_diagrams_vbs} corresponds to the electroweak background component. A large destructive interference is present in the off-shell regime between the vector boson fusion signal and the background components.\label{fig:ew_diagrams}}
\end{figure}

The same concepts can be applied to the \ew\ production of \ewprod. In this case, the \textit{signal} scales as $g_V^4$, the \textit{interference} scales as $g_V^2$, and the \textit{background} component is independent of $g_V$. Similar to the case of ggF production, a negative interference is present in the off-shell regime between the electroweak signal background components. The LO Feynman diagrams for the production of \ew \ewprod are shown in \Fig{\ref{fig:ew_diagrams}}.

\begin{figure}[tbp]
\centering
\subfloat[]{\label{fig:ni_diagrams_qqzz}
\raisebox{0.4cm}{\includegraphics[width=0.30\textwidth,trim={0 1cm 0 1cm},clip]{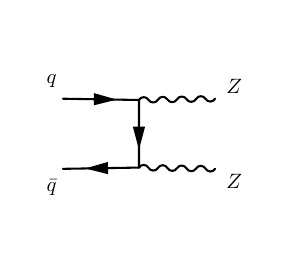}}
}\hspace*{1cm}
\subfloat[]{\label{fig:ni_diagrams_vvv}
\includegraphics[width=0.25\textwidth,trim={0 0.05cm 0 0cm},clip]{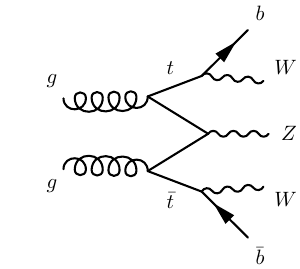}
}
\caption{Illustrative leading-order Feynman diagrams for non-interfering background processes. Diagram \protect\subref{fig:ni_diagrams_qqzz} shows the leading $q\bar{q}ZZ$ process and diagram \protect\subref{fig:ni_diagrams_vvv} shows the top-quark induced sub-leading $VVV$ process. The leptonic decays of the $Z$ and $W$ bosons are not shown.}
\end{figure}

The scaling of the cross-sections with the effective couplings $g_g$ and $g_V$ defines uniquely each component (signal, interference, and background) in the ggF and EW production of off-shell Higgs bosons. The probability density model used to measure the off-shell Higgs boson production is defined as function of the coupling modifiers $\kappag = g_g/g_{g,\text{SM}}$ and $\kappaV=g_V/g_{V,\text{SM}}$, independently of the Higgs boson virtuality, which are used to define the signal strengths:

\begin{equation}
\muggf=\okappag^2\okappaV^2,\qquad \muew=\okappaV^4,
\end{equation}
where the subscript \textit{off-shell} indicates that the modifier only affects processes with virtuality sufficiently above the pole mass. These signal strengths are used in the definition of a probability density model that is used to interpret the collected data:

\begin{equation}\label{eq:measurement_model}
\begin{split}
p(x|\muggf,\muew)=&\frac{1}{\nu(\muggf,\muew)}\times\\
\Bigg[&\muggf \nu_{\text{S}}^{\text{ggF}}p_{\text{S}}^{\text{ggF}}(x) + \sqrt{\muggf}\nu_{\text{I}}^{\text{ggF}} p_{\text{I}}^{\text{ggF}}(x) +\nu_{\text{B}}^{\text{ggF}} p_{\text{B}}^{\text{ggF}}(x)\; +\\
&\muew \nu_{\text{S}}^{\text{EW}}p_{\text{S}}^{\text{EW}}(x) + \sqrt{\muew}\nu_{\text{I}}^{\text{EW}} p_{\text{I}}^{\text{EW}}(x) +\nu_{\text{B}}^{\text{EW}} p_{\text{B}}^{\text{EW}}(x)+\nu_{\text{NI}}p_{\text{NI}}(x)\Bigg],
\end{split}
\end{equation}
where $x$ is a vector of reconstructed observables, which are defined in \Sect{\ref{sec:event}}, and where \probx\ and \yieldx\ are the probability density function and the expected yield for process $\text{X}$, respectively. The expected number of events $\nu(\muggf,\muew)$ can be written as a function of the expected number of events \yieldx for each process. The term $p_{\text{NI}}(x)$ represents the probability density for processes that do not interfere with the ggF and EW processes described above. The leading non-interfering process is $\qqbar\to ZZ\to 4\ell$ production via a $t$-channel exchange at LO\@. Non-interfering triboson $VVV$ processes, including those from top-quark decays $t\bar{t}V$, are subleading processes, but are also included in the analysis. The exception is the interfering $ZZZ\to 4\ell+\text{jets}$ process, which is modeled as part of the \ewb\ process. \Fig{\ref{fig:ni_diagrams_qqzz}} shows the LO Feynman diagram of the leading non-interfering \qqzzprod process (\qqzz) and \Fig{\ref{fig:ni_diagrams_vvv}} shows the corresponding LO Feynman diagram of the subleading non-interfering top-quark-induced $VVV$ process.

Monte Carlo (MC) simulated samples are used to describe the expected event yields $\nu_{\text{X}}$ and probability densities $p_{\text{X}}(x)$ in \Eqn{\ref{eq:measurement_model}}. Due to the technical challenges associated with the efficient production of interference-only MC simulations~\cite{Mattelaer:2021xdr}, the interference terms in \Eqn{\ref{eq:measurement_model}} are not generated separately, but inferred from samples generated with signal, interference, and background terms (SBI sample, where $\text{SBI}=\text{S}+\text{I}+\text{B}$). For ggF production, a single SBI sample is generated and the interference term $\nu_I^{\text{ggF}}p_I^{\text{ggF}}(x)$ is calculated as the difference of the SBI sample and the S and B samples ($\text{I}=\text{SBI}-\text{S}+\text{B}$):

\begin{equation}\label{eq:gg_SBI}
\begin{split}
\nu_{\text{I}}^{\text{ggF}}p_{\text{I}}^{\text{ggF}}(x) &= \nu_{\text{SBI}}^{\text{ggF}}p_{\text{SBI}}^{\text{ggF}}(x) - \nu_{\text{S}}^{\text{ggF}}p_{\text{S}}^{\text{ggF}}(x) - \nu_{\text{B}}^{\text{ggF}}p_{\text{B}}^{\text{ggF}}(x).
\end{split}
\end{equation}

In the simulation of \ew\ production, it is impossible to generate an off-shell signal-only sample. Due to the diagram shown in \Fig{\ref{fig:ew_diagrams_t}}, where an $s$-channel Higgs boson propagator is absent, and in \Fig{\ref{fig:ew_diagrams_vh}}, where the Higgs boson can decay as $H^*\to ZZ\to 2\ell 2q$, there is always contamination of on-shell events. Any on-shell contamination is formally part of the EW background process, since it does not scale with $\mu_{\text{off-shell}}$. Instead of generating pure signal and interference samples, two linear combinations ($\text{EWSBI}_1$ and $\text{EWSBI}_{10}$) are used to model the EW processes:

\begin{equation}\label{eq:SBI}
\begin{split}
\nu_{\text{SBI}_{1}}^{\text{EW}}p_{\text{SBI}_1}^{\text{EW}}(x) &= \nu_{\text{S}}^{\text{EW}}p_{\text{S}}^{\text{EW}}(x) + \nu_{\text{I}}^{\text{EW}}p_{\text{I}}^{\text{EW}}(x) + \nu_{\text{B}}^{\text{EW}}p_{\text{B}}^{\text{EW}}(x),\\[1ex] \nu_{\text{SBI}_{10}}^{\text{EW}}p_{\text{SBI}_{10}}^{\text{EW}}(x) &= 10\nu_{\text{S}}^{\text{EW}}p_{\text{S}}^{\text{EW}}(x) + \sqrt{10}\nu_{\text{I}}^{\text{EW}}p_{\text{I}}^{\text{EW}}(x) + \nu_{\text{B}}^{\text{EW}}p_{\text{B}}^{\text{EW}}(x).
\end{split}
\end{equation}
These samples contain the signal, interference, and background components, including any possible on-shell contamination, albeit with different fractions. An additional sample is generated exclusively modeling the diagram in \Fig{\ref{fig:ew_diagrams_vbs}} to describe the \ewb\ process $\nu^{\text{EW}}_{\text{B}}p_{\text{B}}^{\text{EW}}(x)$. While this sample does not contain the on-shell contamination from diagrams \Fig{\ref{fig:ew_diagrams_t}} and \Fig{\ref{fig:ew_diagrams_vh}} discussed above, it was found that considering the contribution of these terms only in $\text{EWSBI}_1$ and $\text{EWSBI}_{10}$ is a good approximation.

The \ewsbiten sample is simulated by choosing the effective coupling between the Higgs boson and the EW vector bosons to achieve an EW signal cross-section ten times larger than the SM value ($\kappa_V^4=10$). The change of the effective coupling scales the interference component by a factor $\sqrt{10}$, yielding the expression in \Eqn{\ref{eq:SBI}}. When simulating the \ewsbiten sample, the Higgs boson width is modified to keep the on-shell effective coupling unchanged at $g_V^4/\Gamma_H$. Linear combinations of the \ewb, \ewsbi, and \ewsbiten samples can be inverted to obtain the EW signal, interference and background processes separately, as shown in \Tab{\ref{tab:model_processes}}.

As described earlier, the non-interfering background process $\nu_{\text{NI}}p_{\text{NI}}(x)$ in \Eqn{\ref{eq:measurement_model}} contains both the leading $\qqbar ZZ$ and the subleading $VVV$ processes:

\begin{equation}
\nu_{\text{NI}}p_{\text{NI}}(x) = \nu_{\qqbar ZZ}p_{\qqbar ZZ}(x)+\nu_{VVV}p_{VVV}(x).
\end{equation}
All $VVV$ processes are considered separately as a single non-interfering background process, except for the $ZZZ\to 4\ell+\text{jets}$ sample is modeled as part of the EW background process.

A data-driven normalization is introduced for the leading \qqzzprod process as a function of the number of reconstructed jets, which is one of the observables in the vector $x$ defined in \Sect{\ref{sec:event}}:

\begin{equation}\label{eq:ni_background}
\begin{split}
\nu_{\text{NI}}p_{\text{NI}}(x)= \qqzzzj\:\nu_{\qqbar ZZ}^{0j}\:p_{\qqbar ZZ}^{0j}(x)+\qqzzzj\:\qqzzoj\:\nu_{\qqbar ZZ}^{1j}\:p_{\qqbar ZZ}^{1j}(x)\\+\;\qqzzzj\:\qqzzoj\:\qqzztj\:\nu_{\qqbar ZZ}^{2j}\:p_{\qqbar ZZ}^{2j}(x)+\nu_{VVV}\:p_{VVV}(x).
\end{split}
\end{equation}
The parameter $\qqzzzj$ provides a data-driven normalization for the total $\qqbar\to ZZ\to 4\ell$ observed yield. The parameter $\qqzzoj$ ($\qqzztj$) provides a data-driven normalization for the ratio of the observed yield of $\qqbar\to ZZ\to 4\ell$ events with one (at least two) reconstructed jet and that with zero (one) reconstructed jets. The process referred to as $\nu_{\qqbar ZZ}^{2j}p_{\qqbar ZZ}^{2j}(x)$ includes all events with at least two reconstructed jets.

\Tab{\ref{tab:model_processes}} summarizes the processes that are used in the model after all transformations, as well as the simulated samples used to describe them.

\begin{table}[tbp]
\centering
\caption{Definition of processes in the probability model used to interpret the data and the simulated samples used to describe them. The multipliers define both the measured off-shell Higgs boson production signal strength and the data-driven normalization of leading backgrounds. The $(H^*\to)$ notation is used for the $\text{SBI}$ process. The multipliers are obtained after substituting the interference terms of \Eqns{\ref{eq:gg_SBI}}{\ref{eq:SBI}} into \Eqn{\ref{eq:measurement_model}}, since interference-only samples are not simulated. For instance, in the case of $\text{ggF}$ production: $\mu\nu_{\text{S}}+\sqrt{\mu}\nu_{\text{I}}+\nu_{\text{B}}$$=(\mu-\sqrt{\mu})\nu_{\text{S}}+\sqrt{\mu}(\nu_{\text{S}}+\nu_{\text{I}}+\nu_{\text{B}})+(1-\sqrt{\mu})\nu_{\text{B}}$.}
\begin{tabular}{ccc}\hline\hline
\toprule\textbf{Process} & \textbf{Multipliers} & \textbf{Samples}\\
\midrule\ggs\ & $\muggf - \sqrt{\muggf}$ & \ggsprod\\
\midrule\ggsbi\ & $\sqrt{\muggf}$ & \makecell{\ggsbiprod \\ ($\kappaV^2=1$)}\\
\midrule\ggb\ & $1-\sqrt{\muggf}$ & \makecell{$gg\to ZZ\to 4\ell$\\($\kappaV^2=0$)}\\
\midrule\ewb\ & $\dfrac{(1-\sqrt{10})\muew+9\sqrt{\muew}-10+\sqrt{10}}{-10+\sqrt{10}}$ & \makecell{\ew \ewprod\\ ($\kappaV^4=0$)\\$ZZZ\to 4\ell+2j$}\\
\midrule\ewsbi\ & $\dfrac{\sqrt{10}\muew-10\sqrt{\muew}}{-10+\sqrt{10}}$ & \makecell{\ew \ewsbiprod \\($\kappaV^4=1$)\\$ZZZ\to 4\ell+2j$}\\
\midrule\ewsbiten\ & $\dfrac{-\muew+\sqrt{\muew}}{-10+\sqrt{10}}$ & \makecell{\ew \ewsbiprod \\($\kappaV^4=10$)\\$ZZZ\to 4\ell + 2j$}\\
\midrule$\qqbar ZZ$ $n_{\text{jets}}=0$ & $\qqzzzj$ & $\qqbar \to ZZ\to 4\ell$\\
\midrule$\qqbar ZZ$ $n_{\text{jets}}=1$ & $\qqzzzj\qqzzoj$ & $\qqbar \to ZZ\to 4\ell$\\
\midrule$\qqbar ZZ$ $n_{\text{jets}}\geq 2$ & $\qqzzzj\qqzzoj\qqzztj$ & $\qqbar \to ZZ\to 4\ell$\\
\midrule$VVV$ & -- & \makecell{$WWZ\to 4\ell$\\ $WZZ\to 4\ell$\\ $\ttbar Z\to 4\ell$}\\\bottomrule\hline\hline
\end{tabular}
\label{tab:model_processes}
\end{table}



%

\section{Data and simulated event samples}

The analysis uses \pp\ collision data collected with the ATLAS detector in \RunTwo of the LHC, at a center-of-mass energy of $\sqrt{s} = 13\,\TeV$, corresponding to a total integrated luminosity of $140\,\text{fb}^{-1}$~\cite{DAPR-2021-01} after data-quality requirements~\cite{DAPR-2018-01}. Events were recorded using a combination of single-lepton, dilepton and trilepton triggers~\cite{TRIG-2018-05,TRIG-2018-01,TRIG-2019-03} with either a low transverse momentum, \pt, threshold and a lepton isolation requirement, or a higher threshold but a looser identification criterion and without any isolation requirement. The overall trigger efficiency for the ggF signal process is more than 98\% in each final state after object selection and after imposing the $180 < m_{4\ell} < 2000\,\GeV$ requirement.

The \ggprod samples (\ggs, \ggb, and \ggsbi) are generated with \SHERPA~{v2.2.2}~\cite{Bothmann:2019yzt} and  \textsc{OpenLoops}~\cite{Buccioni:2019sur,Cascioli:2011va,Denner:2016kdg} at LO accuracy in quantum chromodynamics (QCD), with up to one additional parton in the final state, using the \NNPDF[3.0]NNLO parton distribution function (PDF) set~\cite{Ball:2014uwa}. The merging with the parton shower was performed using the MEPS@NLO prescription~\cite{Hoeche:2012yf,Cascioli:2013gfa} and the \SHERPA{} built-in algorithm was used for parton showering and hadronization. Next-to-leading-order ($\text{NLO}$) QCD corrections are included as a function of the invariant mass of the two $Z$ bosons, $m_{ZZ}$, separately for the \ggb, $\text{S}$, and $\text{SBI}$ processes~\cite{Caola2015}. Fully differential next-to-next-to-leading-order ($\text{NNLO}$) corrections to the $gg\to H\to ZZ$ signal process are known~\cite{Anastasiou:2005qj,Catani:2007vq,Grazzini:2017mhc}, but not for the interference and background components. A common, average $\text{NNLO}/\text{NLO}$ correction of 1.2 is applied to the signal, interference, and background components  of the $gg\to ZZ$ process. Inclusive next-to-next-to-next-to-leading-order ($\text{N}^3\text{LO}$) corrections to the $gg\to H\to ZZ$ signal process are known~\cite{Anastasiou:2015vya}, dominated by the on-shell Higgs boson contribution. Currently, $\text{N}^3\text{LO}$ corrections in the off-shell region and for the interference and background components are not available. An average, common $\text{N}^3\text{LO}/\text{NNLO}$ correction of 1.1 is extrapolated to the off-shell region and applied to the signal, interference, and background components of the $gg\to ZZ$ process~\cite{Passarino:2013bha}.

The \ew \ewprod samples (\ewb, \ewsbi, and \ewsbiten) are generated with \MGNLO~\cite{Alwall:2014hca} at LO QCD and LO EW accuracy using the NNPDF3.0 NLO PDF set~\cite{Ball:2014uwa}. The \PYTHIA[8.2]~\cite{Sjostrand:2014zea} program was used for parton showering and hadronization with the A14 set of tuned parameters (A14 tune) for the underlying event~\cite{ATL-PHYS-PUB-2014-021} and \NNPDF[2.3]LO PDF set~\cite{Ball:2012cx}.

The \qqzzprod sample is generated with \SHERPA~{v2.2.2} and \textsc{OpenLoops} using the \NNPDF[3.0] NNLO PDF set. The matrix elements (ME) are calculated to NLO accuracy in QCD for 0- and 1-jet final states, and to LO accuracy for 2- and 3-jet final states. The merging with the \SHERPA parton shower was performed using the MEPS@NLO prescription. The NLO EW corrections are included as a function of $m_{ZZ}$~\cite{Biedermann:2016yvs,Biedermann:2016lvg}.

The triboson samples $ZZZ$, $WZZ$, and $WWZ$ with fully leptonic decays were modeled with \SHERPA~{v2.2.2} at NLO QCD accuracy.  The $ZZZ\rightarrow 4\ell +2j$ process is included in the EW $\qqbar\to ZZ +2j$ sample described above. The simulation of $t\bar{t}V$ production with at least one of the top quarks decaying leptonically and the vector boson decaying inclusively into either quarks or leptons was performed with \MGNLO~interfaced to \PYTHIA[8.2] for parton showering and hadronization with the A14 tune. The total cross-sections for the $t\bar{t}V$ backgrounds were normalized to the NLO QCD and EW predictions from Ref.~\cite{Frixione:2015zaa}.

\Tab{\ref{tab:model_samples}} summarizes the order in perturbation theory with which each simulated sample was generated and the source of higher-order correction ($K$-factors) used to improve the modeling. All simulated samples are processed with the ATLAS detector simulation~\cite{SOFT-2010-01} based on \GEANT~\cite{Agostinelli:2002hh}. The effects of multiple inelastic interactions in the same and neighboring bunch crossings (\pileup) were modeled by overlaying each simulated hard-scattering event with inelastic $pp$ events generated with \PYTHIA[8.186]~\cite{Sjostrand:2007gs} using the \NNPDF[2.3] LO PDF set and the A3 tune~\cite{ATL-PHYS-PUB-2016-017}. Simulated events are reweighted to match the \pileup conditions observed in the full \RunTwo dataset. Simulated events are reconstructed with the same algorithms and analysis chain as the data.

\begin{table}[tbp]
\centering
\caption{Matrix element generator, parton shower (PS), and higher-order corrections used to describe the different samples in the measurement. The notation $[Xj]$ indicates that up to $X$ jets are included in the hard-scatter matrix element calculation.\label{tab:model_samples}}
\begin{tabular}{cccc}\hline\hline
\toprule\textbf{Sample} & \textbf{ME generator} & \textbf{PS} & \textbf{Higher-order correction}\\
\midrule$gg\to ZZ$ & \makecell{\SHERPA[2.2.2]\\(LO [1j])} & \SHERPA[2.2.2] & \makecell{NLO QCD ($m_{ZZ}$ dependent)~\cite{Caola2015} \\ Approx. NNLO QCD (global)~\cite{Anastasiou:2005qj,Catani:2007vq,Grazzini:2017mhc}\\ Approx. N3LO QCD (global)~\cite{Passarino:2013bha} }\\
\midrule \ew $\qqbar \to ZZ+2j$ & \AtlasMC[2.3.3]{MG5\_aMC@NLO} & \PYTHIA[8.244] & -\\
\midrule\qqzzprod & \makecell{\SHERPA[2.2.2]\\(NLO [1j], LO [3j])} & \SHERPA[2.2.2] & NLO EW ($m_{ZZ}$ dependent)~\cite{Biedermann:2016yvs,Biedermann:2016lvg}\\\hline
$WWZ$, $WZZ$, $ZZZ$ & \SHERPA[2.2.2] & \SHERPA[2.2.2]	& --\\
\midrule\ttbar Z & \AtlasMC[2.3.3]{MG5\_aMC@NLO} & \PYTHIA[8.210] & NLO QCD + NLO EW (global)~\cite{Frixione:2015zaa}\\\bottomrule\hline\hline
\end{tabular}
\end{table}



%

\section{Object reconstruction, event selection and description}

The measurement of the off-shell Higgs boson production is performed in the $H^*\to ZZ\to 4\ell$ decay channel, where $\ell$ is either an electron or a muon. The object and event selections aim to identify opposite-charged electron or muon pairs consistent with the decay of a $Z$ boson. A quadruplet is formed from two pairs with a common production vertex and invariant mass above the $ZZ$ threshold. The selection criteria are optimized to increase the acceptance of $H^*\to ZZ\to 4\ell$ events while maintaining negligible levels of non-prompt background events. The object and event reconstruction used in this analysis is exactly the same as in the previous result~\cite{HIGG-2018-32} and only a short summary is provided here.


\subsection{Object reconstruction}

Muons are identified by tracks or segments reconstructed in the MS and matched to tracks reconstructed in the ID, with exceptions in areas where the MS lacks coverage. In the region $2.5<|\eta|<2.7$, muons can also be identified by tracks from the MS alone. In the central gap region ($|\eta|<0.1$) of the MS, muons can be identified by a track from the ID associated with a compatible calorimeter energy deposit (calorimeter-tagged muons).  Candidate muons are required to have $\pt > 5\,\GeV$ and $|\eta| < 2.7$, except calorimeter-tagged muons for which the \pt threshold is raised to 15~\GeV.
Muons must satisfy the \textit{loose} identification criterion~\cite{MUON-2018-03} with at most one standalone or calorimeter-tagged muon allowed per Higgs boson candidate. Electrons are reconstructed from energy deposits in the electromagnetic calorimeter matched to a track in the ID\@. Candidate electrons must have $\pT > 7$~\GeV{} and $|\eta| < 2.47$, and satisfy the \textit{loose} identification criteria~\cite{EGAM-2021-01}.

All electrons and muons used in both channels must be isolated, satisfying a \textit{loose} isolation criteria~\cite{MUON-2018-03,EGAM-2021-01}.
Furthermore, electrons (muons) are required to have associated tracks satisfying $|d_0/\sigma_{d_0}|<5~(3)$ and $|z_0\sin\theta|<0.5~\text{mm}$, where $d_0$ is the transverse impact parameter relative to the beam line, $\sigma_{d_0}$ is its uncertainty, and $z_0$ is the $z$ coordinate of the $r$-$\phi$ impact point, defined relative to the primary vertex. The event is rejected if the minimum angular separation between two leptons is $dR_{\ell\ell}< 0.1$, where $dR_{\ell\ell}=\sqrt{(\Delta\phi_{\ell\ell})^2 + (\Delta\eta_{\ell\ell})^2}$.

Jets are reconstructed from particle-flow objects~\cite{PERF-2015-09} using the anti-$k_t$ algorithm~\cite{Cacciari:2008gp,Fastjet} with radius parameter $R = 0.4$. The jet-energy scale is calibrated using simulation and further corrected with in situ methods~\cite{JETM-2018-05}. Reconstructed jets are required to have \pt $>$ 30 \GeV~and $|\eta| < 4.5$. A jet-vertex tagger~\cite{PERF-2014-03} is applied to jets with $\pT<60~\GeV$ and $|\eta|<2.4$ to suppress jets that originate from pileup. In the forward region, for jets with $\pt<50\,\GeV$ and $2.5<|\eta|<4.5$, another tagger based on jet shapes and topological jet correlations~\cite{ATL-PHYS-PUB-2019-026} is used to suppress pileup jets.


\subsection{Event selection and description}
\label{sec:event}

The selection of candidate events follows that described in Ref~\cite{HIGG-2018-32}. Events with at least four leptons (electrons or muons) are used in this analysis. The \pT\ thresholds for the three leading leptons are 20, 15 and 10 \GeV, respectively. The four-lepton invariant mass is required to be above the on-shell $ZZ$ production threshold, $180 < m_{4\ell} <  2000~\GeV$. Candidate lepton quadruplets are formed by selecting two opposite-charge, same-flavor dilepton pairs in each event.  In the 4e and 4$\mu$ channels, in which there are two possible pairings, the one that includes the lepton pair with mass closest to that of the $Z$ boson mass is chosen.  In each quadruplet, the lepton pair with mass closest to the $Z$ boson mass, $m_{Z_1}$, is referred to as the leading pair and required to have $50 <m_{Z_1} < 106$ \GeV.  The sub-leading pair mass, $m_{Z_2}$, must satisfy $50 < m_{Z_2} < 115$~\GeV~when $m_{4\ell}>190~\GeV$.
Due to the increased probability of one $Z$ boson being off-shell at lower values of $m_{4\ell}$, the lower threshold for $m_{Z_2}$ decreases linearly from 50 \GeV\ at $m_{4\ell} = 190~\GeV$ to 45 \GeV\ at $m_{4\ell} = 180~\GeV$.

Events are described in the analysis by 14 observables, summarized in \Tab{\ref{tab:event_vars}}. The 14 observables provide a complete description of the reconstructed final state phase space. The three-momentum of the fermion (anti-fermion) in the $Z_1$ decay is defined as $\mathbf{q}_{11}$ ($\mathbf{q}_{12}$). Similarly, the three-momentum of the fermion (anti-fermion) in the $Z_2$ decay is defined as $\mathbf{q}_{21}$ ($\mathbf{q}_{22}$). The three-momentum of $Z_1$ ($Z_2$) is defined as $\mathbf{q}_1$ ($\mathbf{q}_2$). All three-momenta are defined in the rest frame of the quadruplet. Jets are ordered in \pt\ and their momenta are defined in the laboratory reference frame.

The observables in \Tab{\ref{tab:event_vars}} are the components of the vector $x$ in \Eqn{\ref{eq:measurement_model}}. The observables $m_{jj}$, $\eta_{jj}$, and $\phi_{jj}$ related to the leading dijet system, \textit{i.e.}, the two jets with highest $\pt$ in the event, are only well-defined for events with at least two jets. For events with fewer jets, the value of these observables are chosen as the median of the corresponding distribution for events with at least two jets. The observable $n_{\text{jets}}$ is used for classification of the non-interfering background in \Eqn{\ref{eq:ni_background}}, where all events with more than two jets are described by $n_{\text{jets}}=2$.

\begin{table}[tbp]
\centering
\caption{Definition of the observables to describe an event. The observables are defined relative to the ATLAS coordinate system. The vectors $\mathbf{q}_1$, $\mathbf{q}_2$, $\mathbf{q}_{11}$, $\mathbf{q}_{21}$, $\mathbf{n}_1$, $\mathbf{n}_2$, and $\mathbf{n}_{\text{sc}}$ are defined in \Sect{\ref{sec:event}}.}
\begin{tabular}{cc}\hline\hline\toprule
\textbf{Variable} & \textbf{Definition}\\\hline
$m_{4\ell}$ & quadruplet mass\\
$m_{Z_1}$ & $Z_1$ mass \\
$m_{Z_2}$ & $Z_2$ mass\\
$\cos\theta^*$ & cosine of the Higgs boson decay angle [$\mathbf{q}_1\cdot\mathbf{n}_z/|\mathbf{q}_1|$]\\
$\cos\theta_1$ & cosine of the $Z_1$ decay angle [$-(\mathbf{q}_{2})\cdot\mathbf{q}_{11}/(|\mathbf{q}_{2}|\cdot|\mathbf{q}_{11}|)$]\\
$\cos\theta_2$ & cosine of the $Z_2$ decay angle [$-(\mathbf{q}_{1})\cdot\mathbf{q}_{21}/(|\mathbf{q}_{1}|\cdot|\mathbf{q}_{21}|)$]\\
$\Phi_1$ & $Z_1$ decay plane angle [$\cos^{-1}(\mathbf{n}_1\cdot\mathbf{n}_{\text{sc}})\,(\mathbf{q}_1\cdot(\mathbf{n}_1\times\mathbf{n}_{\text{sc}})/(|\mathbf{q}_1|\cdot|\mathbf{n}_1\times\mathbf{n}_{\text{sc}}|)$]\\
$\Phi$ & angle between $Z_1,Z_2$ decay planes [$\cos^{-1}(\mathbf{n}_1\cdot\mathbf{n}_2)\,(\mathbf{q}_1\cdot(\mathbf{n}_1\times\mathbf{n}_2)/(|\mathbf{q}_1|\cdot|\mathbf{n}_1\times\mathbf{n}_2|)$]\\
$p_T^{4\ell}$ & quadruplet transverse momentum\\
$y^{4\ell}$ & quadruplet rapidity\\
$n_{\text{jets}}$ & number of jets in the event\\
$m_{jj}$ & leading dijet system mass\\
$\Delta\eta_{jj}$ & leading dijet system pseudorapidity\\
$\Delta\phi_{jj}$ & leading dijet system azimuthal angle difference\\\bottomrule
\hline\hline
\end{tabular}
\label{tab:event_vars}
\end{table}


\begin{figure}[tbp]
\centering
\includegraphics[width=0.7\textwidth]{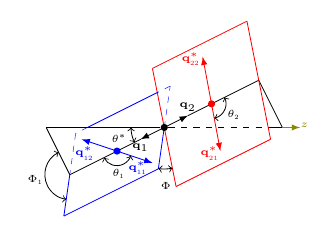}
\caption{Representation of the angular observables used to describe the probability density ratio of each event. The three-momenta of the fermions ($\mathbf{q}_{11}^*$, $\mathbf{q}_{21}^*$) and anti-fermions ($\mathbf{q}_{12}^*$, $\mathbf{q}_{22}^*$) are shown in their parent rest-frames, and the three-momenta of the vector bosons ($\mathbf{q}_{1}$, $\mathbf{q}_{2}$) are shown in the quadruplet rest frame.}
\label{fig:mela}
\end{figure}

\begin{figure}[tbp]
\centering
\subfloat[]{\includegraphics[width=0.46\linewidth]{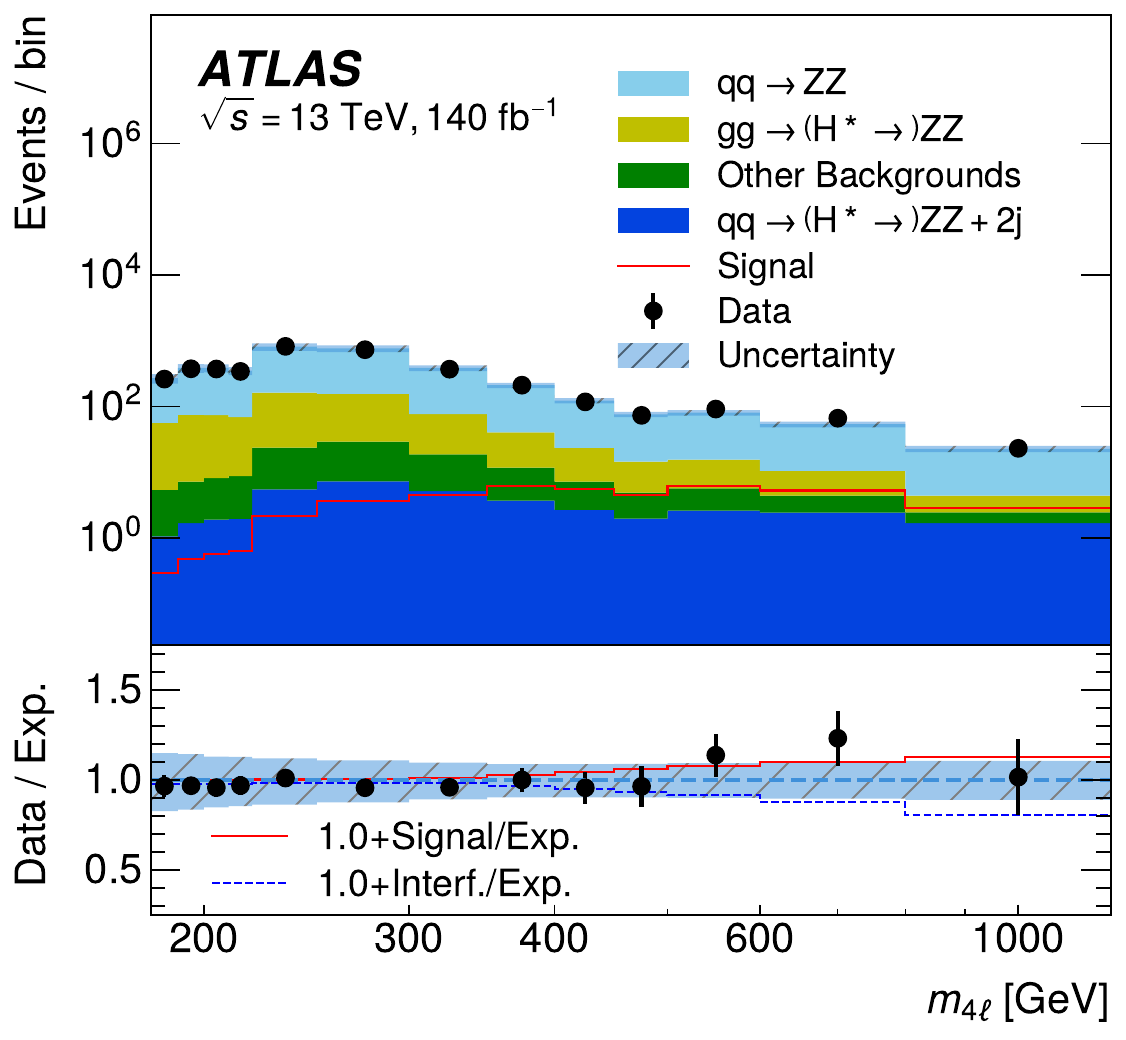}\label{fig:example_vars_m4l}}\hfill
\subfloat[]{\includegraphics[width=0.477\linewidth]{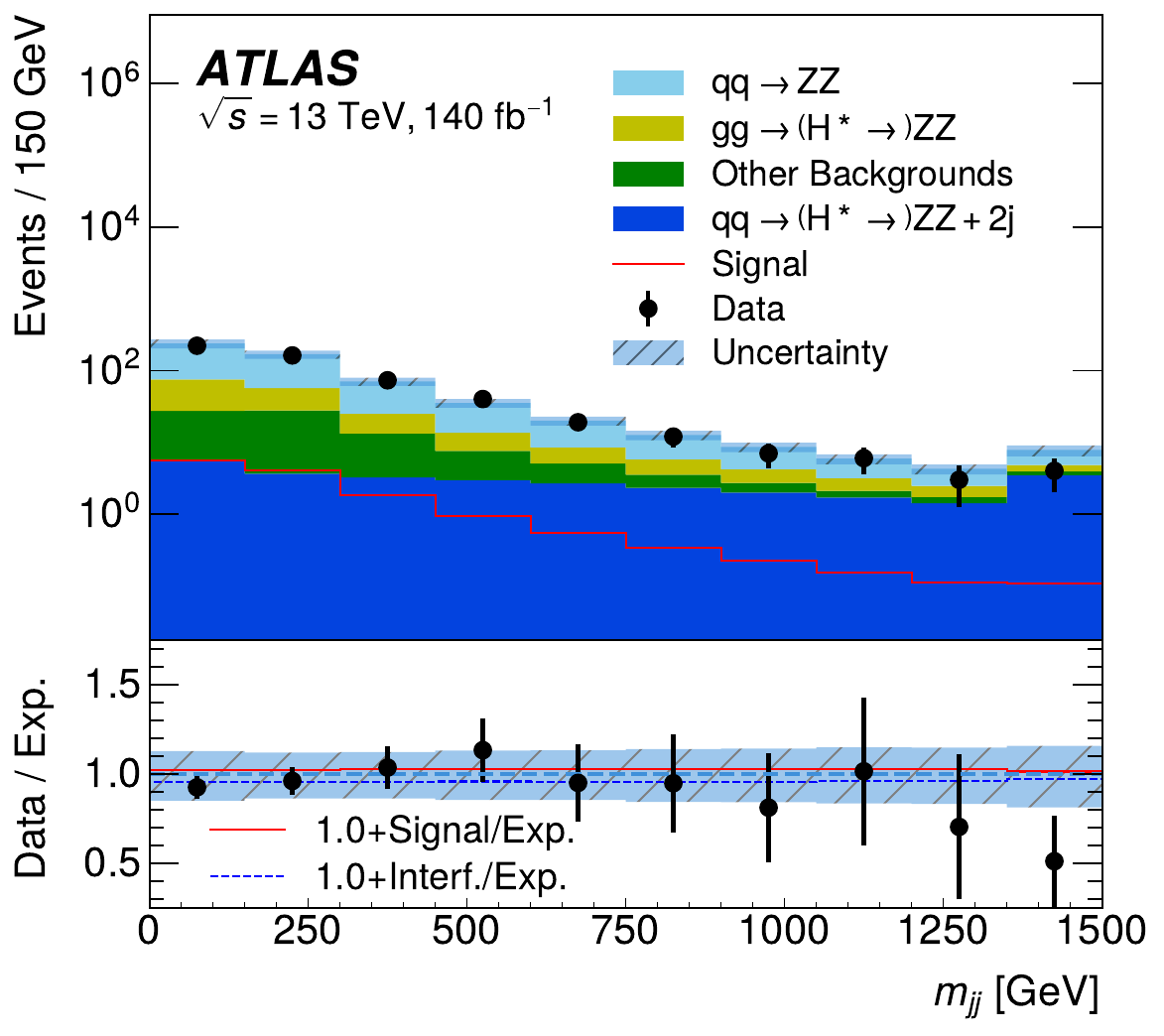}\label{fig:example_vars_mjj}}
\caption{Comparison between observed and expected distribution of two observables used to describe events: \protect\subref{fig:example_vars_m4l} the four-lepton invariant mass and the \protect\subref{fig:example_vars_mjj} invariant mass of the two leading jets showing good agreement for two of the observables used in this analysis. The expected distributions for $\qqbar \to ZZ$, \ggsbi ($gg\to (H^*\to ) ZZ$), \ewsbi  ($qq\to (H^*\to ) ZZ + 2j$) and other backgrounds are shown as stacked histograms, and the expected signal (interference) is shown as a red solid (blue dashed) line. The background is estimated under the SM hypothesis (post-fit, $\muoff=1$).  The lower panels show the ratio of data to expectation. The hatched band shows the total systematic uncertainty in the expected distribution. The last bin contains overflow events.}
\label{fig:example_vars}
\end{figure}

The normal vectors $\mathbf{n}_1$ and $\mathbf{n}_2$ to the $Z_1$ and $Z_2$ decay planes and the normal vector $\mathbf{n}_{\text{sc}}$ to the Higgs boson decay plane are defined as:

\begin{equation}
\mathbf{n}_1=\frac{\mathbf{q}_{11}\times\mathbf{q}_{12}}{|\mathbf{q}_{11}\times\mathbf{q}_{12}|},\qquad
\mathbf{n}_2=\frac{\mathbf{q}_{21}\times\mathbf{q}_{22}}{|\mathbf{q}_{21}\times\mathbf{q}_{22}|},\qquad
\mathbf{n}_{\text{sc}}=\frac{\mathbf{n}_{z}\times\mathbf{q}_{1}}{|\mathbf{n}_{z}\times\mathbf{q}_{1}|},
\end{equation}
where $\mathbf{n}_z$ is the unit vector in the $z$ direction. These vectors are used to build angles that are sensitive to the spin and parity of the quadruplet. Their geometrical visualization~\cite{Bolognesi:2012mm} is given in \Fig{\ref{fig:mela}}, where the three-momenta are shown in the rest frame of each particle's parent for clarity. Figure~\ref{fig:example_vars} shows the comparison between observed and expected background distributions for the $m_{4\ell}$ and $m_{jj}$ observables showing good agreement for two of the observables used in the analysis.

%

\section{Neural simulation-based inference}
\label{sec:neural}

Several analyses at the LHC use multinomial probability densities (histograms) of a single observable to describe each component \probx of the probability model used to interpret the data. A commonly used framework for histogram-based analyses in ATLAS is described in detail in Ref.~\cite{Cranmer:2012sba} and was used in the previous result on the off-shell Higgs boson production~\cite{HIGG-2018-32}. In analyses with non-linear signal models, like the measurements of processes with large quantum interference, a single observable cannot optimally capture the information needed to measure all possible signal strength values and complete dimensional reduction cannot be achieved without information loss. In addition, the multinomial modeling treats all events inside a bin as indistinguishable, which leads to a loss in statistical power. These losses can be partially mitigated by using approximations to optimal observables~\cite{Brehmer:2018hga}, by increasing the dimensionality of the histograms, and by reducing the bin width. Recent measurements of the off-shell Higgs boson production in the \ohtollll\ channel by the CMS Collaboration~\cite{CMS-HIG-18-002} have used all of these strategies to improve the significance of the result.

These mitigation strategies are limited by the finite number of simulated events and by the so-called curse of dimensionality. Practical considerations may reduce the accuracy of the multinomial approximation and reduce the power of the statistical inference, especially in regions with high signal significance (signal regions). Regions with low signal significance (control regions), traditionally used for the description of backgrounds and systematic uncertainties, are less sensitive to these limitations.

Neural network approximations of probability densities and probability density ratios can outperform histogram approximations when high-dimensional parameter spaces are considered. The use of NNs for statistical inference is known as \textit{neural simulation-based inference} (NSBI)~\cite{Brehmer:2018hga,Brehmer:2018eca,Brehmer:2019xox,BrehmerPNAS}. This analysis uses a particular version of NSBI adapted to the type of parameter inference done at the LHC to model events in the signal region. A self-contained description of the method is given below, and more details can be found in Ref.~\cite{atlasML}.

\subsection{Signal and control regions}

\begin{figure}[tbp]
\centering
\subfloat[]{\includegraphics[width=0.473\linewidth]{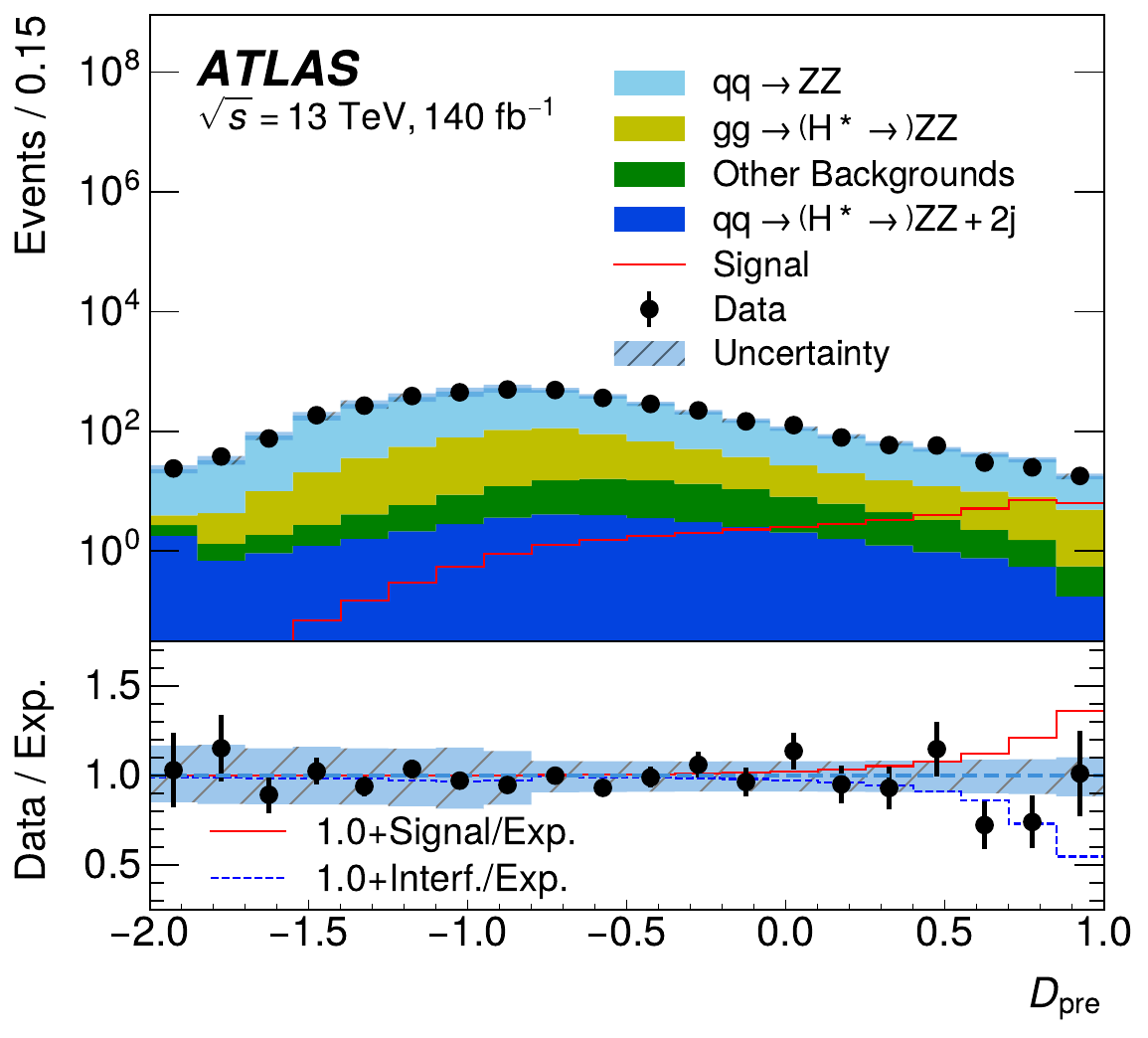}\label{fig:preselection_dpre}}\hfill
\subfloat[]{\includegraphics[width=0.46\linewidth]{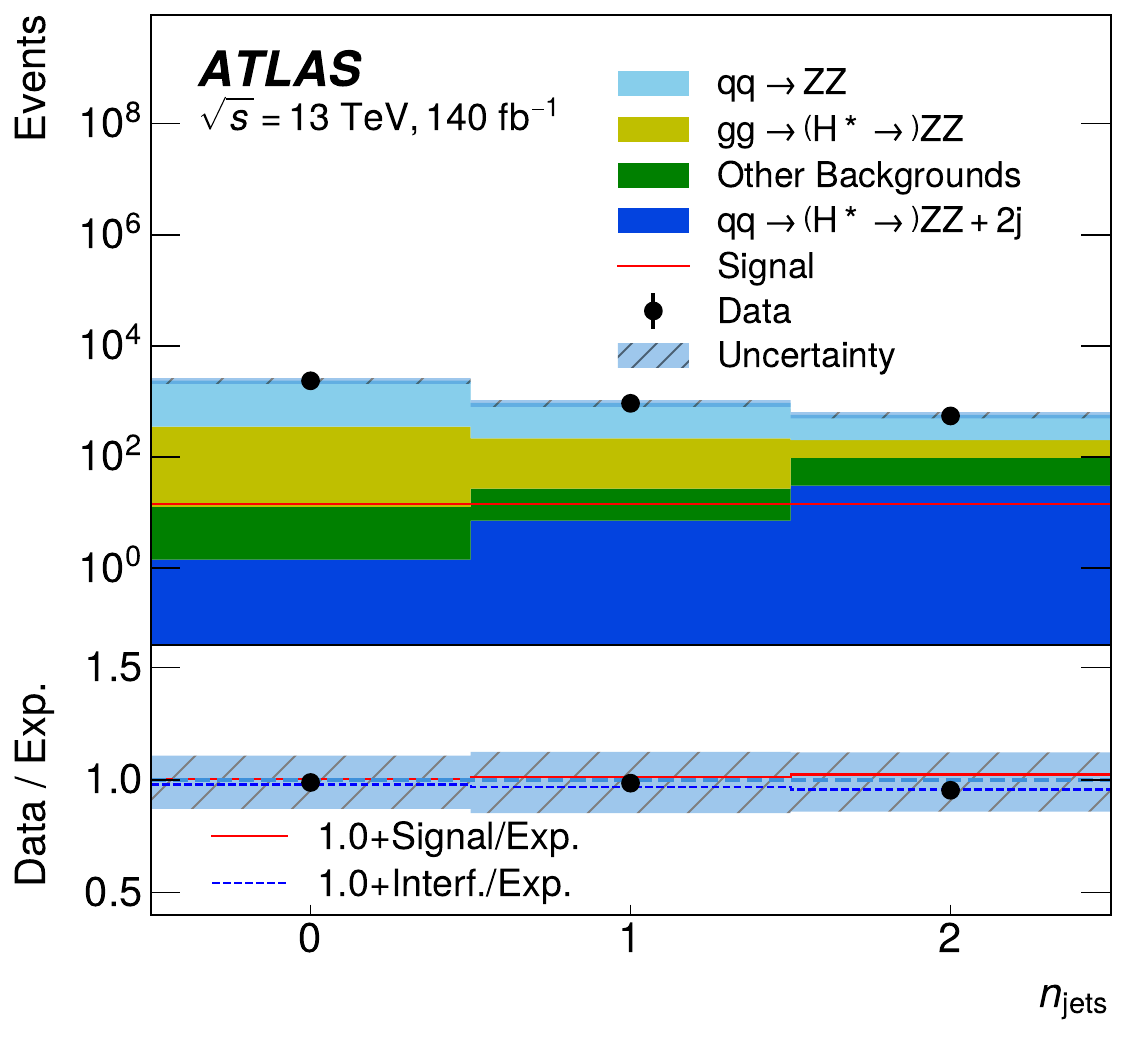}\label{fig:preselection_njets}}
\caption{\protect\subref{fig:preselection_dpre} Comparison between observed and expected distribution of the preselection discriminant \predisc showing a good description of the acceptance times efficiency in each region. \protect\subref{fig:preselection_njets} Comparison between observed and expected distribution of $n_{\text{jets}}$ showing a good description of the normalization obtained in each bin with the data-driven background normalization parameters. The expected distributions for $\qqbar \to ZZ$, \ggsbi ($gg\to (H^*\to ) ZZ$), \ewsbi  ($qq\to (H^*\to ) ZZ + 2j$) and other backgrounds are shown as stacked histograms, and the expected signal (interference) is shown as a solid red (dashed blue) line. The background is estimated under the SM hypothesis (post-fit, $\muoff=1$). The lower panels show the ratio of data to expectation. The hatched band shows the total systematic uncertainty in the expected distribution. The last bin of panel \protect\subref{fig:preselection_njets} contains overflow events.}
\end{figure}

An initial multi-class classification NN is trained to split the events into signal and control regions using the observables defined in \Tab{\ref{tab:event_vars}}. This NN has five hidden layers each with $1\,000$ neurons and a \textit{swish} activation function~\cite{DBLP:journals/corr/abs-1710-05941}. The output layer has five neurons with a \textit{softmax} $[e^{-x_i}/\sum_j e^{-x_j}]$ activation function. The NN is trained with a multi-class cross-entropy loss corresponding to the five processes used in the training: \ggs, \ggb, $\text{VBF}$, \ewb, and \qqzzprod. The vector boson fusion ($\text{VBF}$) process is obtained from a dedicated MC sample that uses only the EW VBF diagram in \Fig{\ref{fig:ew_diagrams_vbf}}. While this process does not provide a full description of the \ews production of off-shell Higgs bosons, it provides a sufficiently good approximation to define control and signal regions. The preselection discriminant \predisc is defined as:

\begin{equation}
\predisc  = \log\frac{s_{\text{pre}}^{\text{ggF S}}(x)+s_{\text{pre}}^{\text{VBF}}(x)}{s_{\text{pre}}^{\text{ggF B}}(x)+s_{\text{pre}}^{\text{EW B}}(x)+s_{\text{pre}}^{\qqbar ZZ}(x)},
\end{equation}
where $s(x)$ denotes the score function of the NN\@. The signal region (SR) is defined as events satisfying $\predisc >-0.85$. Events failing this condition define the control region (CR). The CR is largely dominated by $\qqbar\to ZZ$ events and can be used to constrain the parameters $\theta_{\qqbar ZZ}^{0j,1j,2j}$ in \Eqn{\ref{eq:ni_background}}. \Fig{\ref{fig:preselection_dpre}} shows a comparison between the observed and expected distribution of the observable $\predisc $ used to define the SR and the CR indicating a good description of the acceptance times efficiency in each region. \Fig{\ref{fig:preselection_njets}} depicts a comparison of the observed and expected $n_{\text{jets}}$ distribution, showing a good description of the normalization obtained in each bin with the data-driven background normalization parameters.

In the CR, only the $n_{\text{jets}}$ observable is used to describe the probability model, while in the SR all 14 observables listed in \Tab{\ref{tab:event_vars}} are used to create an NSBI model. For the NSBI model, a probability density ratio is formed from each process $\text{X}$ (where $\text{X}$ is one of the processes in \Tab{\ref{tab:model_processes}}) and a fixed reference process, $\probx/\probref$, and each ratio is estimated with a separate NN. The reference process is chosen as a mixture of the ggF signal and \ewsbiten processes:

\begin{equation}
\nu_{\text{ref}}p_{\text{ref}}(x) = \nu^{\text{ggF}}_{\text{S}}p_{\text{S}}^{\text{ggF}}(x)+\nu^{\text{EW}}_{\text{SBI}_{10}}p_{\text{SBI}_{10}}^{\text{EW}}(x),
\end{equation}
and does not depend on any parameter. While machine learning methods exist that allow to directly estimate the probability densities \probx~\cite{papamakarios2021normalizingflowsprobabilisticmodeling} in \Eqn{\ref{eq:measurement_model}}, the estimation of probability density ratios~\cite{Cranmer:2015bka} is a simpler numerical problem and is used in this analysis. As shown in \Sect{\ref{sec:result}}, estimating these probability density ratios for the different hypotheses is sufficient for a frequentist statistical data analysis~\cite{Cranmer:2014lly}, as \probref cancels out in the likelihood ratio test statistic. The choice of reference sample is driven by the phase space where the method is applied~\cite{atlasML}. The density ratio $\probx/\probref$ can only reliably be estimated when $\probref > 0$, which is ensured by the preselection condition $\predisc >-0.85$ without significantly reducing the power of the analysis.

\subsection{Probability density ratio estimation}

The probability density ratios are estimated as functions of the 14 observables in \Tab{\ref{tab:event_vars}} by using fully-connected NNs, \textit{i.e.}, where all neurons in a layer are connected to the neurons in the next layer. The architecture of this NN is the same as the one used for the preselection discriminant, but additional steps are taken to improve the accuracy of the score, which are described below.

The events in the simulated sample for each process $\text{X}$ are split into ten disjoint sets for use in ten-fold cross validation~\cite{hastie2009elements}. For each cross-validation set, the events in the other nine sets are used to train an ensemble of NNs, each with the structure described above. The ten-fold cross validation ensures that the NNs are never evaluated using events used in their training, which would otherwise generate over-confident estimates of the probability density ratios for rare events. An ensemble member is trained with 80\% of the events in the training set of the cross-validation splitting, randomly sampled without replacement. The total number of ensemble members varies between 10 and 70 for each ten-fold cross-validation set, depending on the process $\text{X}$, and resulting in 100 to 700 NNs in total. The larger ensembles are required for processes $\text{X}$ that are very different from the reference process. The estimate of a probability density ratio is taken as its ensemble mean. Using ensembles sampled without replacement helps to minimize both the bias and variance of the final result, and the ensembles are also used for an uncertainty estimate, as is explained below.

The NNs are trained to minimize the binary cross-entropy loss between the normalized simulated sample \probx , with truth label $s_{\text{truth}}\equiv 1$, and the normalized simulated sample \probref, with truth label $s_{\text{truth}}\equiv 0$. For balanced training samples ($\yieldx=\yieldref$), the NN converges to the optimal classifier score function $s_{\text{X}}(x) = \probx/(\probref+\probx)$ and the probability density ratio can be written as
$\probx /\probref = \scorex /(1-\scorex )$~\cite{GARRIDO199825,Cranmer:2015bka,hastie2009elements,Rizvi:2023mws}.

The batch size used for the gradient descent step of the training process is optimized separately for each process, but has to be kept sufficiently large to ensure that the NN extrapolates well and remains representative for other samples not used in the training. Larger batch sizes are also necessary when the training samples have a large fraction of negatively-weighted events to ensure convexity of the loss function. Events with negative weights come from simulations with higher-order corrections in perturbation theory and multijet merging with parton showers. An average batch size of $1\,024$ was used for the NNs in this analysis. The \textit{NAdam} adaptive learning rate algorithm~\cite{DBLP:journals/corr/KingmaB14,dozat.2016} is used for all trainings. In each training, 10\% of the available events are used for loss function validation. A summary of the NN structure and training is given in \App{\ref{sec:nn_parameters}}.

The NN-estimated probability density ratios $p_{\text{X}}(x)/p_{\text{ref}}(x)$ for each process $\text{X}$ are used to construct the signal strength-dependent probability density ratio:

\begin{equation}
\frac{p(x|\mu,\theta)}{p_{\text{ref}}(x)} = \frac{1}{\nu(\mu,\theta)}\sum_{\text{processes X}}f_X(\mu,\theta)\;\nu_X\;\frac{p_{\text{X}}(x)}{p_{\text{ref}}(x)},
\end{equation}
where $f_X(\mu,\theta)$ are the sample-dependent multipliers listed in \Tab{\ref{tab:model_processes}}.

\begin{figure}[tbp]
\centering
\subfloat[]{\includegraphics[width=0.46\linewidth]{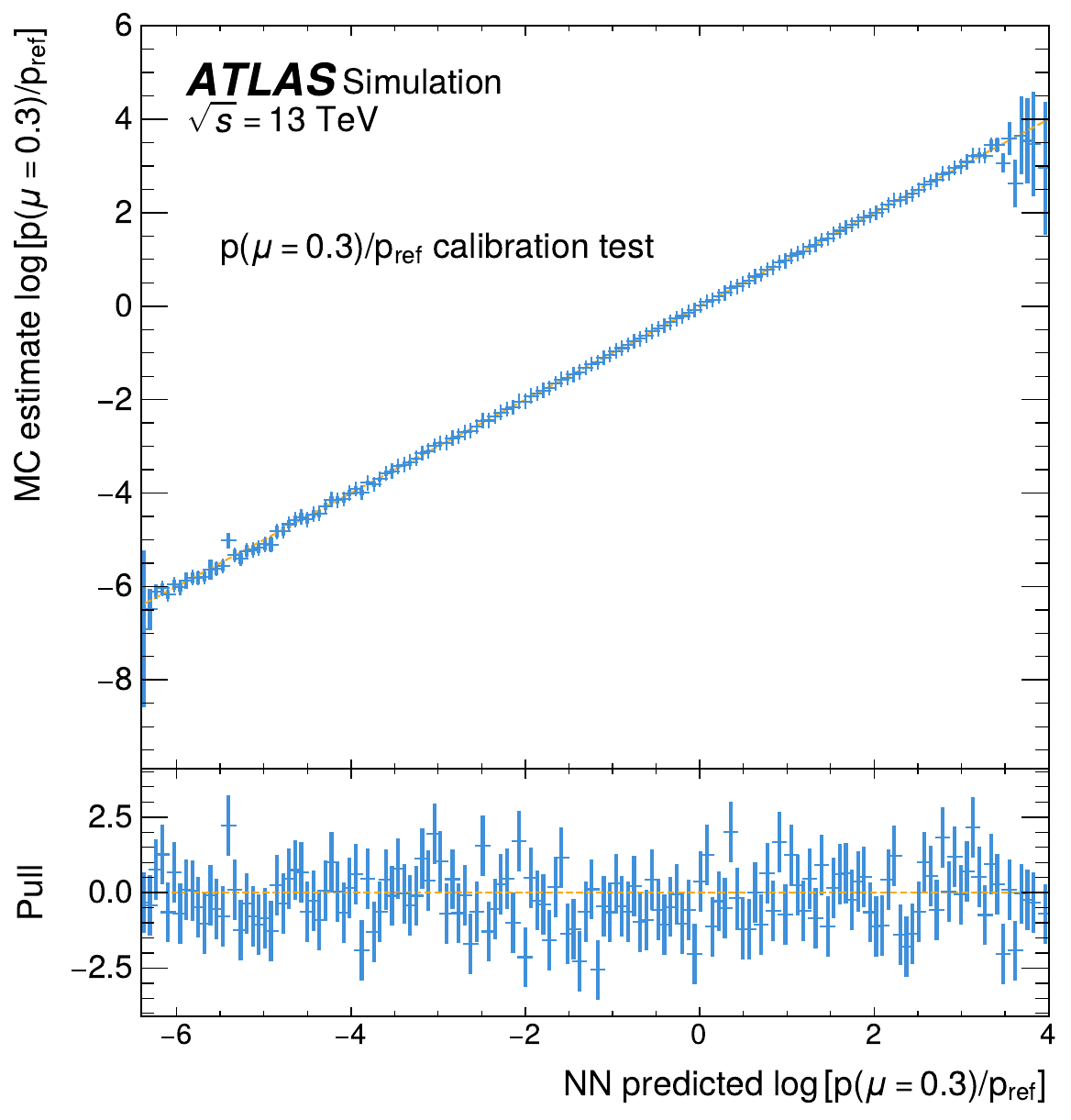}\label{fig:calib_test_0p3}}\hfill
\subfloat[]{\includegraphics[width=0.46\linewidth]{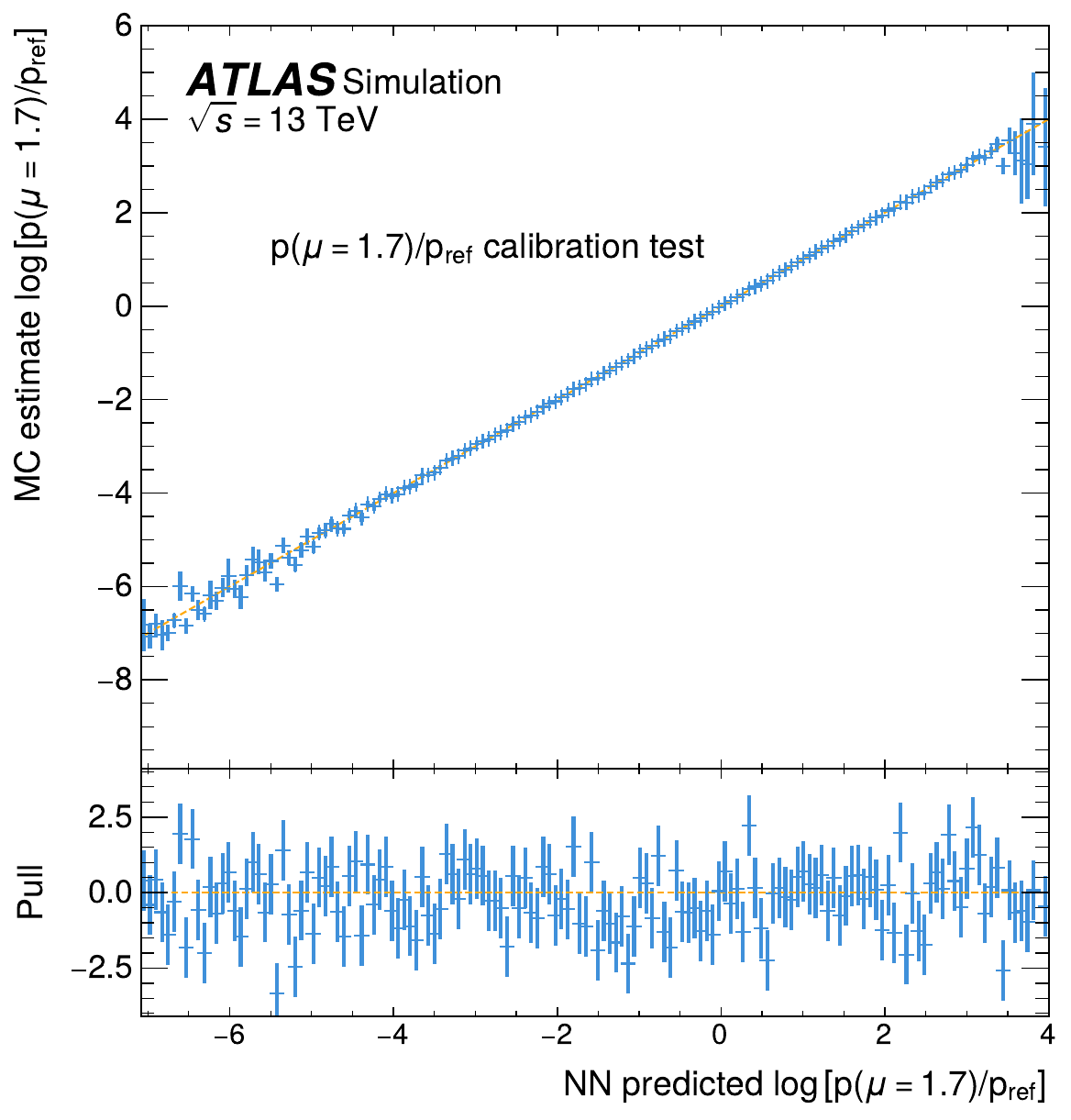}\label{fig:calib_test_1p7}}
\caption{Comparison between the NN-based and histogram-based estimate of the density ratio $p(x|\muoff )/p_{\text{ref}}(x)$ for \protect\subref{fig:calib_test_0p3} $\muoff=0.3$ and \protect\subref{fig:calib_test_1p7} $\muoff=1.7$. The comparison is done as a function of $\log\left[p(x|\muoff )/\probref\right]$ to separate the comparison for events that are very signal-like and reference-like. The lower panels show the pull, defined as the difference between the NN and MC-based estimate divided by the statistical uncertainty due to the finite number of MC events. The error bars indicate the uncertainty due to the finite number of simulated events in the histogram-based estimate of the density ratio.}
\label{fig:calib_test}
\end{figure}

The different hyperparameters described above, including the preselection threshold, the width and depth of each individual NN, and the size of the NN ensembles were optimized to obtain density ratio estimates as accurate as possible given the number of events in the simulated samples available for training. Several tests are performed to assess the accuracy of the NN training. A complete description of all tests performed can be found in Ref.~\cite{atlasML}, and only a summary is given here. \Fig{\ref{fig:calib_test}} compares the NN-based and histogram-based estimates of the density ratio $p(x|\muoff )/\probref$ for two high-statistics Asimov samples\footnote{An Asimov dataset is one for which the application of any unbiased estimator for all parameters will provide the true values~\cite{Cowan:2010js}. In unbinned analyses, an approximation of such a dataset can be constructed using a large number of simulated events with appropriate event weights.} with signal strengths $\muoff=0.3$ and $1.7$. As can be seen with \Eqn{\ref{eq:measurement_model}}, these two samples validate the NN-based procedure in a regime that is dominated by the interference component ($\muoff < 1$), and in a regime that is dominated by the signal component ($\muoff > 1$), respectively. In both cases, the NNs show excellent probability calibration. Note that the result is obtained without a NN calibration layer~\cite{10.1145/1102351.1102430}.

\begin{figure}[tbp]
\centering
\subfloat[]{\includegraphics[width=0.46\linewidth]{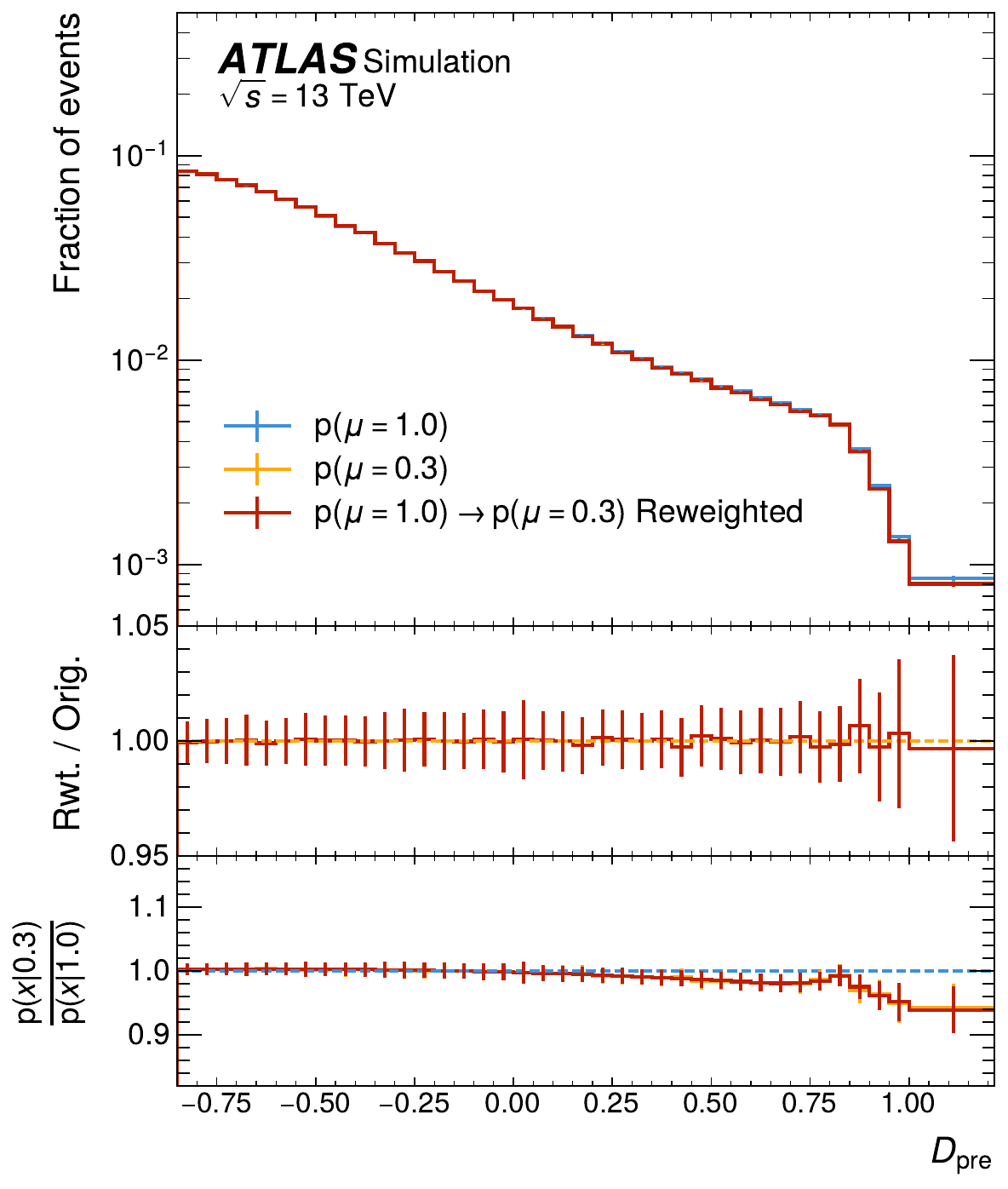}\label{fig:reweight_test_0p3}}\hfill
\subfloat[]{\includegraphics[width=0.46\linewidth]{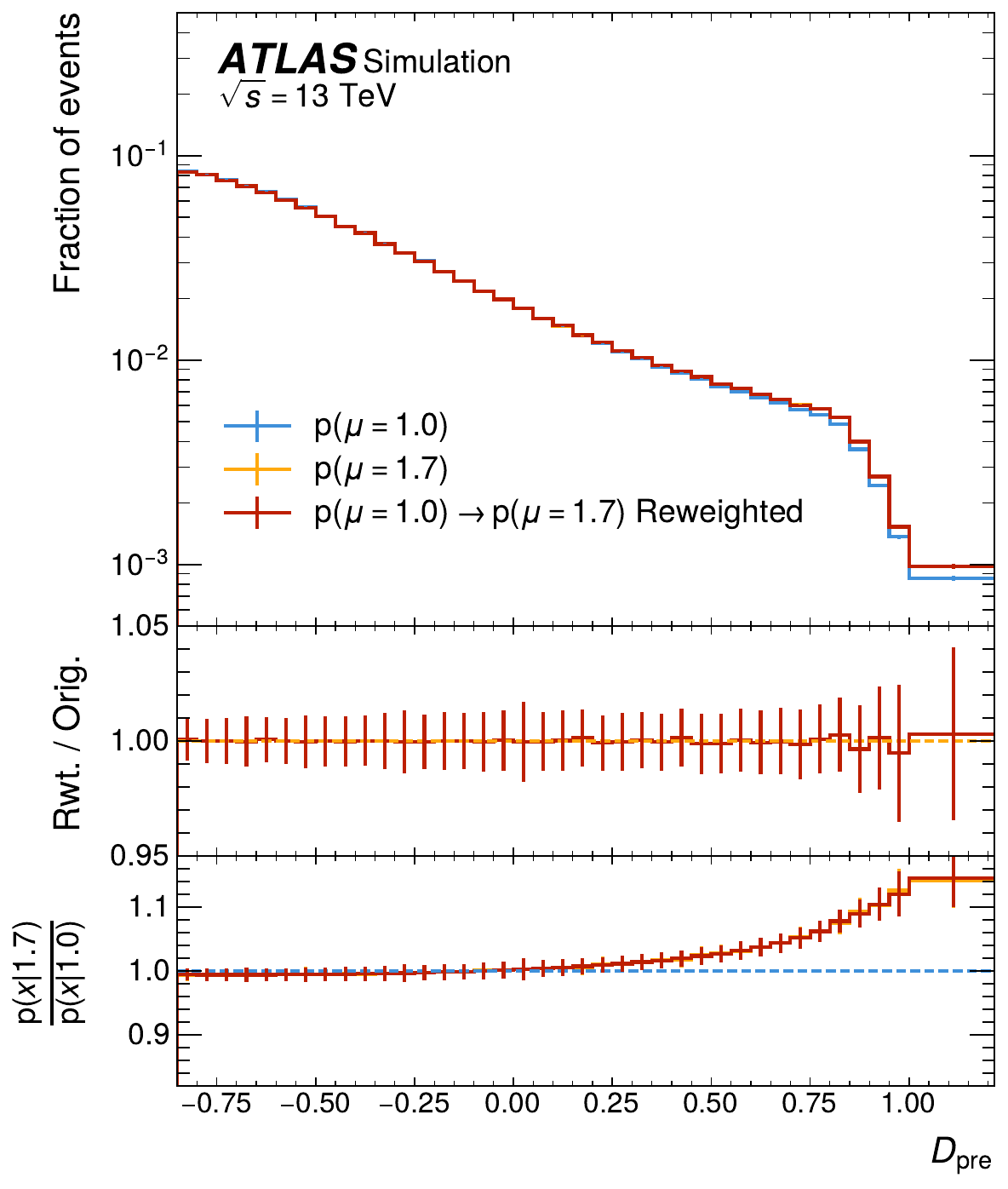}\label{fig:reweight_test_1p7}}
\caption{Comparison between the $\predisc $ normalized distribution obtained using an Asimov sample with \protect\subref{fig:reweight_test_0p3}  $\mu_{\text{off-shell}}=0.3$ and \protect\subref{fig:reweight_test_1p7}  $\mu_{\text{off-shell}}=1.7$, and the same distributions obtained through reweighting of a SM ($\mu_{\text{off-shell}}=1.0$) Asimov sample. The ratio plots provide a comparison between the two estimates (middle panel) and between the distributions at the given value of $\mu_{\text{off-shell}}$ and the SM expectation (bottom panel).	\label{fig:reweight_test}}
\end{figure}

If the NNs are unbiased, the probability density ratio $p(x|\muoff)/\probref$ can be used to reweight distributions from one value of \muoff\ to another, where the reweighting factor is obtained from the NN-based estimate of the probability density ratio using:

\begin{equation}
\frac{p(x|\muoff)}{p(x|\muoff=1)}=\left(\frac{p(x|\muoff)}{\probref}\right)\left(\frac{p(x|\muoff=1)}{\probref}\right)^{-1}.
\end{equation}
Comparisons between distributions estimated directly from an Asimov sample with a known value of \muoff\ with those obtained through reweighting of an Asimov sample with a different value of \muoff\ allows to test for possible bias in different regions of phase space. This test is demonstrated in \Fig{\ref{fig:reweight_test}} where histograms of distributions of \predisc\ obtained from high-statistics Asimov samples with signal strengths $\muoff=0.3$ and $1.7$ are compared with the same distributions obtained through reweighting of a SM ($\muoff=1.0$) Asimov sample. The results shown in \Figs{\ref{fig:calib_test}}{\ref{fig:reweight_test}} indicate that the NNs are trained with low bias and low variance. A separate multidimensional test is performed by training a second NN to discriminate between an Asimov sample with known \muoff\ and another sample obtained from reweighting the reference sample with $p(x|\muoff)/\probref$. No discriminating power is observed in this second NN, which indicates that it cannot distinguish between the original and reweighted samples, and that no significant biases in the estimate of the density ratios are present. The multidimensional test probes a limited range of the complete phase space and only confirms what thorough reweighting tests performed with many different observables already show.

\subsection{Systematic uncertainties}
\label{sec:systematics}

Systematic uncertainties include experimental uncertainties in the reconstructed objects and modeling uncertainties in the simulated samples. The systematic uncertainty model follows closely that reported in Ref.~\cite{HIGG-2018-32} and only a summary is presented here. Modeling uncertainties and uncertainties in the jet energy scale and resolution dominate the systematic uncertainty of the measurement. Jet-related observables used to describe the events provide important information for discrimination between hypotheses, especially for EW off-shell Higgs boson production. However, they also make the result sensitive to modeling of jet energy scale and resolution.

The uncertainty in the integrated luminosity for the full \RunTwo dataset is 0.83\%~\cite{DAPR-2021-01}, obtained using the LUCID-2 detector~\cite{LUCID2} for the primary luminosity measurements. Experimental uncertainties include momentum scale and resolution uncertainties for muons, electrons, and jets. These uncertainties are estimated by using calibrations performed for each individual object and by comparing them with different simulation models. Uncertainties in the reconstruction, identification and trigger efficiency of electrons~\cite{TRIG-2018-05, EGAM-2021-01} and muons~\cite{TRIG-2018-01,MUON-2018-03} are determined from tag-and-probe efficiency measurements using $Z\to \ell^+\ell^-$ and $J/\psi\to \ell^+\ell^-$ events. uncertainties in the jet energy scale are derived by combining information from test-beam data, LHC collision data and simulation~\cite{JETM-2018-05}. Uncertainties in the jet energy resolution are estimated as a function of jet $\pt$ and rapidity using dijet events, based on a similar method as in \Refn{\cite{JETM-2018-05}}.

Modeling uncertainties arise from the choice of PDF, missing higher-order corrections in both QCD and EW perturbative calculations, the merging of additional partons to the hard-scatter ME, and the description of the parton shower. The methods used to estimate modeling uncertainties are summarized in \Tab{\ref{tab:systSummary}}

The PDF uncertainties are evaluated using the \NNPDF prescription with 100 replicas from the \NNPDF[3.0] set~\cite{Ball:2014uwa}. The uncertainties due to missing higher-order QCD corrections are estimated by varying the renormalization and factorization scales independently, by factors one-half and two.

For the \ggprod processes, the renormalization and factorization scales variations are evaluated on the NLO/LO $K$-factors and propagated to the measurement as a function of $m_{ZZ}$ only~\cite{Caola2015}. The \ggprod NLO scale variation is found to be approximately uniform throughout the phase space, with a 10\% relative magnitude, and independent of the process (\ggs, \ggb, \ggsbi). The NLO $K$-factors and scale variations used for the $gg\to ZZ\to 4\ell$ processes do not contain the complete top-quark mass dependency, which has only been recently calculated~\cite{Agarwal:2024pod}. Therefore, the $gg\to ZZ\to 4\ell$ uncertainties due to scale variations are increased by 50\% for $m_{ZZ}$ in the $\ttbar$ threshold region, and doubled in the phase space containing a jet with $\pt>150~\GeV$~\cite{HIGG-2018-32}.

For all other processes, the two variations of the renormalization and factorization scales with largest impact in the expected value \yieldx were taken as representative of this uncertainty. An additional uncertainty in missing higher-order QCD corrections is estimated for the \ggprod and \qqzzllllprod processes by varying the soft-gluon resummation scale (QSF) in \SHERPA by half and twice the nominal value used in the simulation. The QSF scale variation is evaluated separately for each \ggprod component (\ggs, \ggb, and \ggsbi) and can induce variations as large as 40\% on the expected yield~\cite{Cascioli:2013gfa}.

Jet merging uncertainties are evaluated by varying the matching scale (based on the Catani-Krauss-Kuhn-Webber CKKW prescription~\cite{Catani:2001cc}) for the processes simulated with the \SHERPA~generator. Parton-shower uncertainties are evaluated by varying the \SHERPA  showering scheme~\cite{Cascioli:2013gfa}. For those processes simulated with the \PYTHIA~shower program, the uncertainty is assessed by varying the \PYTHIA~configurations, such as the parameter values of the A14 tune, the multi-parton models, and the initial and final-state radiation scales.

The uncertainties due to missing higher-order EW corrections are considered for the main \qqzzllllprod process. The largest missing higher-order component comprises mixed NLO QCD + NLO EW corrections, which are relevant in regions of phase space with large NLO QCD corrections. Following the procedure from Ref.~\cite{Gieseke:2014gka}, the full magnitude of the NLO EW correction is taken as uncertainty for events in which the quadruplet has large recoil. The same prescription was used in histogram-based analysis and further discussion can be found in Ref.~\cite{HIGG-2018-32}. These uncertainties are subleading when compared to the dominant modeling uncertainties related to missing higher-order corrections in $gg\to ZZ$, soft-gluon resummation in $gg\to ZZ$, and jet matching in \qqzzprod.

\begin{table}[tbp]
\centering
\caption{Description of the different sources of modeling systematic uncertainties considered for each process and the method used to estimate it.}
\begin{tabular}{cp{4cm}p{7cm}}
\hline\hline\toprule
\textbf{Process} & \textbf{Uncertainty} & \textbf{Method}\\
\hline
$\qqbar\to ZZ$ & Missing higher-order QCD & Renormalization and factorization scales\\
$\qqbar\to ZZ$ & Soft-gluon resummation & QSF resummation scale\\
$\qqbar\to ZZ$ & Jet merging & CKKW merging scale\\
$\qqbar\to ZZ$ & Parton shower & \SHERPA showering scheme\\
$\qqbar\to ZZ$ & Missing higher-order EW & NLO/LO $K$-factor in regions of high recoil\\
$\qqbar\to ZZ$ & PDF & NNPDF MC replicas\\
EW $qq\to ZZ + 2j$ & Missing higher-order QCD & Renormalization and factorization scales\\
EW $qq\to ZZ + 2j$ & Parton Shower & A14 tune parameters, ISR and FSR scales\\
EW $qq\to ZZ + 2j$ & PDF & NNPDF MC replicas\\
$gg\to ZZ$ & Missing higher-order QCD & Renormalization and factorization scales\\
$gg\to ZZ$ & Soft-gluon resummation & QSF resummation scale\\
$gg\to ZZ$ & Jet merging & CKKW merging scale\\
$gg\to ZZ$ & Parton shower & \SHERPA showering scheme\\
$gg\to ZZ$ & PDF & NNPDF MC replicas\\\bottomrule
\hline
\hline
\end{tabular}
\label{tab:systSummary}
\end{table}


Systematic uncertainties are introduced in the measurement model by nuisance parameters (NP) $\alpha_m$, which modify both the expected event rates $\nu_{\text{X}}(\alpha_m)= G_{X,m}(\alpha_m)\nu_{\text{X}}$ and the probability densities $p_{\text{X}}(x|\alpha_m)= g_{X,m}(x|\alpha_m)p_{\text{X}}(x)$. The different sources of systematic uncertainties are considered to be independent, and lead to the following probability model for each process $\text{X}$:

\begin{equation}
\nu_{\text{X}}(\alpha)p_{\text{X}}(x_i|\alpha) = \left[\prod_m G_{\text{X},m}(\alpha_m)\right]\nu_{\text{X}}\left[\prod_m g_{\text{X},m}(x_i,\alpha_m)\right]p_{\text{X}}(x_i).
\end{equation}

The values of the NPs are constrained by auxiliary measurements that are defined by two auxiliary observables (AO): the central value $a_m$ and the uncertainty $\delta_{\alpha,m}$. The functions $G_{\text{X},m}(\alpha_m)$ are polynomial-exponential interpolations of the expected number of events $\nu_{\text{X}}$ based on simulated samples with NPs varied between $a_{m}+\delta_{\alpha,m}$ and $a_m-\delta_{\alpha,m}$~\cite{Cranmer:2014lly}. The same samples are used to describe the probability density ratios $g_{\text{X},m}(x,\alpha_m=a_m\pm \delta_{\alpha,m})$ using NN estimates for the density ratios $p_{\text{X}}(x|\alpha_m=a_m\pm \delta_{\alpha,m})/p_{\text{X}}(x|\alpha_m=0)$~\cite{atlasML}. In the CR, where no per-event density ratios are used, only per-bin $G$ functions are used. In the SR, the per-bin $G$ functions parametrize the normalization uncertainty, while the per-event $g$ functions parametrize the shape uncertainty as a function of the 14 observables used to describe the event.

For the theoretical modeling uncertainties of $gg\to ZZ\to 4\ell$ production, common NPs are introduced for the \ggs, \ggb, and \ggsbi processes. The same is done for the theoretical modeling uncertainties of EW $qq \to ZZ + 2j\to 4\ell + 2j$ production, where common NPs are used for \ewsbi, \ewsbiten, and \ewb. In the description of theoretical uncertainties in $gg\to ZZ\to 4\ell$, the normalization and shape components are described with different NPs. Since separate normalization parameters are introduced for the $\qqbar\to ZZ\to 4\ell$ processes with $n_{\text{jets}}=0,1,\geq 2$, separate NPs are also used to describe the corresponding modeling systematic uncertainty for the three different processes.

Dedicated NPs are introduced to account for the uncertainty due to the finite number of simulated events in each process and to the natural stochasticity of the NN training. These uncertainties are evaluated by bootstrapping the ensemble members used to estimate the density ratios. Each bootstrapped ensemble is used to infer the signal strengths $\hat{\mu}$ using the maximum likelihood methods described in \Sect{\ref{sec:result}}. The standard deviation $\delta\hat{\mu}$ of the maximum likelihood estimators (MLE) is used as a proxy for the uncertainty arising from the finite number of simulated events, with a correction factor to account for the partial overlap of events in each member of the ensemble~\cite{efron1994introduction}. The associated NP is introduced as a spurious signal uncertainty~\cite{ATL-PHYS-PUB-2020-028} by introducing a new NP $\alpha_{\text{MCstat}}$ and shifting $\mu\to \mu+\alpha_{\text{MCstat}}\delta\hat{\mu}$.


%

\section{Statistical analysis and results}
\label{sec:result}

\begin{figure}[tbp]
\centering
\subfloat[]{\includegraphics[width=0.32\linewidth]{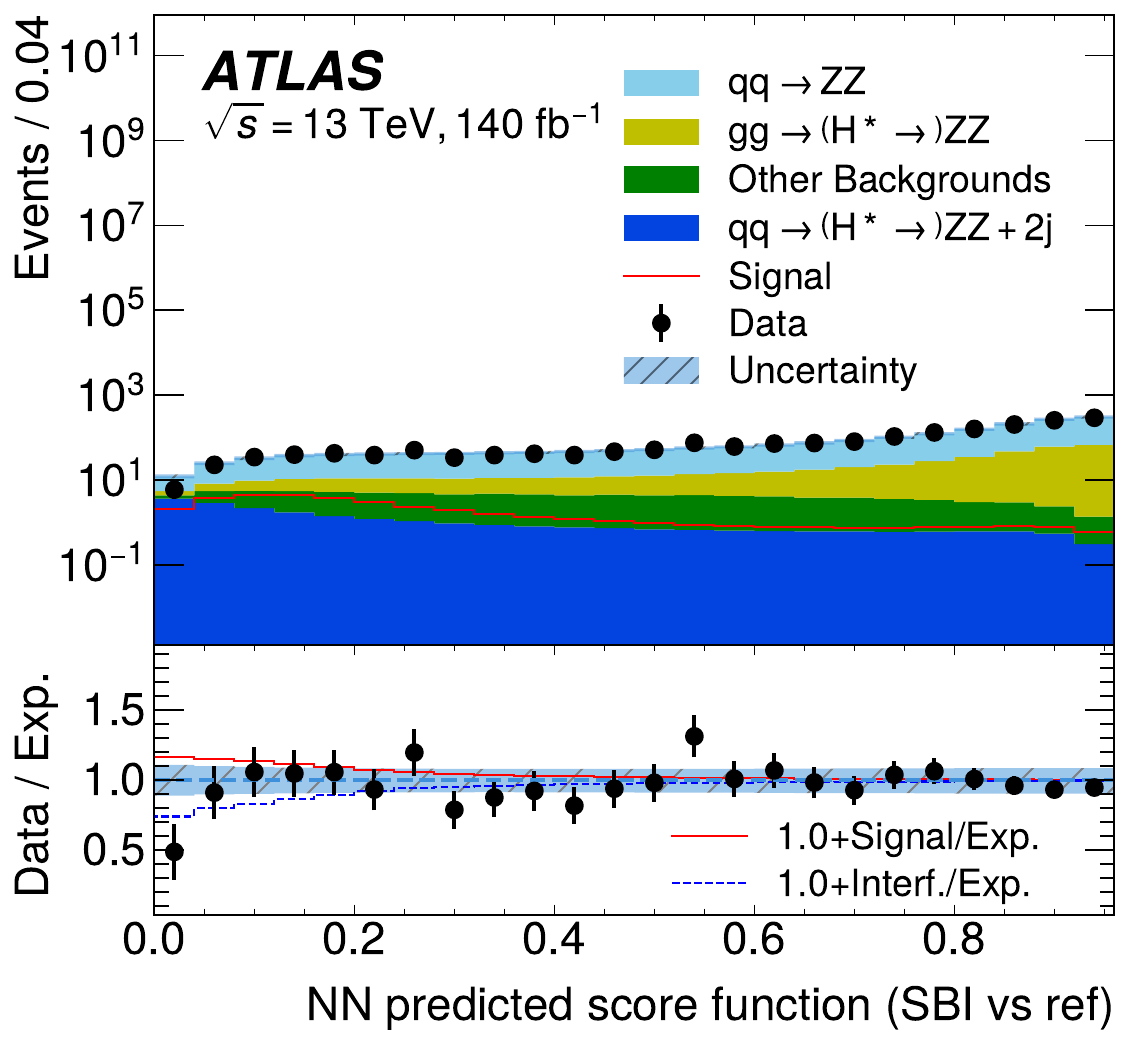}\label{fig:datamc_ratio_sbi}}\hfill
\subfloat[]{\includegraphics[width=0.32\linewidth]{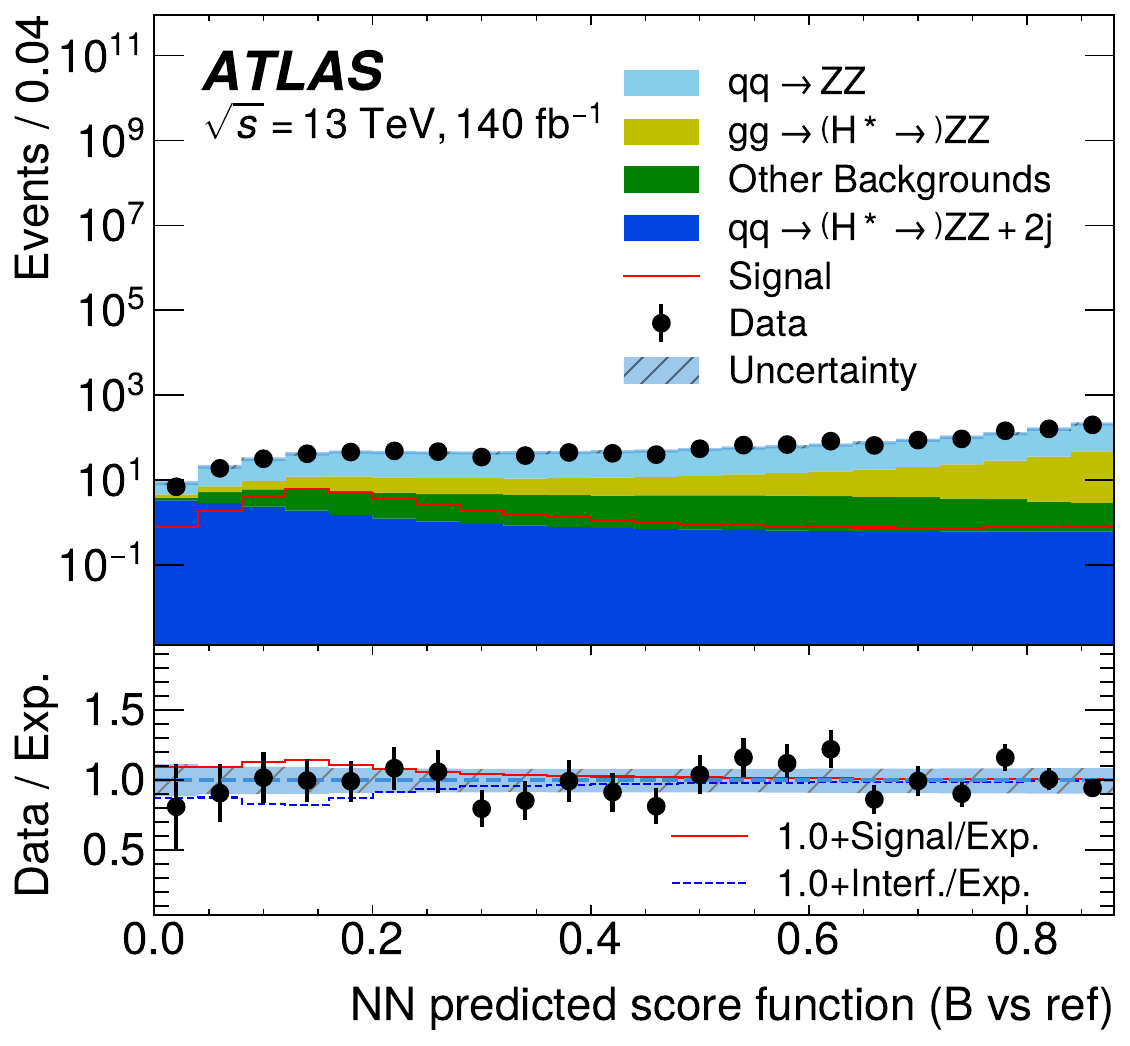}\label{fig:datamc_ratio_s}}\hfill
\subfloat[]{\includegraphics[width=0.32\linewidth]{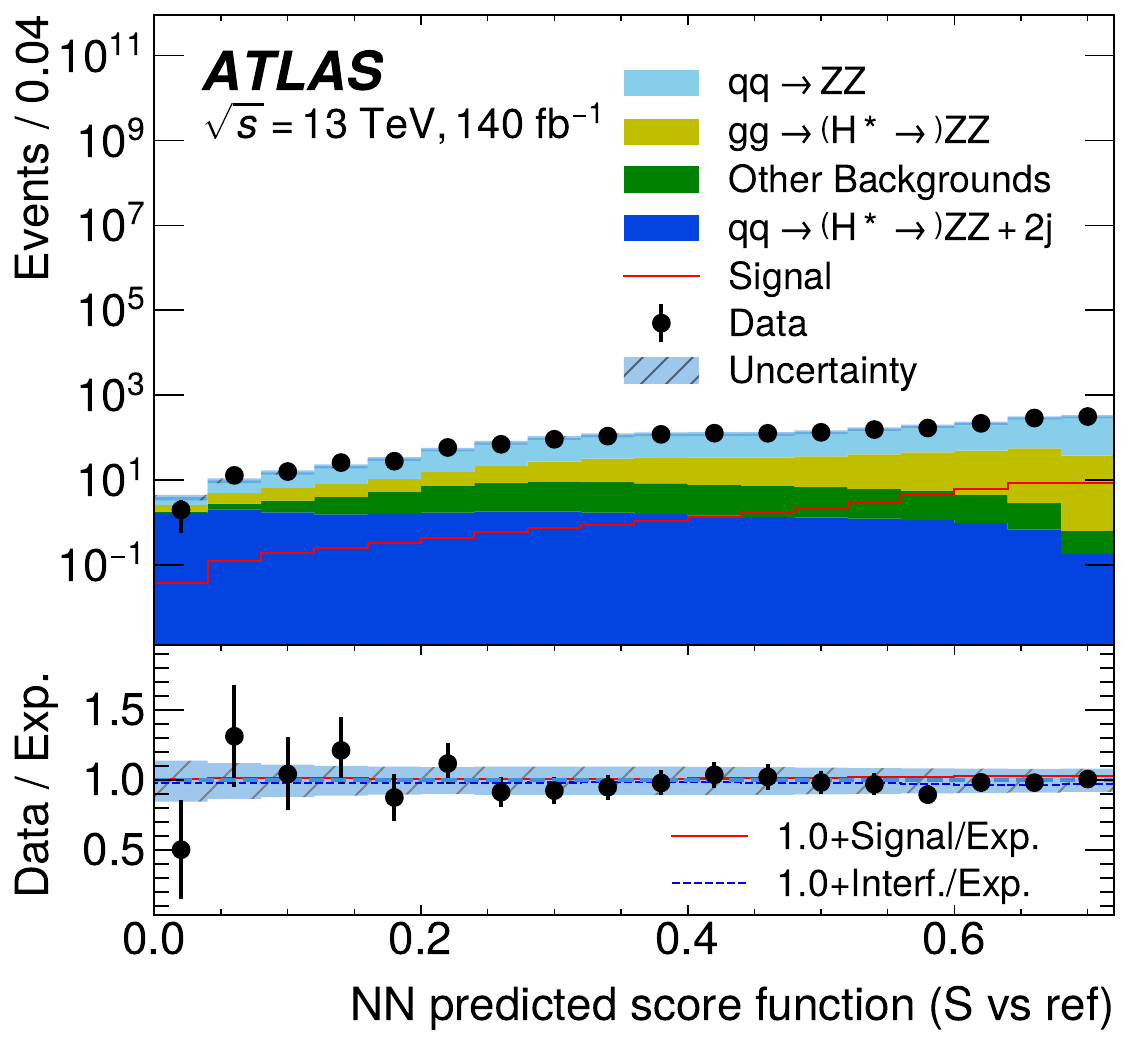}\label{fig:datamc_ratio_b}}\\
\subfloat[]{\includegraphics[width=0.32\linewidth]{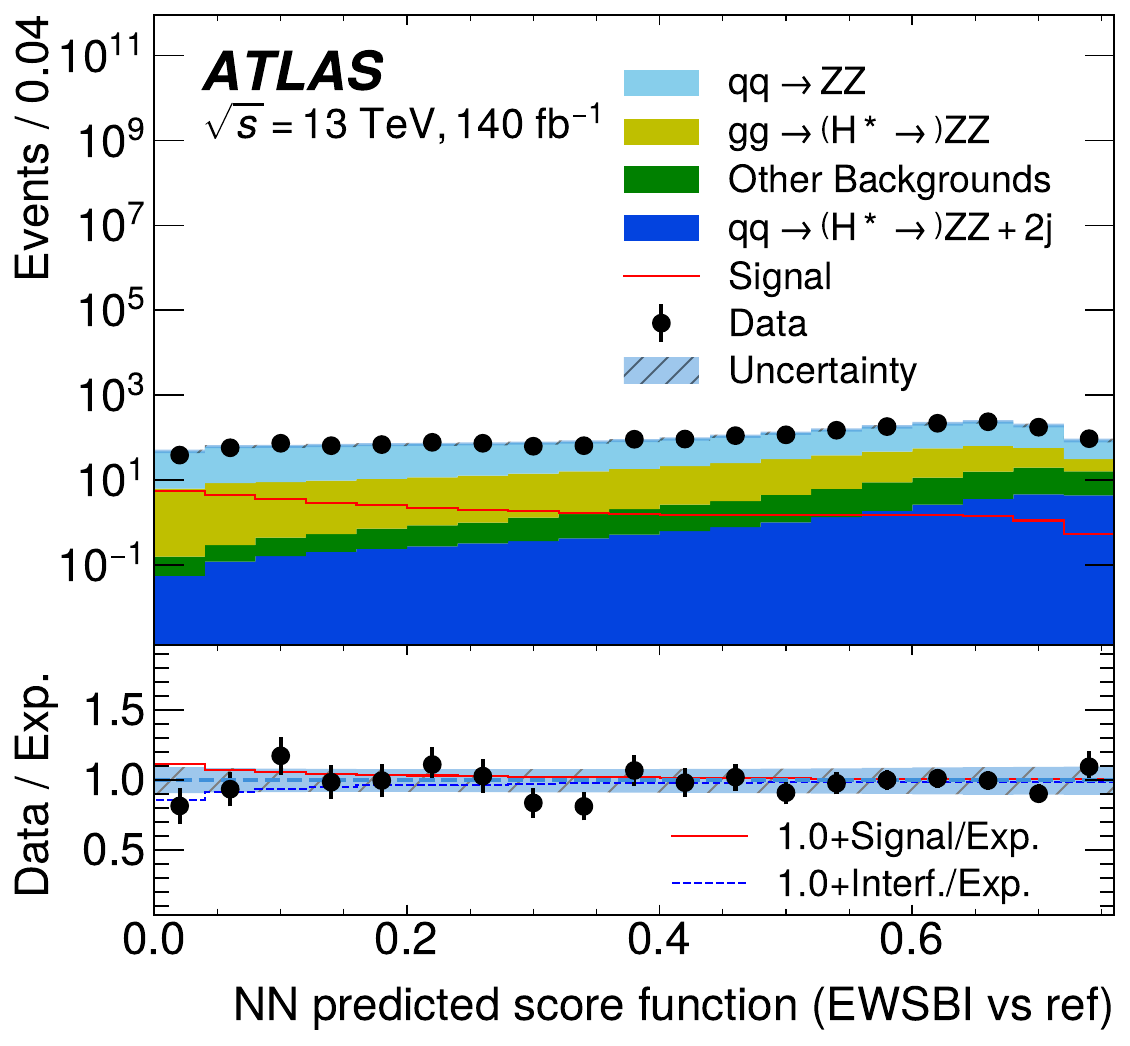}\label{fig:datamc_ratio_ewsbi}}\hfill
\subfloat[]{\includegraphics[width=0.32\linewidth]{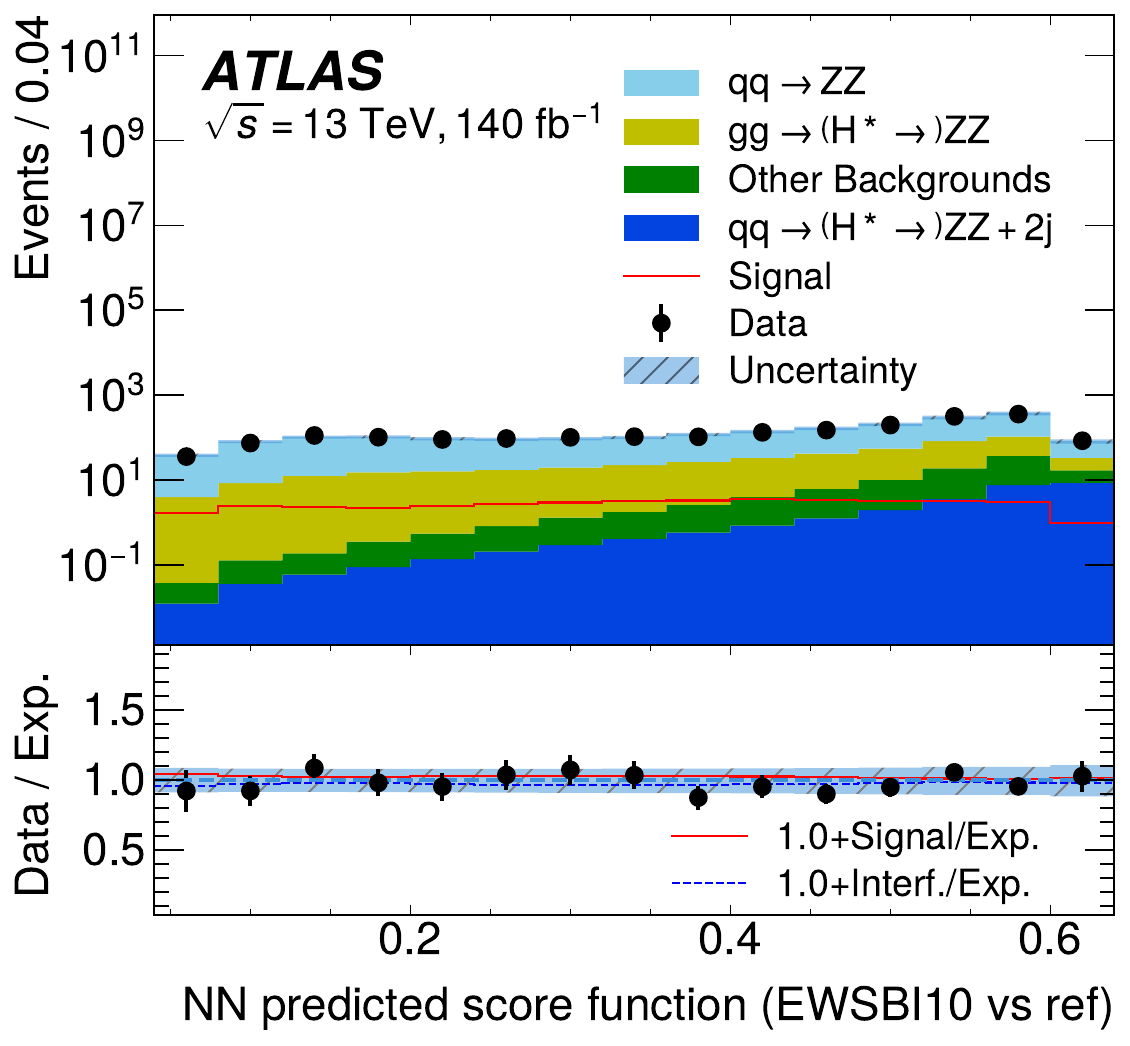}\label{fig:datamc_ratio_ewsbiten}}\hfill
\subfloat[]{\includegraphics[width=0.32\linewidth]{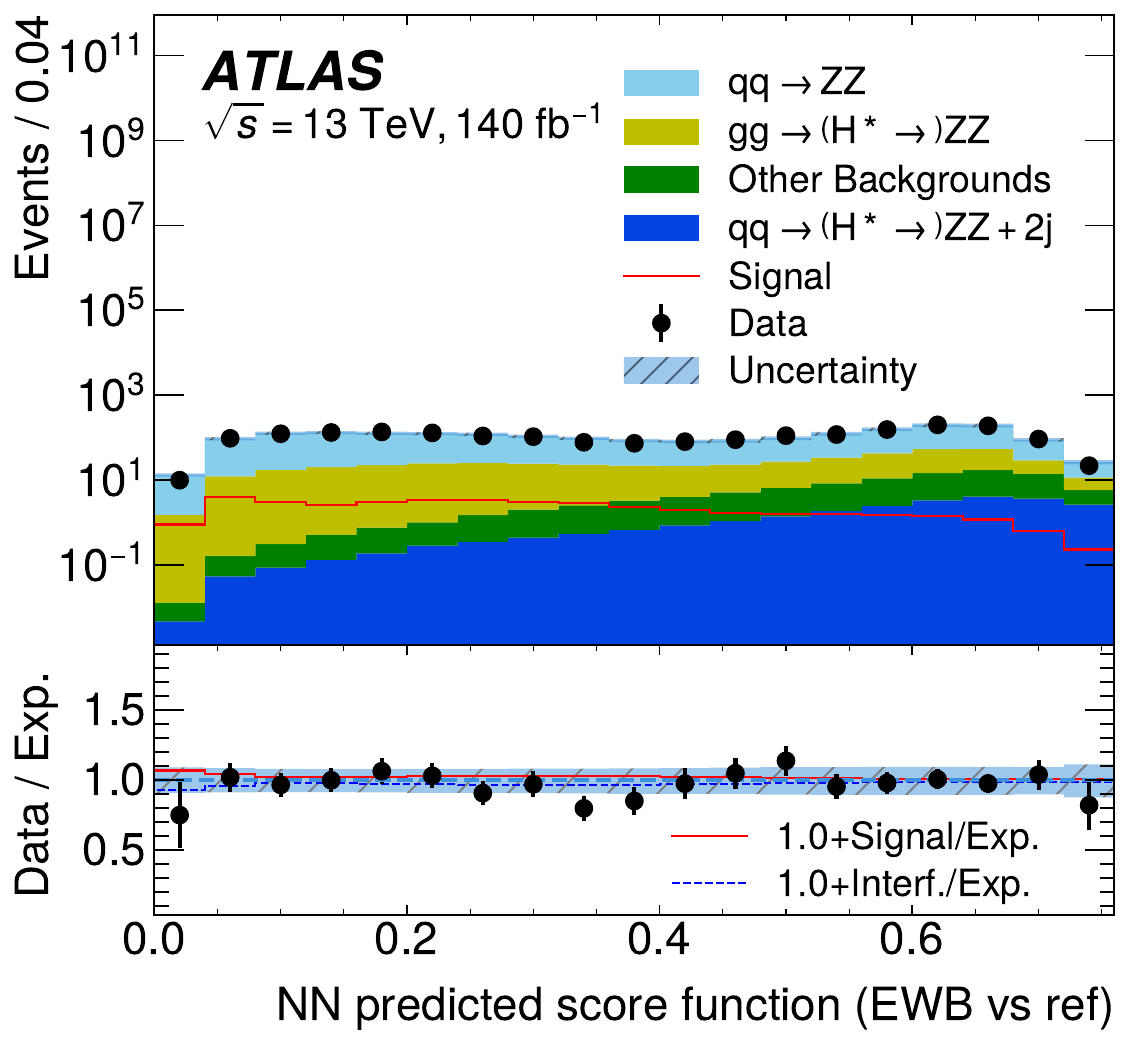}\label{fig:datamc_ratio_ewb}}\\
\subfloat[]{\includegraphics[width=0.32\linewidth]{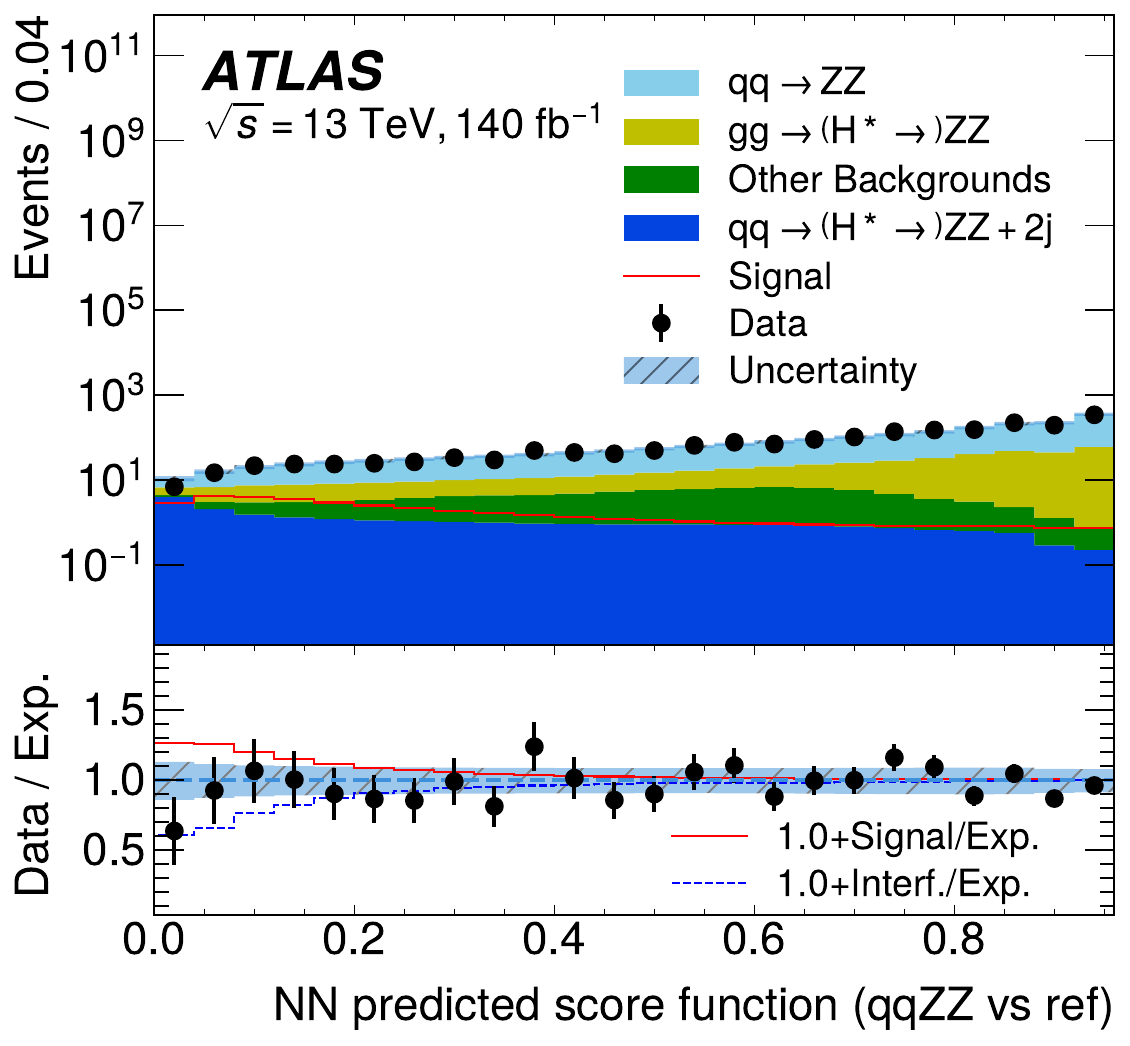}\label{fig:datamc_ratio_qqzz}}\hspace{1cm}
\subfloat[]{\includegraphics[width=0.32\linewidth]{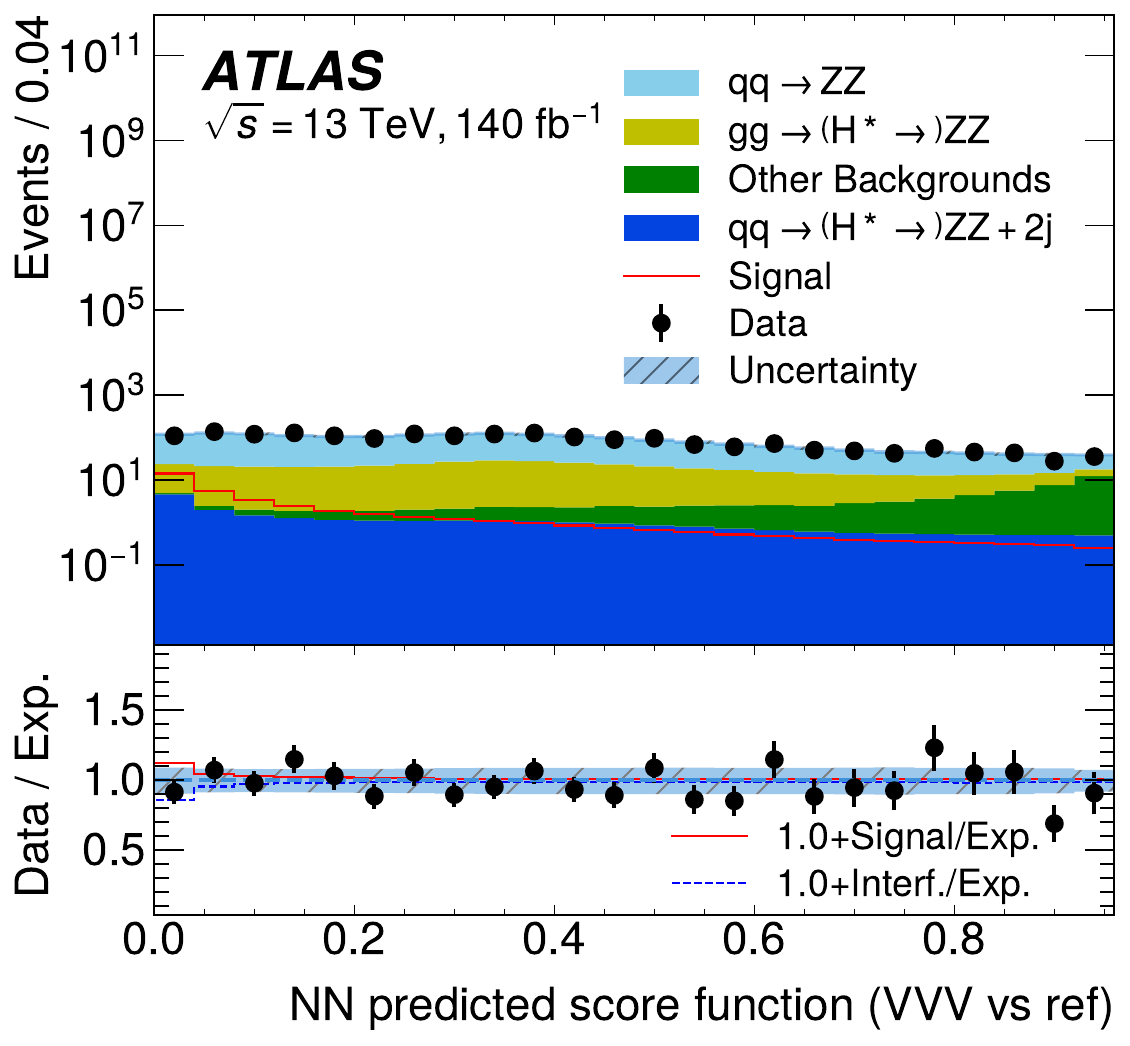}\label{fig:datamc_ratio_vvv}}\\
\caption{Comparison between data and expectation in the SR ($\predisc >-0.85$) of the NN scores of the all major samples used to describe the probability density ratio of the reference process and \protect\subref{fig:datamc_ratio_sbi} $\text{ggF SBI}$, \protect\subref{fig:datamc_ratio_s} $\text{ggF S}$, \protect\subref{fig:datamc_ratio_b} $\text{ggF B}$, \protect\subref{fig:datamc_ratio_ewsbi} \ewsbi, \protect\subref{fig:datamc_ratio_ewsbiten} \ewsbiten,
\protect\subref{fig:datamc_ratio_ewb} \ewb, \protect\subref{fig:datamc_ratio_qqzz} $\qqbar\to ZZ$, and \protect\subref{fig:datamc_ratio_vvv} $VVV$. The comparisons show a good modeling of the probability density ratios throughout the SR\@. The lower panels show the ratio of data to expectation. The expected SM distributions are shown as stacked histograms and the expected signal (interference) is shown as a solid red (dashed blue) line. The background is estimated under the SM hypothesis (post-fit, $\muoff=1$). The hatched band shows the total systematic uncertainty in the expected distribution. The last bin contains overflow events. \label{fig:datamc_ratio}}
\end{figure}

Histograms of the scores \scorex used to estimate the density ratios in the SR are shown in \Fig{\ref{fig:datamc_ratio}} for all processes in this analysis indicating a good modeling of the probability density ratios throughout the SR\@. A single histogram is shown for $s_{q\bar{q}ZZ}$ since a single NN ensemble is used for the three $q\bar{q}ZZ$ processes (with $n_{\text{jets}}=0$, $1$, and $\geq 2$). This is possible without loss of information because the observable that distinguishes the three processes ($n_{\text{jets}}$) is part of the vector $x$ that describes the event, and therefore:

\begin{equation}
\nu_{\qqbar ZZ}^{n_{\text{jets}}=i}\;\frac{p_{\qqbar ZZ}^{n_{\text{jets}}=i}(x)}{p_{\text{ref}}(x)}=\delta_{n_{\text{jets}}, i}\;\nu_{\qqbar ZZ}\;\frac{p_{\qqbar ZZ}(x)}{p_{\text{ref}}(x)},
\end{equation}
where $\delta_{n_{\text{jets}}, i}$ is the Kronecker delta.

The NN-based density ratios allows an unbinned analysis to be performed. The data are analyzed using the
likelihood function $\lambda(\mu,\theta,\alpha)$:
\begin{equation}\label{eq:llratio}
\begin{split}
-2\ln\lambda(\mu,\theta,\alpha) =-2&\sum_{\text{regions }(I)}\ln\left[\text{Pois}(N_I|\nu_I(\mu,\theta,\alpha))\right]\\&\qquad\qquad -2\sum_{\text{events }(i)}\ln \left[\frac{p(x_i|\mu,\theta,\alpha)}{p_{\text{ref}}(x_i)}\right]
+\sum_{\text{systematics }(m)} (\alpha_m-a_m)^2.
\end{split}
\end{equation}
The first term in the sum corresponds to the Poisson probability of observing a total of $N_I$ events in a region $I$ with $\nu_I(\mu,\theta,\alpha)$ expected events. The second term is the ratio of the probability density of the dataset $\{x_i\}$ for a hypothesis given by parameters $(\mu,\theta,\alpha)$ and the probability density of the same dataset for the fixed reference hypothesis. The last term in the sum represents Gaussian constraints for the NPs $\alpha_m$ from auxiliary measurements with value given by the AOs $a_m$. The values of the NPs and AOs are considered to be normalized by the value of the uncertainty of the corresponding auxiliary measurement, so that the width of the Gaussian density can always be taken as unity.

Four regions are considered in the sum of Poisson terms. The SR (defined as $\predisc  > -0.85$), and the CR (defined by $\predisc  \leq -0.85$) that is further divided into three $n_{\text{jets}}$ bins: 0, 1, and $\geq 2$ to provide enough constraining power for the $\theta_{\qqbar ZZ}^{\text{incl},1,2j}$ background normalization parameters, which are treated as unconstrained NPs, i.e., they have no corresponding auxiliary measurements that constrain them.

\begin{figure}[tbp]
\centering
\includegraphics[width=0.5\textwidth]{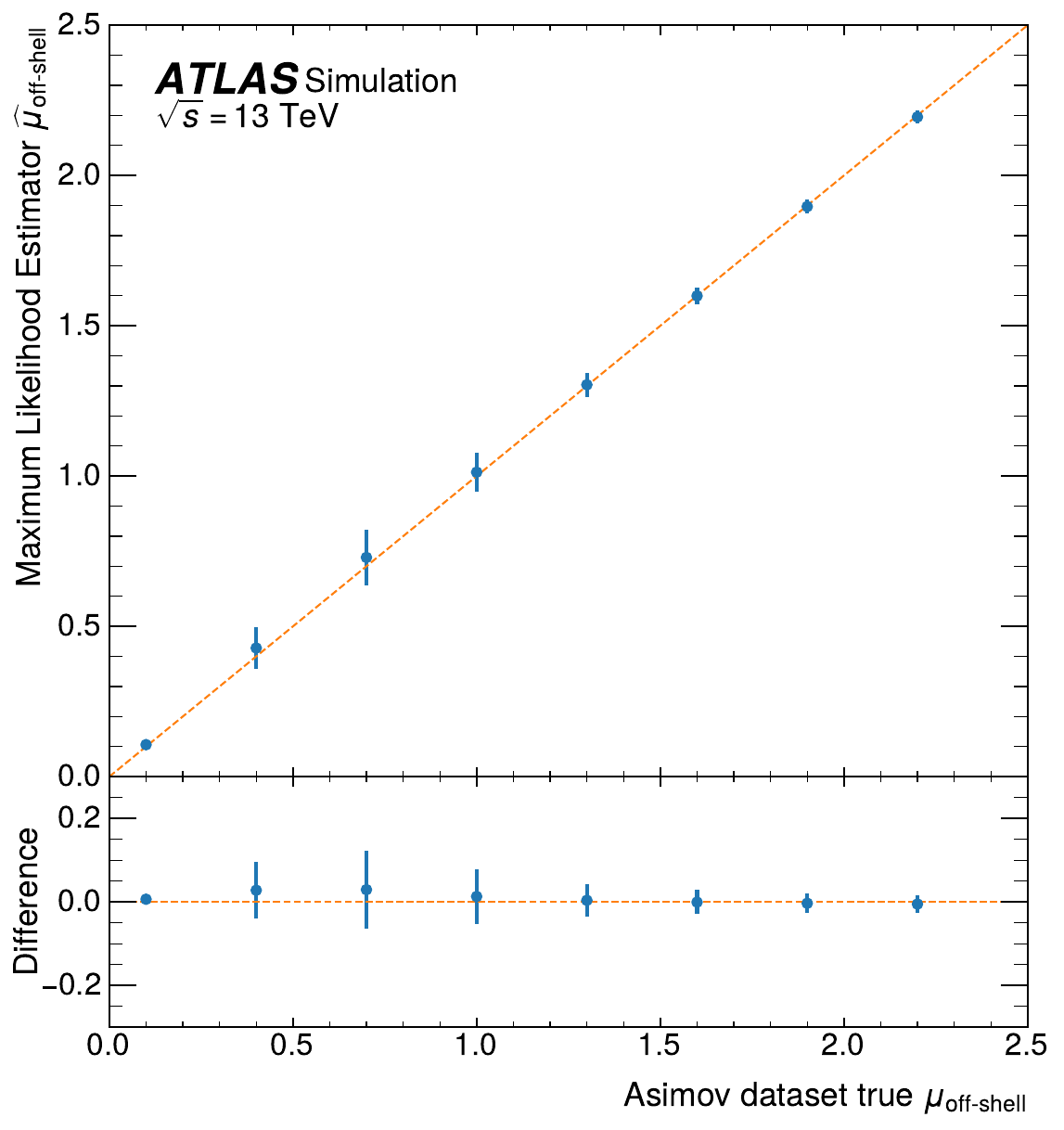}
\caption{Maximum likelihood estimator $\widehat{\mu}_{\text{off-shell}}$ for different Asimov pseudo-data with known value of $\mu^{\text{truth}}_{\text{off-shell}}$. The error bar shows the estimate of the uncertainty from the limited number of MC events in the Asimov dataset, which is introduced as a spurious signal uncertainty.}
\label{fig:mleclosure}
\end{figure}

Only events in the SR are used in the sum over events, as defined in \Eqn{\ref{eq:llratio}}. To simplify the notation, all constrained and unconstrained NPs are collectively denoted by $\alpha$ hereafter. The test statistic used in this measurement is the profile log-likelihood ratio, traditionally used in LHC measurements~\cite{Cranmer:2014lly}:

\begin{equation}
t_{\mu} = -2\ln\frac{\lambda(\mu,\dhatalpha(\mu))}{\lambda(\hatmu,\hatalpha)},
\end{equation}
where $(\hatmu,\hatalpha)$ are the parameters estimates that maximize the function $\lambda(\mu,\alpha)$ and $\dhatalpha(\mu)$ is a parameter estimate that conditionally maximizes the function $\lambda(\mu,\alpha)$ for a given $\mu$:

\begin{equation}    (\widehat{\mu},\widehat{\alpha})=\underset{\mu,\alpha}{\text{argmax }} \lambda(\mu,\alpha);\qquad
\dhatalpha(\mu)=\underset{\alpha}{\text{argmax }} \lambda(\mu,\alpha).
\end{equation}
The term $\sum_{\text{events}}\ln(\probref)$ in the denominator of \Eqn{\ref{eq:llratio}} cancels in the ratio used to define the test statistic \tmu since it is independent of any parameter $(\mu,\alpha)$. \Fig{\ref{fig:mleclosure}} shows the results of a closure test of the \tmuoff\ test statistic: the value of the estimate $\hatmu_{\text{off-shell}}$ for an Asimov sample with a true value of \muoff\, for a wide range of \muoff\ values. Closure is observed for all values of \muoff, within the statistical uncertainty of the simulation samples used. An additional test was performed replacing the $\qqbar \to ZZ$ \SHERPA sample in the Asimov data by an alternative simulation done with \POWHEG. The hard scattering ME in both samples is calculated at the same perturbative order in QCD, but the samples have different parton shower matching and simulation. Closure of the \muoff\ MLE is also observed with the alternative Asimov sample.

\begin{figure}[tbp]
\centering
\subfloat[]{\includegraphics[width=0.46\linewidth]{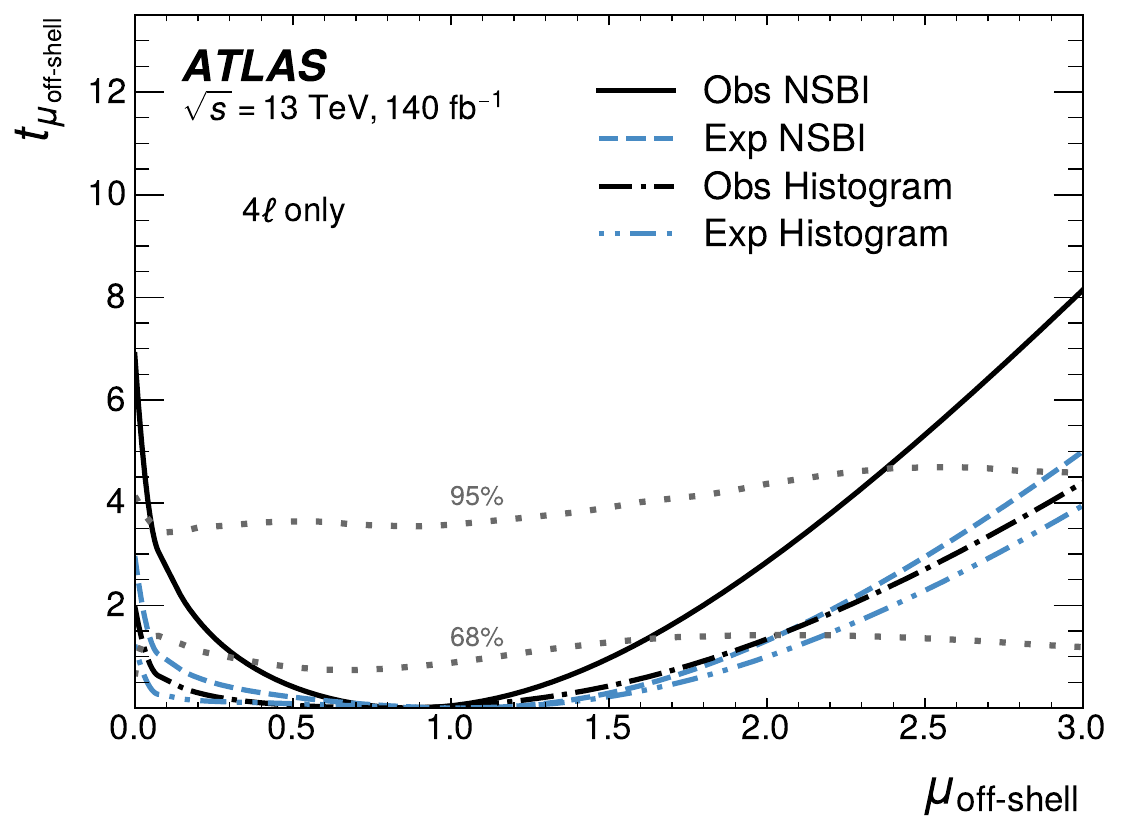}\label{fig:muoffshell_4l_results_histo}}\hfill
\subfloat[]{\includegraphics[width=0.46\linewidth]{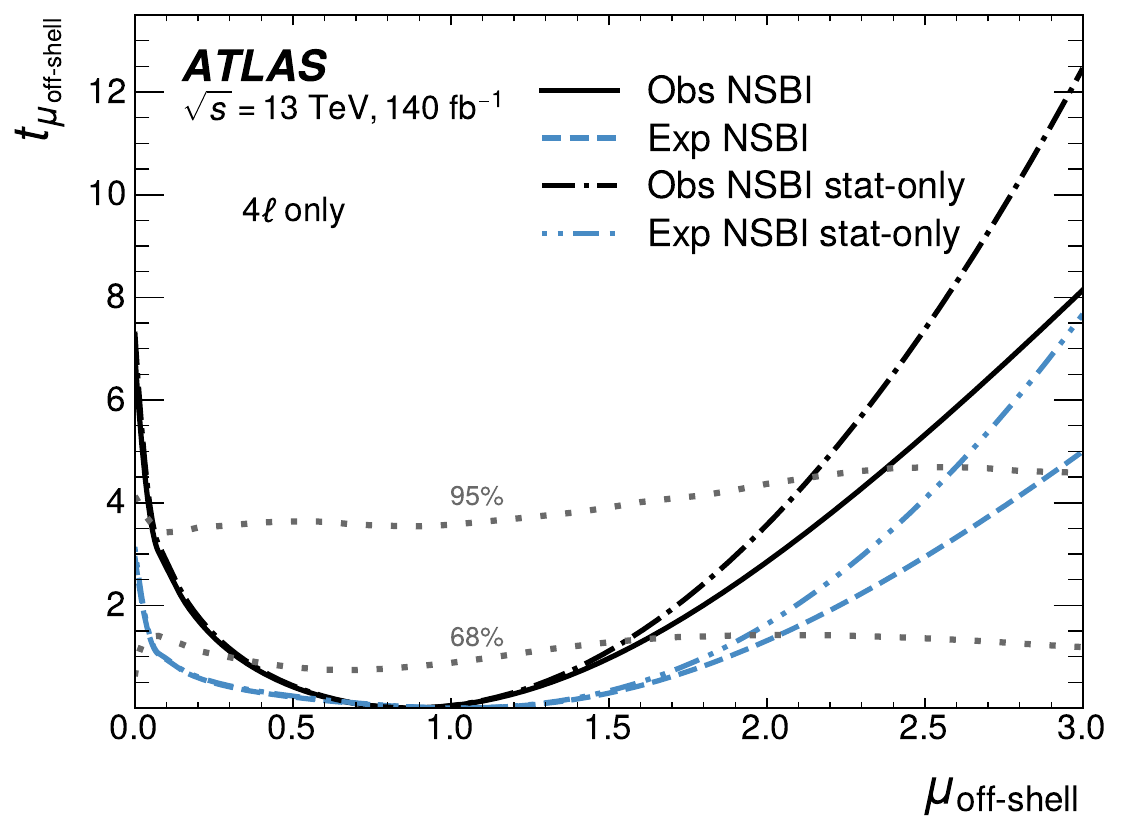}\label{fig:muoffshell_4l_results_stat}}\\
\caption{\protect\subref{fig:muoffshell_4l_results_histo} Values of the test statistic \tmuoff\ assuming a single parameter of interest \muoff\ obtained with an Asimov dataset (expected, dashed blue) and with data (observed, solid black) in the \ohtollll decay channel. The values from the histogram-based analysis~\cite{HIGG-2018-32} are added in dash-dotted lines for comparison. The dotted gray lines show the 68\% and 95\% confidence belt, obtained from the Neyman construction. \protect\subref{fig:muoffshell_4l_results_stat} Same values obtained with data (observed, solid black) and Asimov dataset (expected, dashed blue) compared with the \textit{statistics-only} case with all NP fixed at their best-fit values \hatalpha.\label{fig:muoffshell_4l_results}}
\end{figure}

A total of 127 NPs are used in the measurement, and no significant differences between the MLE estimates $\hatalpha_m$ and the values of the auxiliary measurements $a_m$ are observed. The observed and expected values of the profile likelihood ratios \tmuoff\ are shown in \Fig{\ref{fig:muoffshell_4l_results}} as functions of \muoff\ for a variety of scenarios, assuming $\muoff = \okappag^2\okappaV^2=\okappaV^4$. \Fig{\ref{fig:muoffshell_4l_results_histo}} compares the profile likelihood ratio for the histogram-based~\cite{HIGG-2018-32} and the NSBI-based analyses, and shows the improved constraints on \muoff\ obtained with the latter. Additional comparisons with the histogram-based analysis are shown in \App{\ref{sec:histo_comparison}}. \Fig{\ref{fig:muoffshell_4l_results_stat}} compares the profile likelihood ratio for the NSBI-based analysis to a variant where the NPs are fixed to the best-fit value $\widehat{\alpha}$, reflecting only the statistical uncertainty on the data. The comparison indicates that systematic uncertainties are more important for tests of signal-dominated hypotheses ($\muoff > 1$) than for tests of interference-dominated hypotheses ($\muoff < 1$).

As is the case for the histogram-based analysis, the test statistic is not distributed as a $\chi^2$ probability density, due to the double minima created by the interference terms and due to the constraint $\mu_{\text{off-shell}}>0$ imposed by the model of \Eqn{\ref{eq:measurement_model}}. Confidence intervals are built using the Neyman construction~\cite{10.2307/91337} (NC) instead of relying on the asymptotic approximation. In the NC, pseudo-experiments are built by performing a Poisson bootstrapping on the high-statistic reference sample~\cite{poisboot,ATL-PHYS-PUB-2020-028}. In each pseudo-experiment, the bootstrapped weight of an event $x_i$ with weight $w_i$ in the reference sample is sampled from a Poisson distribution with expected value $w_i(\nu(\mu,\hatalpha)/\yieldref)(p(x_i|\mu,\hatalpha)/\probref)$. The value of the AO $a_m$ associated with each constrained NP $\alpha_m$ is sampled from a Gaussian probability density $\text{Gaus}(a_m|\widehat{\alpha}_m,1)$~\cite{Cranmer:2014lly}.

\begin{figure}[tbp]
\centering
\includegraphics[width=0.53\linewidth]{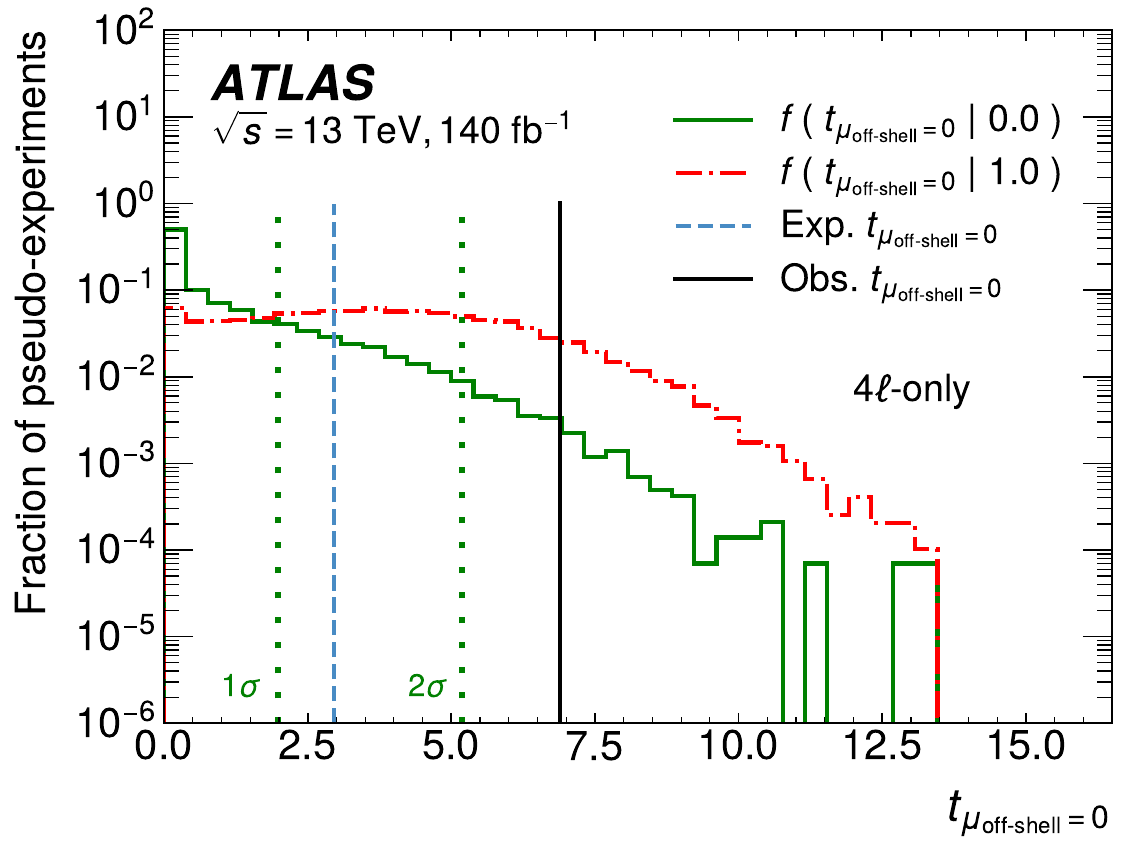}
\caption{Expected distribution of $t_{\mu_{\text{off-shell}=0}}$ estimated with pseudo-experiments for the case of $\muoff^{\text{truth}}=0$ (solid green, no off-shell Higgs boson production hypothesis) and $\muoff^{\text{truth}}=1$ (dash-dotted red, SM hypothesis). The vertical solid black (dashed blue) line shows the observed (expected) value of $t_{\mu_{\text{off-shell}=0}}$. The dotted vertical green lines at $t_{\mu_{\text{off-shell}=0}}=1.99$ and $5.18$ represent the one-sided $1\sigma$ and $2\sigma$ significance thresholds, respectively, under the $\muoff^{\text{truth}}=0$ hypothesis. These values differ slightly from the ones in the asymptotic approximation, in which the thresholds would be $1.97$ and $5.19$.}
\label{fig:results_nc}
\end{figure}

\Fig{\ref{fig:results_nc}} shows the expected distribution of $t_{\mu_{\text{off-shell}}=0}$ for the SM hypothesis ($\mu^{\text{truth}}_{\text{off-shell}}=1$) and the \textit{no off-shell Higgs boson hypothesis} ($\mu^{\text{truth}}_{\text{off-shell}}=0$). The black and dashed blue vertical lines indicate the observed and expected values of $t_{\mu_{\text{off-shell}}=0}$, respectively.The red dash-dotted curve in \Fig{\ref{fig:results_nc}} shows the expected distribution of $t_{\mu_{\text{off-shell}}=0}$ assuming the SM hypothesis ($\mu_{\text{off-shell}}^{\text{truth}}=1$). The $p$-value of the observed value of $t_{\mu_{\text{off-shell}}=0}$ under the SM hypothesis is 0.11, corresponding to one-sided significance of $1.2\sigma$. The green solid curve shows the shows the expected distribution of $t_{\mu_{\text{off-shell}}=0}$ assuming $\mu_{\text{off-shell}}^{\text{truth}}=0$ (no off-shell Higgs boson hypothesis). The green dotted lines show the $p$-value thresholds corresponding to the one-sided significance of $1\sigma$ and $2\sigma$ under this hypothesis. The evidence for off-shell Higgs boson production has an observed (expected) significance of $\muofffourlobssig\sigma$ ($\muofffourlexpsig\sigma$) using only the \ohtollll decay channel. The evidence for off-shell Higgs boson hypothesis has a larger significance than the one observed (expected) in the previous histogram-based analysis~\cite{HIGG-2018-32} of the same dataset, which had a value of $\muofffourlhistoobssig\sigma$ ($\muofffourlhistoexpsig\sigma$).

\begin{figure}[tbp]
\centering
\subfloat[]{\includegraphics[width=0.477\linewidth]{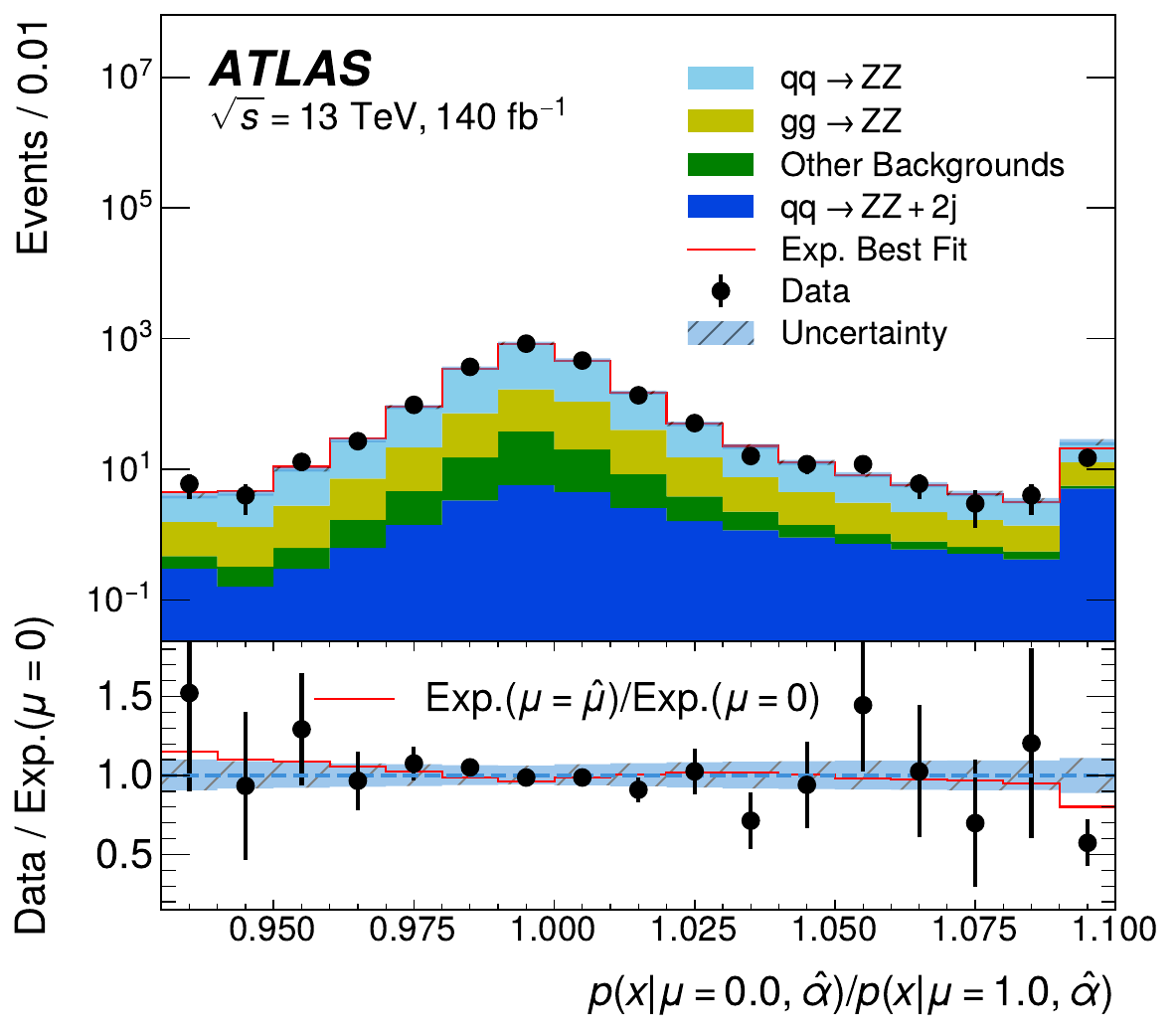}\label{fig:results_dist_oo}}\hfill
\subfloat[]{\includegraphics[width=0.457\linewidth]{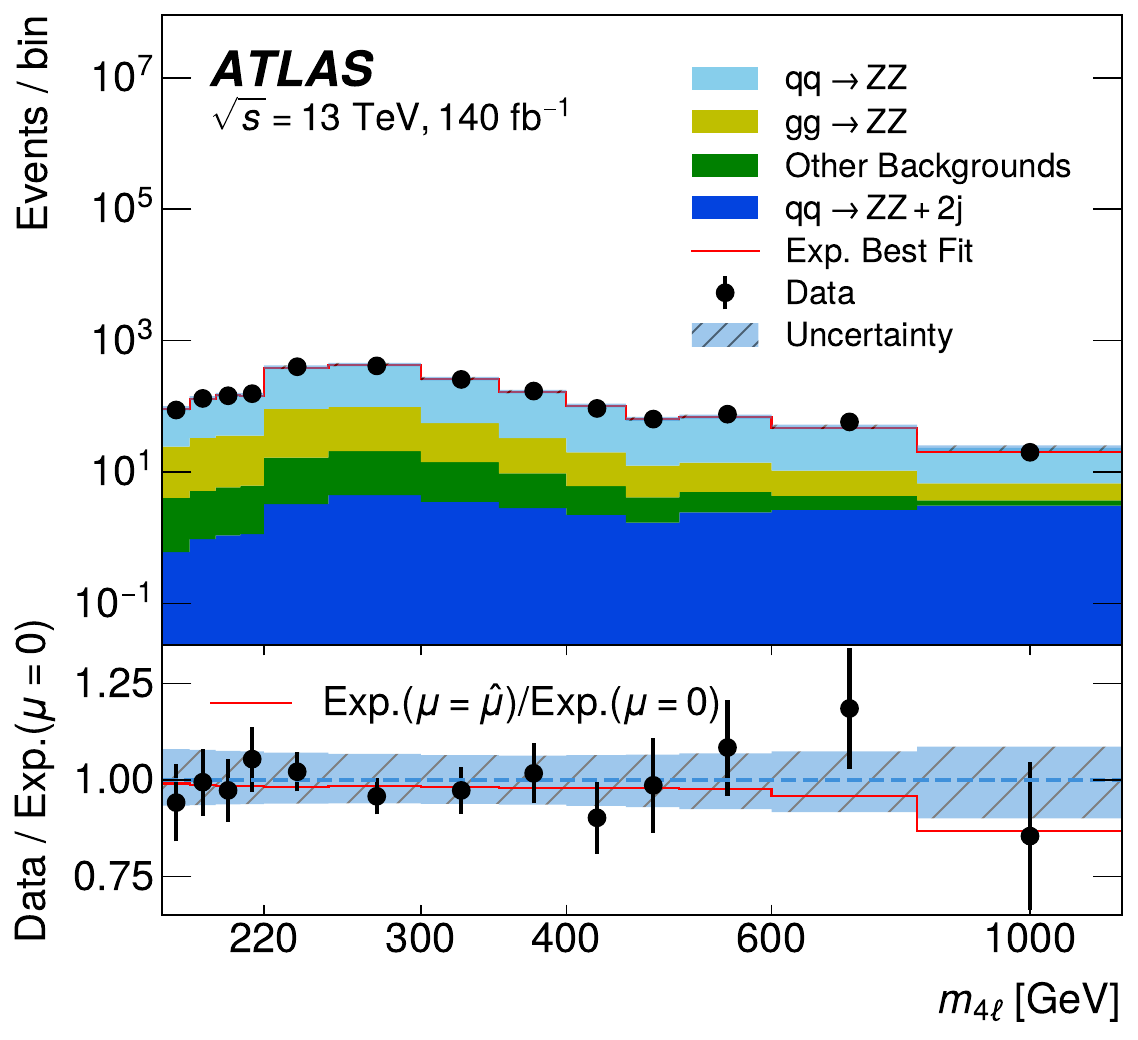}\label{fig:results_dist_m4l}}
\caption{Comparison between observed and the expected background distributions of \protect\subref{fig:results_dist_oo} the optimal observable at $\muoff=0$ and \protect\subref{fig:results_dist_m4l} the quadruplet mass $m_{4\ell}$. The solid red lines shows the expected distribution of the best-fit hypothesis $\hatmu$. The lower panel shows a comparison between the distribution of the background-only and best-fit hypotheses. The background is estimated under the SM hypothesis (post-fit, $\muoff=1$). A comparison between the two distributions indicate that the optimal observable built with the NSBI method provides better evidence for off-shell Higgs boson production than only the $m_{4\ell}$ distribution. The hatched area corresponds to the total systematic uncertainty in the expected distributions. The first and last bins contain overflow events.}
\end{figure}

\Fig{\ref{fig:results_dist_oo}} shows the distribution of the probability density ratio $p(x|\mu_{\text{off-shell}}=0,\widehat{\alpha})/p(x|\mu_{\text{off-shell}}=1,\widehat{\alpha})$, which is a optimal observable for $\mu_{\text{off-shell}}=0$. The lower panel shows a comparison with the distribution from the best-fit hypothesis depicting the data behavior that leads to the observed exclusion of the no off-shell Higgs boson hypothesis. The difference between the observed and expected values of $t_{\muoff=0}$ indicates that there are regions of phase-space where the data are more interference-like than signal-like, as can be seen by the rightmost bins in \Fig{\ref{fig:results_dist_oo}} where the observed deficit of events is larger than expect. Given the small significance of the difference between expected and observed values of $t_{\muoff=0}$, it is difficult to isolate a specific region of phase-space with this behavior.

\Fig{\ref{fig:results_dist_m4l}} shows the distribution of the quadruplet mass $m_{4\ell}$ and, in the lower panel, the comparison with the best-fit hypothesis.  This indicates that the quadruplet mass information alone would not be enough to obtain evidence of off-shell Higgs boson production in this channel and illustrates the importance of the ME-based analysis performed with the NSBI method. Further descriptions of the optimal observables can be found in \App{\ref{sec:oo_construction}}.

Two methods are used to estimate the sensitivity of the measurement to different systematic uncertainties. The two methods differ in what is varied: either the NPs $\alpha_m$, or of the AOs $a_m$ associated with these NPs~\cite{Pinto:2023yob}. When using the variations of NPs, each parameter $\alpha_m$ is varied by its uncertainty and the conditional maximum likelihood estimate $\dhatmu$ is re-derived with only that NP fixed. On the other hand, when varying AOs, each observable $a_m$ is set to $\pm 1$ and the unconditional maximum likelihood estimate $\widehat{\mu}$ is re-derived without fixing any NP. In both methods, the difference between the new and original estimate $\widehat{\mu}$ is taken as the propagated uncertainty.

The results of both methods are summarized in \Tab{\ref{tab:syst_impact}}. The propagated uncertainties are summed in quadrature within each group. The sum in quadrature does not take into account correlations between the different parameters $\widehat{\alpha}_m$ when using the method based on the variation of NPs. The same problem does not exist when varying AOs: the sum in quadrature can be made without loss of information and it can be used to provide a consistent decomposition of the total uncertainty into statistical and systematic uncertainties. In both methods, the statistical uncertainty is obtained by the square root of the difference of the total uncertainty squared and the systematic uncertainty squared.

The largest contributions to the measurement uncertainty are the statistical uncertainty on the data, the theoretical modeling uncertainties, and uncertainties on the jet energy scale and resolution. The contribution of MC statistical uncertainty to the total uncertainty is less than 0.01. The observed (expected) value of $\mu_{\text{off-shell}}$ at 68\% CL is:

\begin{equation}\nonumber
\mu_{\text{off-shell}} = \muofffourlnsbi^{+\muofffourlnsbiobsoneup}_{-\muofffourlnsbiobsonedn} \;\;(1.00^{+\muofffourlnsbiexponeup}_{-\muofffourlnsbiexponedn}).
\end{equation}
The fitted values of $\qqzzzj$, $\qqzzoj$, and $\qqzztj$ are $1.12\pm 0.04$, $0.85\pm 0.05$, and $0.90\pm 0.07$, respectively. The result presented here has a reduced uncertainty when compared with the previous histogram-based analysis observed value at 68\% CL of $\muofffourlhisto^{+\muofffourlhistoobsoneup}_{-\muofffourlhistoobsonedn}$ (expected $\muoff<\muofffourlhistoexponeup$).

\begin{table}[tbp]
\centering
\caption{Absolute systematic uncertainties in the measurement of $\mu_{\text{off-shell}}$ in the \htollll decay channel. Two methods of estimation are presented: based on the variation of nuisance parameters and on the variation of auxiliary observables. Uncertainties are given using the auxiliary observables methods since it allows variations to be summed in quadrature. The total uncertainty is independent of the method used to estimate systematic uncertainties.}
\begin{tabular}{lcc}\hline\hline
\toprule\textbf{Uncertainty source} &  \multicolumn{2}{c}{\textbf{Absolute impact on \muoff }}\\
& \textbf{Nuisance Parameter} & \textbf{Auxiliary Observable}\\\hline  && \\[-2.0ex]
Electron uncertainties & $(-0.05, +0.06)$ & $(-0.05, +0.06)$  \\[1.1ex]
Muon uncertainties & $(-0.03, +0.03)$ & $(-0.02, +0.03)$  \\[1.1ex]
Jet uncertainties & $(-0.10, +0.10)$ & $(-0.09, +0.11)$  \\[1.1ex]
Luminosity & $(-0.01, +0.01)$ & $(-0.01, +0.01)$ \\[1.2ex]\hline
Total experimental & $(-0.12, +0.13)$ & $(-0.11, +0.12)$  \\[1.1ex]\hline
$\qqbar \to ZZ$ modeling & $(-0.06, +0.07)$ & $(-0.06, +0.07)$  \\[1.1ex]
$gg\to ZZ$ modeling & $(-0.08, +0.13)$ & $(-0.07, +0.09)$  \\[1.1ex]
EW $\qqbar\to ZZ+2j$ modeling & $(-0.01, +0.01)$ & $(-0.01, +0.01)$ \\[1.1ex]\hline
Total modeling & $(-0.10, +0.15)$ & $(-0.09, +0.12)$  \\[1.1ex]\hline
\textbf{Systematic uncertainty} & $(-0.16, +0.19)$ & $(-0.14, +0.17)$ \\[1.1ex]
\textbf{Statistical uncertainty} & $(-0.49, +0.72)$ & $(-0.50, +0.73)$ \\\hline
\textbf{Total uncertainty} & \multicolumn{2}{c}{$(-\muofffourlnsbiobsonedn, +\muofffourlnsbiobsoneup)$}\\\bottomrule\hline\hline
\end{tabular}
\label{tab:syst_impact}
\end{table}



\subsection{Combination with the analysis in the \texorpdfstring{$2\ell 2\nu$}{2l2nu} decay channel}
\label{sec:combination}

The new result presented in this paper is combined with the most recent ATLAS off-shell Higgs boson production measurement in the \ohtollnunu decay channel~\cite{HIGG-2018-32}. The histogram-based \ohtollnunu analysis uses the binned transverse mass \mTZZ\ distribution in three different SRs, in addition to the yield of four CRs enriched in $\Zboson+\text{jets}$, non-resonant $e\mu$ events, and $q\bar{q}\to \Wboson\Zboson$ events. The test statistic used for the combination is built from the log-likelihood ratio in \Eqn{\ref{eq:llratio}}, where now the sum over regions also includes the several \mTZZ\ bins and CRs of the analysis in the \ohtollnunu decay channel.

The systematic uncertainty model used here is expanded to include constrained and unconstrained NPs exclusive to the analysis of the \ohtollnunu decay channels. Experimental and common theory uncertainties, i.e., pertaining to $gg\to ZZ$ and $\qqbar \to ZZ$ modeling, are modeled with common NPs that modify both likelihood components. Uncertainties related to missing higher-order EW corrections in the \qqzzprod process, which were estimated with different methods in both analyses, are modeled with separate NPs for the \ohtollll and \ohtollnunu analyses. However, the measured off-shell Higgs boson production is largely insensitive to this modeling choice. Common \qqzzprod data-driven normalization parameters \qqzzzj, \qqzzoj, and \qqzztj are used in the two channels since the phase space probed is similar within each $n_{\text{jets}}$ bin.

\begin{figure}[tbp]
\centering
\subfloat[]{\includegraphics[width=0.48\linewidth]{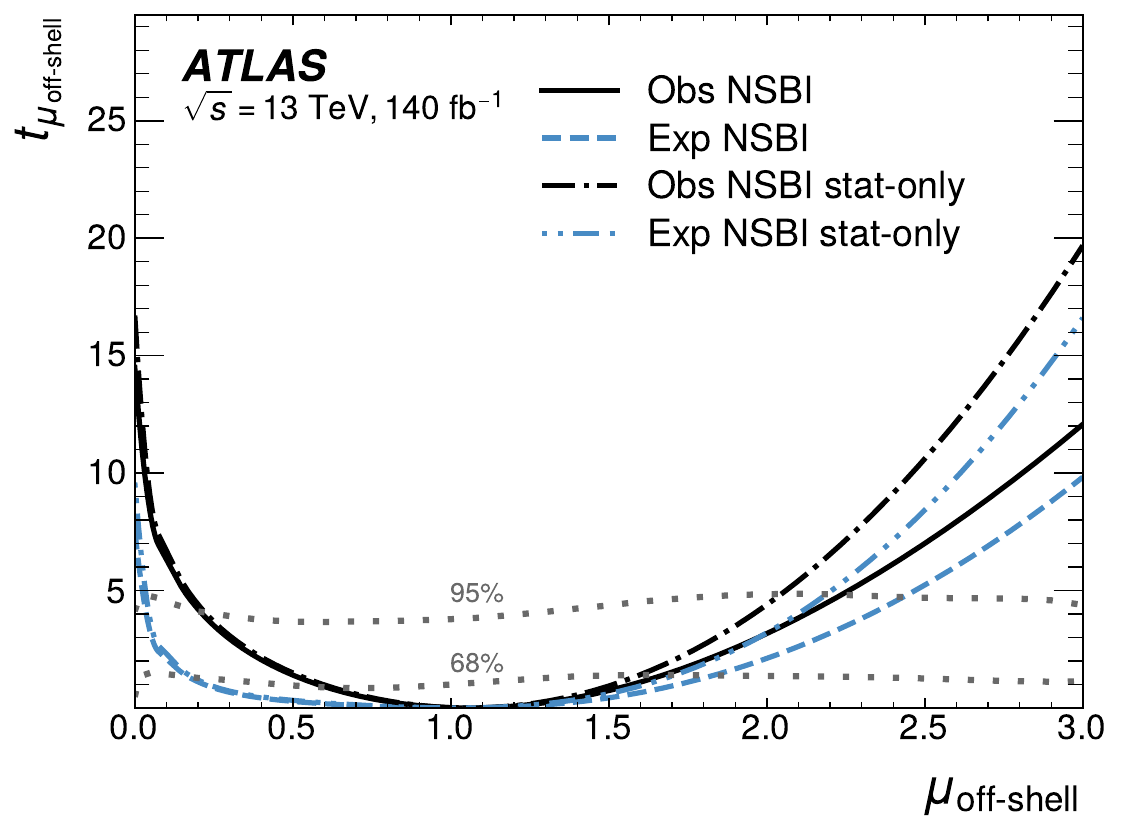}\label{fig:combined_results_muoffshell_scan}}\hfill
\subfloat[]{\includegraphics[width=0.48\linewidth]{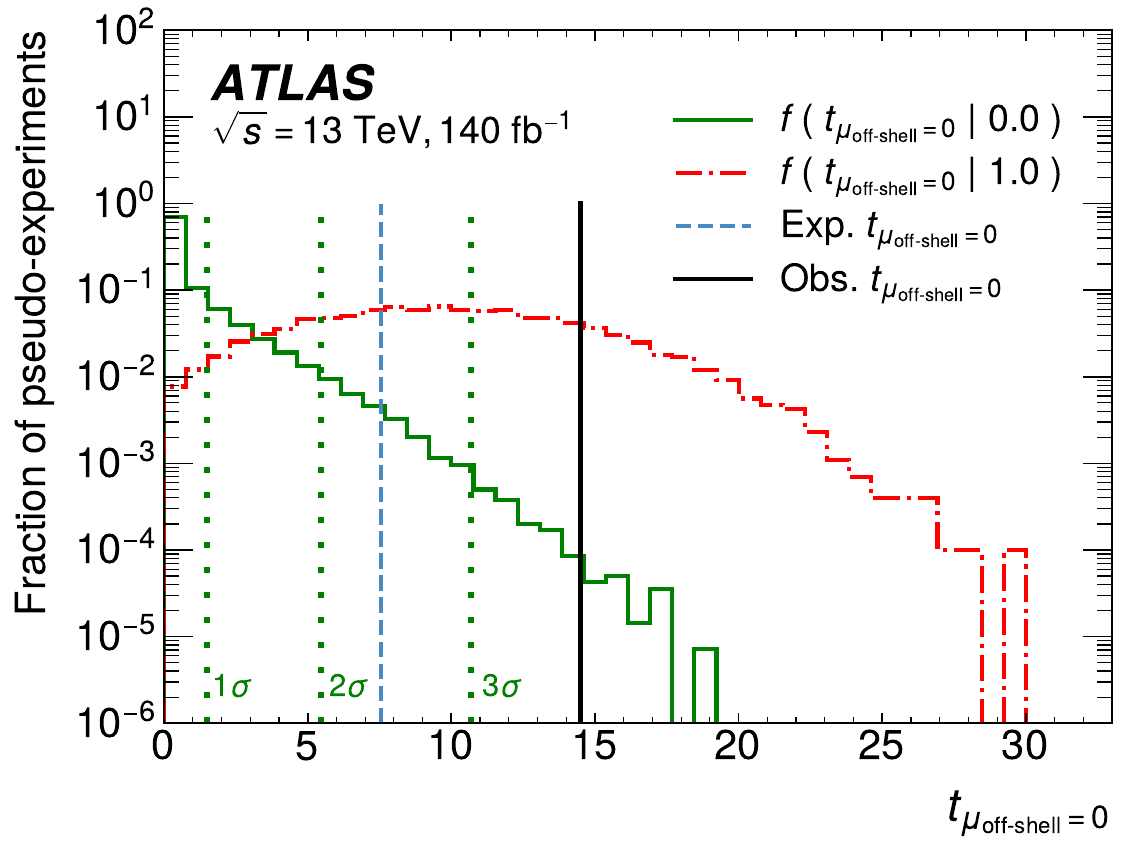}\label{fig:combined_results_muoffshell_nc}}
\caption{\protect\subref{fig:combined_results_muoffshell_scan} Values of the test statistic \tmuoff assuming a single parameter of interest \tmuoff obtained with an Asimov dataset (expected, dashed blue) and with data (observed, solid black) combining the \ohtollll\ and \ohtollnunu\ decay channels. The dash-dotted curves show the \textit{statistics-only} results where all NP are fixed to their best-fit values \hatalpha. The dotted gray lines show the 68\% and 95\% confidence belt, obtained from the Neyman construction. \protect\subref{fig:combined_results_muoffshell_nc} Expected distribution of $t_{\mu_{\text{off-shell}}=0}$ estimated with pseudo-experiments for the case of $\muoff^{\text{truth}}=0$ (solid green, no off-shell Higgs boson production hypothesis) and $\muoff^{\text{truth}}=1$ (dashed-dotted red, SM hypothesis). The vertical solid black (dashed blue) line shows the observed (expected) value of $t_{\mu_{\text{off-shell}=0}}$. The vertical dotted green lines represent the $1\sigma$, $2\sigma$, an $3\sigma$ significance thresholds under the $\muoff^{\text{truth}}=0$ hypothesis.}
\end{figure}

\Fig{\ref{fig:combined_results_muoffshell_scan}} shows the test statistic values as a function of \muoff, for a joint likelihood model with a single common parameter of interest \muoff\ ($\okappag^2\okappaV^2=\okappaV^4$). Similar to the result using only the $ZZ\to 4\ell$ channel, the comparison indicates that systematic uncertainties are more important for tests of signal-dominated hypotheses ($\muoff \gg 1$) than for tests of interference-dominated hypotheses ($\muoff \ll 1$). \Fig{\ref{fig:combined_results_muoffshell_nc}} shows the expected distribution of \tmuoff for the SM hypothesis ($\muoff^{\text{truth}}=1$) and the no off-shell Higgs boson hypothesis ($\muoff^{\text{truth}}=0$) using the bootstrapping technique described above. The observed (expected) value of \muoff\ at 68\% CL is:

\begin{equation}\nonumber
\mu_{\text{off-shell}} = \muoffcomb^{+\muoffcombobsoneup}_{-\muoffcombobsonedn} \;\;(1.00^{+\muoffcombexponeup}_{-\muoffcombexponedn}).
\end{equation}
The result using uncertainties at 95\% CL can be found in \Tab{\ref{tab:paper_summary}}. The evidence for off-shell Higgs boson production has an observed (expected) significance of $\muoffcombobssig\sigma$ ($\muoffcombexpsig\sigma$).

\begin{figure}[tbp]
\centering
\subfloat[]{\includegraphics[width=0.48\linewidth]{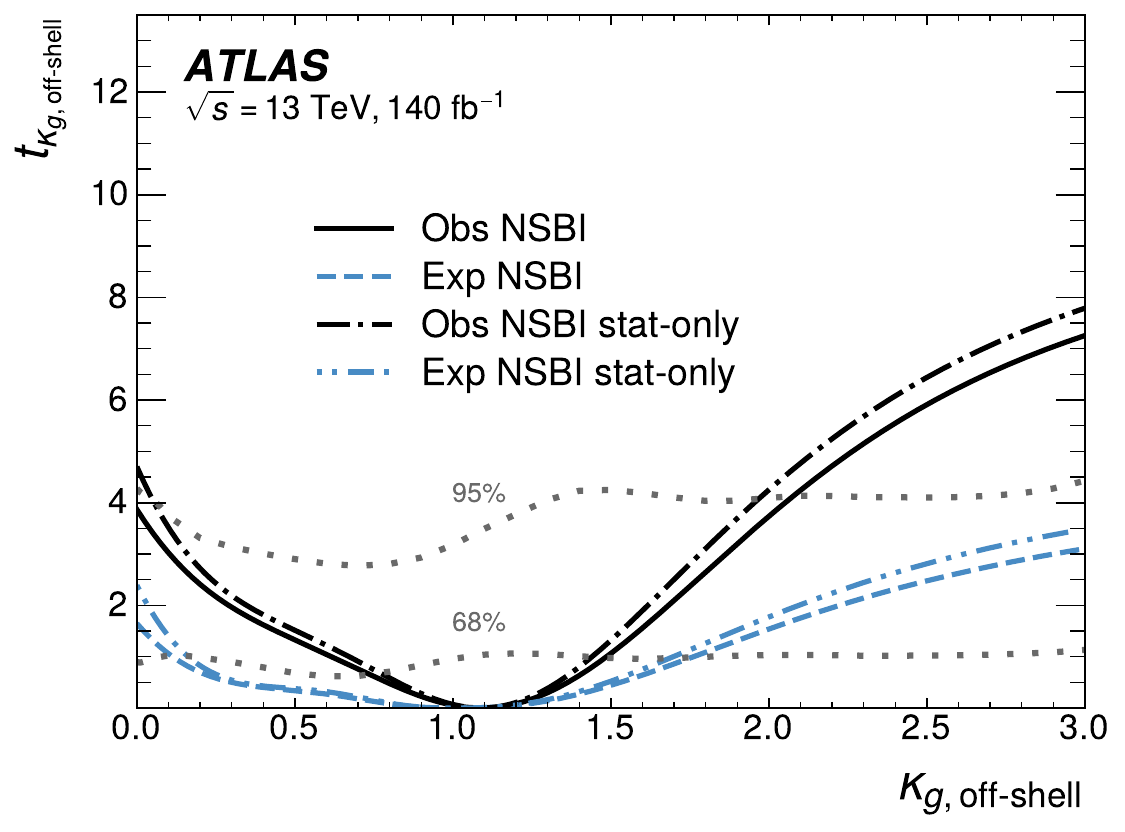}\label{fig:combined_results_kappa_g}}\hfill
\subfloat[]{\includegraphics[width=0.48\linewidth]{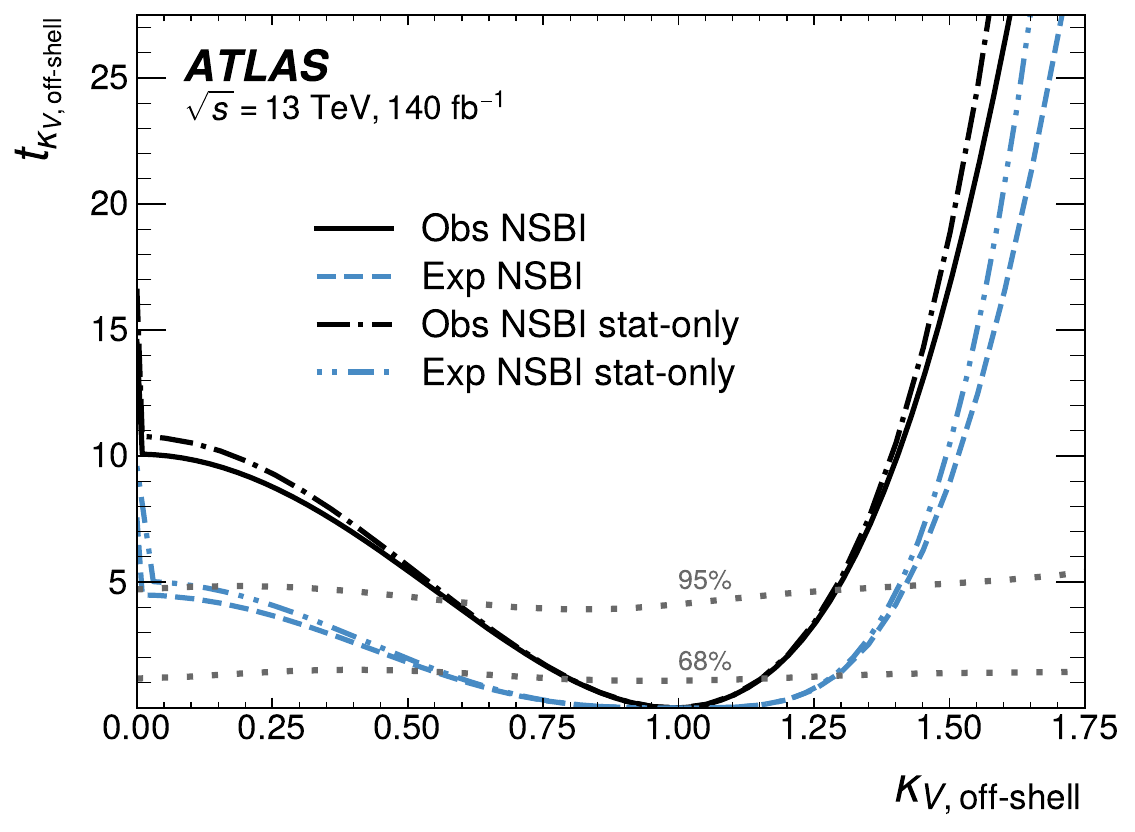}\label{fig:combined_results_kappa_V}}\\
\caption{Values of the \protect\subref{fig:combined_results_kappa_g} test statistic $t_{\kappa_{g,\text{off-shell}}}$ as a function of  $\kappa_{g,\text{off-shell}}$ and \protect\subref{fig:combined_results_kappa_V} the test statistic $t_{\kappa_{V,\text{off-shell}}}$ as a function of $\kappa_{V,\text{off-shell}}$ obtained with an Asimov dataset (expected, dashed blue) and with data (observed, solid black). The dash-dotted curves show the \textit{statistics-only} results where all NP are fixed to their best-fit values \hatalpha. The dotted gray lines show the 68\% and 95\% confidence belt, obtained from the Neyman construction. The $\kappa$ parameter not shown is profiled in both cases. The abrupt change at $\okappaV=0$ comes from the strong $\okappaV^4$ dependency in the EW production of off-shell Higgs bosons.\label{fig:combined_results_kappa}}
\end{figure}

\Fig{\ref{fig:combined_results_kappa}} shows the values of the test statistic $t_{\kappa_{g,\text{off-shell}}}$ and $t_{\kappa_{V,\text{off-shell}}}$ as a function of $\kappa_{g,\text{off-shell}}$ and $\kappa_{V,\text{off-shell}}$, respectively. In both cases, the $\kappa$ parameter not shown is profiled. The resulting confidence intervals in \okappag\ and \okappaV\ provide a measurement of the Higgs boson couplings to gluons and vector bosons without any assumption on the Higgs boson total width~\cite{HIGG-2021-23}. The observed (expected) values at 68\% CL are:

\begin{equation}\nonumber
\okappag =  \kappagcomb^{+\kappagcombobsoneup}_{-\kappagcombobsonedn}\;\; (1.00^{+\kappagcombexponeup}_{-\kappagcombexponedn}),\qquad \okappaV = \kappavcomb^{+\kappavcombobsoneup}_{-\kappavcombobsonedn}\;\; (1.00^{+\kappavcombexponeup}_{-\kappavcombexponedn}).
\end{equation}

\subsection{Combination with on-shell \texorpdfstring{$4\ell$}{4l} analysis and Higgs boson width interpretation}

Measurements of the on-shell Higgs boson production~\cite{HIGG-2018-28} provide constraints on the signal strengths $\mu_{\text{on-shell}}^{\text{ggF}} =\kappa^2_{g,\text{on-shell}}\kappa^2_{V,\text{on-shell}}/\kappa_{H}$ and
$\mu_{\text{on-shell}}^{\text{EW}} =\kappa^4_{V,\text{on-shell}}/\kappa_{H}$, where $\kappa_{H}=\Gamma_H/\Gamma_H^{\text{SM}}$ is the Higgs boson width normalized to the SM expectation.

Assuming that $\kappa^2_{g,\text{on-shell}}\kappa^2_{V,\text{on-shell}}=\kappa^4_{V,\text{on-shell}}=\kappa^2_{g,\text{off-shell}}\kappa^2_{V,\text{off-shell}}=\kappa^4_{V,\text{off-shell}}$, the off-shell Higgs boson production measurement can be combined with the on-shell $H\rightarrow ZZ\rightarrow 4\ell$ production measurement~\cite{HIGG-2018-28} to provide a measurement of the Higgs boson total width. The joint likelihood model for this measurement extends \Eqn{\ref{eq:llratio}} with a common $H\to ZZ$ coupling modifier $\thetaZZ$ and the modifier $\kappa_{H}$ to the Higgs boson width:

\begin{equation}
\begin{split}
-2\ln\lambda(\kappaH, \thetaZZ,\theta,\alpha) = &-2\sum_{\substack{\text{on-shell }4\ell\\\text{regions } (I)}}\ln\left[\text{Pois}(N_I|\nu_I(\thetaZZ/\kappa_{H},\alpha))\right]\\
&-2\sum_{\substack{\text{off-shell }\\\text{regions }(I)}}\ln\left[\text{Pois}(N_I|\nu_I(\thetaZZ,\theta,\alpha))\right]\\
&-2\sum_{\substack{\text{off-shell }4\ell\\\text{SR events }(i)}}\ln \left[\frac{p(x_i|\thetaZZ,\theta,\alpha)}{p_{\text{ref}}(x_i)}\right]
+\sum_{\text{systematics }(m)} (\alpha_m-a_m)^2,
\end{split}
\end{equation}
where $\thetaZZ=\onkappag^2\onkappaV^2=\onkappaV^4=\okappag^2\okappaV^2=\okappaV^4$. This assumption follows the one used in~\cite{HIGG-2018-32}, but the sensitivity could be weakened by allowing $\kappa_V$ and $\kappa_g$ to vary independently and account for BSM contributions. The NP model for common experimental uncertainties
follows the approach described in \Sect{\ref{sec:combination}}. Events from on-shell Higgs boson processes in the off-shell background sample are also scaled as $\theta_{HZZ}/\kappa_H$. Theoretical modeling uncertainties in the on-shell and off-shell measurements are modeled with separate parameters, given the distinctness of the phase-space regions, but the measured Higgs boson width is largely insensitive to this modeling choice.  Background normalization factors are also modeled separately.

\begin{figure}[tbp]
\centering
\includegraphics[width=0.5\linewidth]{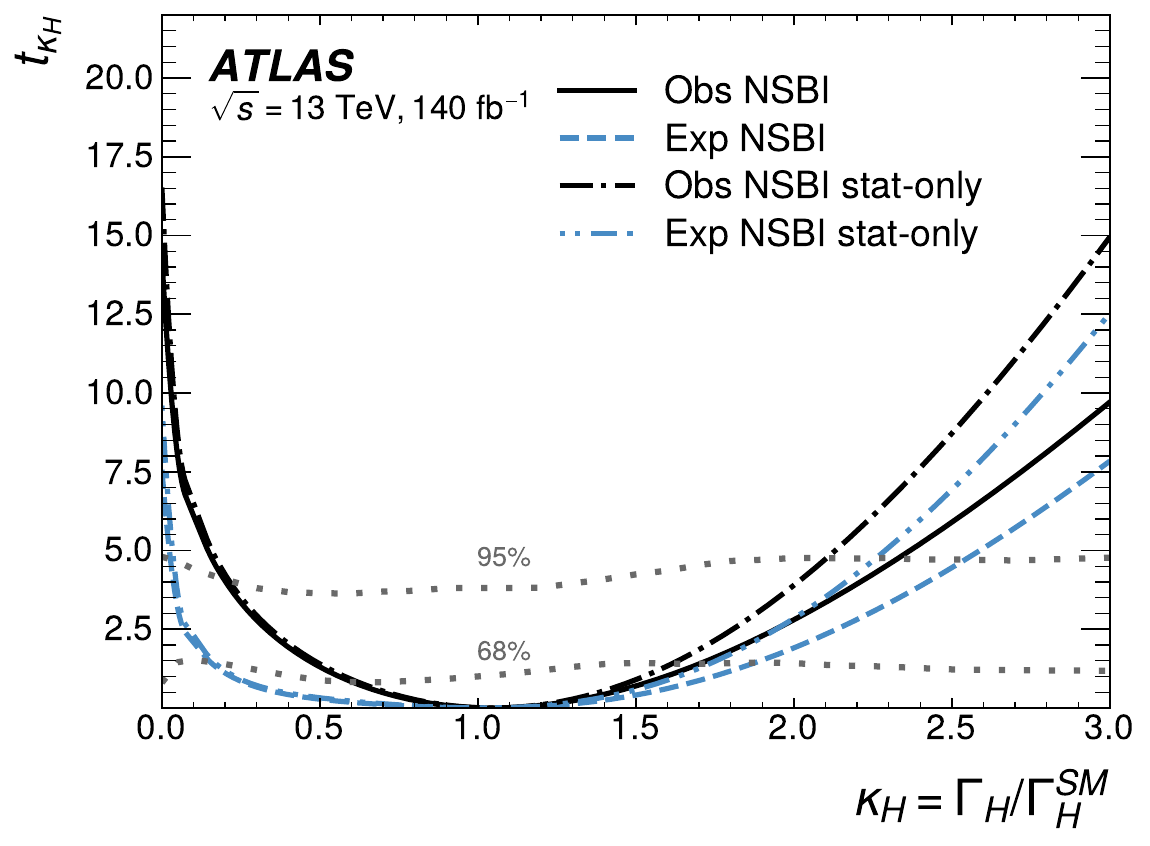}
\caption{Values of the test statistic $t_{\kappaH}$ as a function of $\kappaH=\Gamma_H/\Gamma_H^\text{SM}$ obtained with an Asimov dataset (expected, dotted black) and with data (observed, solid black). The dash-dotted curves show the \textit{statistics-only} results where all NP are fixed to their best-fit values $\hatalpha$. The dotted gray lines show the 68\% and 95\% confidence belt, obtained from the Neyman construction.}
\label{fig:width_results}
\end{figure}

\Fig{\ref{fig:width_results}} shows the test statistic values as a function of \kappaH when profiling $\thetaZZ$. The observed (expected) value of $\kappaH$ and $\Gamma_H$ at 68\% CL are:

\begin{equation}\nonumber
\kappaH = \kappaHcomb^{+\kappaHcombobsoneup}_{-\kappaHcombobsonedn} \;\;(1.00^{+\kappaHcombexponeup}_{-\kappaHcombexponedn}),\qquad
\Gamma_H = \widthcombsimple^{+\widthcombobsoneupsimple}_{-\widthcombobsonednsimple} \;\;(4.1^{+\widthcombexponeupsimple}_{-\widthcombexponednsimple})\;\MeV,
\end{equation}
The result using uncertainties at 95\% CL can be found in \Tab{\ref{tab:paper_summary}}.

\begin{figure}[tbp]
\centering
\subfloat[]{\includegraphics[width=0.48\linewidth]{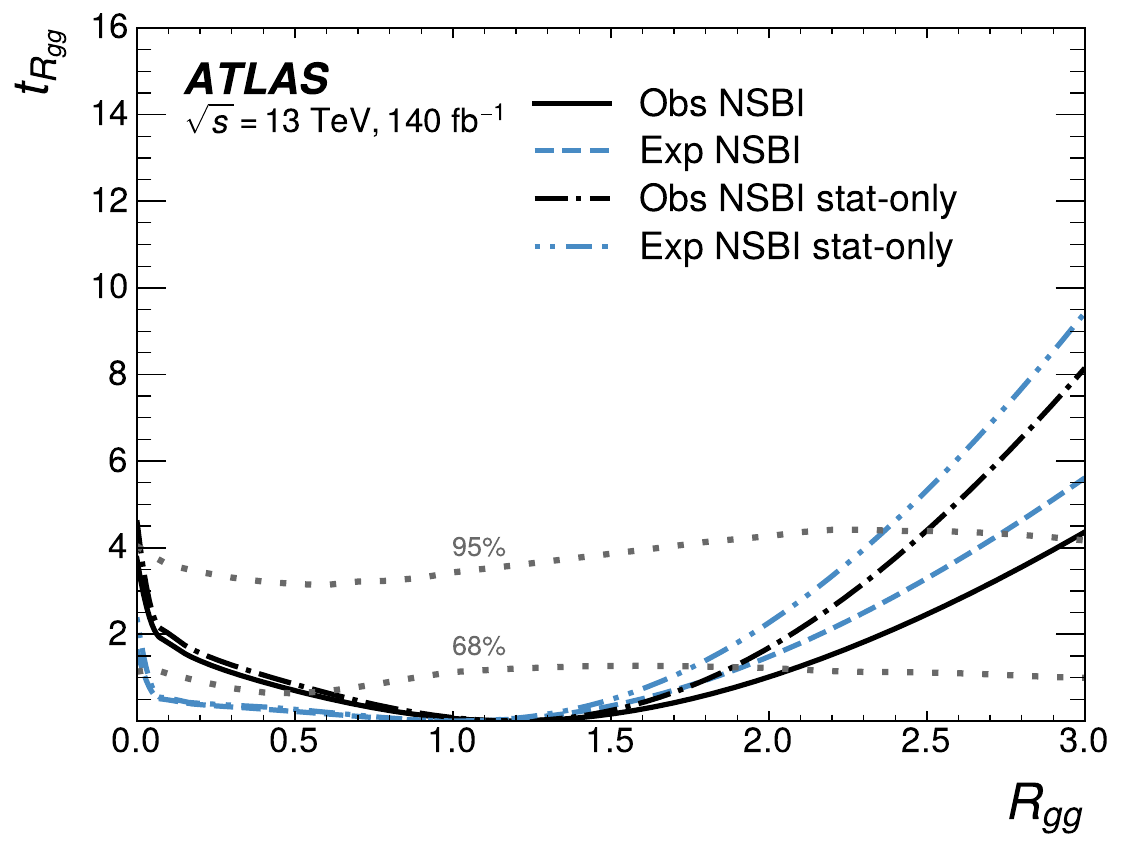}\label{fig:combined_results_Rgg}}\hfill
\subfloat[]{\includegraphics[width=0.48\linewidth]{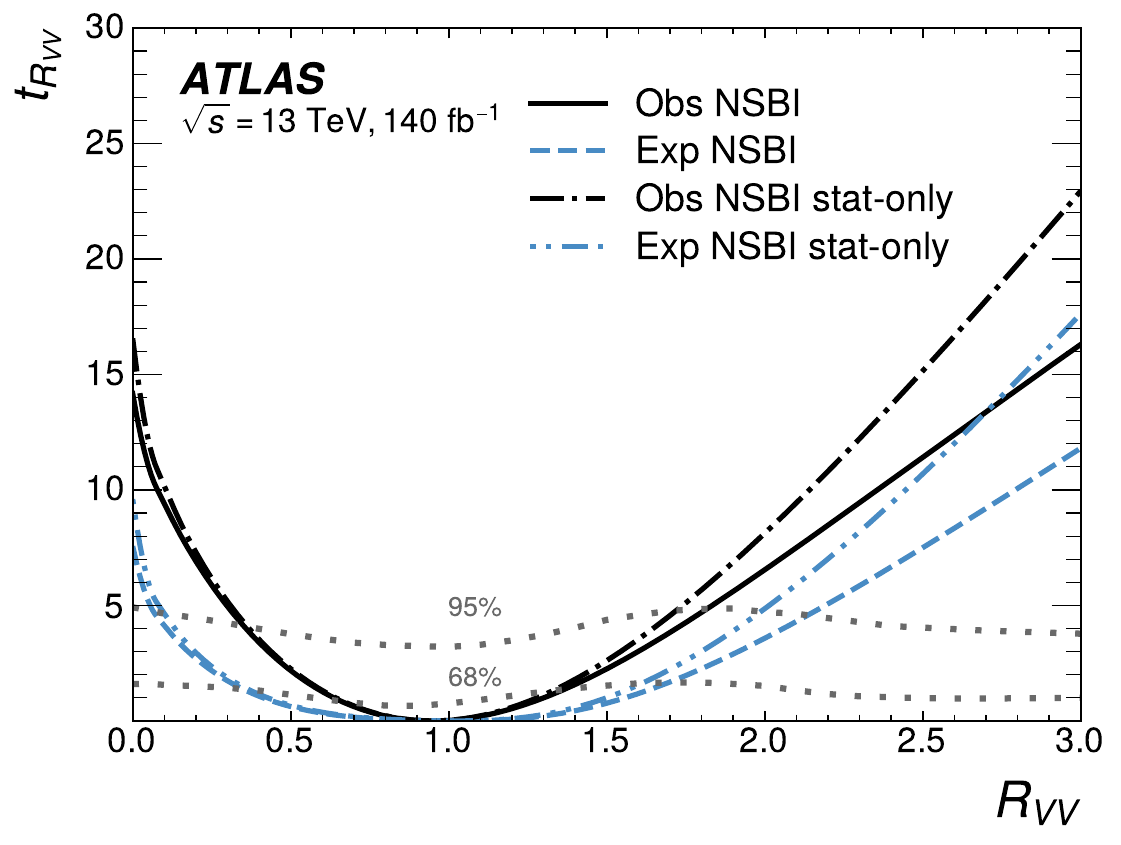}\label{fig:combined_results_RVV}}\\
\caption{Values of \protect\subref{fig:combined_results_Rgg} the test statistic $t_{R_{gg}}$ as a function of $R_{gg}=\onkappag^2/\okappag^2$ and of \protect\subref{fig:combined_results_RVV} the test statistic $t_{R_{VV}}$ as a function of $R_{VV}=\onkappaV^2/\okappaV^2$ obtained with an Asimov dataset (expected, dashed blue) and with data (observed, solid black). The values obtained when all NP are fixed to their best-fit values \hatalpha, corresponding to the \textit{statistics-only} case, are shown in dash-dotted lines for comparison.  The dotted gray lines show the 68\% and 95\% confidence belt, obtained from the Neyman construction. In both figures, the $R$ parameter not shown is profiled and $\kappa_H$ is fixed to 1.}
\label{fig:combined_results_R}
\end{figure}

A similar combined measurement strategy can be used to constrain $R_{gg}=\onkappag^2/\okappag^2$ and $R_{VV}=\onkappaV^2/\okappaV^2$. The value of the two test statistics as a function of $R_{gg}$ and $R_{VV}$ are shown in \Fig{\ref{fig:combined_results_R}}. The observed (expected) values at 68\% CL are:

\begin{equation}\nonumber
R_{gg}= \rggcomb^{+\rggcombobsoneup}_{-\rggcombobsonedn}\;\; (1.00^{+\rggcombexponeup}_{-\rggcombexponedn}),\qquad R_{VV}= \rvvcomb^{+\rvvcombobsoneup}_{-\rvvcombobsonedn}\;\; (1.00^{+\rvvcombexponeup}_{-\rvvcombexponedn}),
\end{equation}
profiling the other $R$ parameter and setting $\kappaH$ to unity. \Tab{\ref{tab:paper_summary}} summarizes all the results presented in this paper.

\begin{table}[tbp]
\centering
\caption{Summary of the results for $\kappa_{g,\text{ off-shell}}$, $\kappa_{V,\text{ off-shell}}$, and $\mu_{\text{off-shell}}$ combining the measurement of the off-shell Higgs boson production analysis in the $\htollll$ decay channel presented in this paper with the analysis in the $\htollnunu$ decay channel from Ref.~\cite{HIGG-2018-32}. The results for $\Gamma_H$, $R_{gg}$ and $R_{VV}$ are obtained combining the off-shell production measurement with the on-shell analysis from Ref.~\cite{HIGG-2018-28}. All results are presented with their 68\% CL and 95\% CL intervals. No expected 95\% CL interval for \okappag\ is obtained because the Asimov sample is unable to break the degeneracy between \okappag\ and \okappaV\ in \ggf\ production. All results use the full Run 2 dataset with $140\,\text{fb}^{-1}$ of integrated luminosity.}
\begin{tabular}{cccccc}\hline\hline
\toprule &  & \multicolumn{2}{c}{\textbf{68\% CL interval}} & \multicolumn{2}{c}{\textbf{95\% CL interval}}  \\
\textbf{Parameter} & \textbf{Value} & \textbf{Observed} & \textbf{Expected} & \textbf{Observed} & \textbf{Expected}  \\
\hline && \\[-2.0ex]
\muoff\ &  \muoffcomb & [\muoffcombobsonednval, \muoffcombobsoneupval] & [\muoffcombexponednval, \muoffcombexponeupval] & [\muoffcombobstwodnval, \muoffcombobstwoupval] & [\muoffcombexptwodnval, \muoffcombexptwoupval]\\[1.5ex]
\okappag\ &  \kappagcomb & [\kappagcombobsonednval, \kappagcombobsoneupval] & [\kappagcombexponednval, \kappagcombexponeupval] & $<\kappagcombobstwoupval$ & - \\[1.5ex]
\okappaV\ &  \kappavcomb & [\kappavcombobsonednval, \kappavcombobsoneupval] & [\kappavcombexponednval, \kappavcombexponeupval] & [\kappavcombobstwodnval, \kappavcombobstwoupval] & [\kappavcombexptwodnval, \kappavcombexptwoupval]\\[1.5ex]
$\Gamma_H$ [MeV] &  \widthcomb & [\widthcombobsonednval, \widthcombobsoneupval] & [\widthcombexponednval, \widthcombexponeupval] & [\widthcombobstwodnval, \widthcombobstwoupval] & [\widthcombexptwodnval, \widthcombexptwoupval]\\[1.5ex]
$R_{gg}$ &  \rggcomb & [\rggcombobsonednval, \rggcombobsoneupval] & [\rggcombexponednval, \rggcombexponeupval] & $<\rggcombobstwoupval$  & $<\rggcombexptwoupval$\\[1.5ex]
$R_{VV}$ &  \rvvcomb & [\rvvcombobsonednval, \rvvcombobsoneupval] & [\rvvcombexponednval, \rvvcombexponeupval] & [\rvvcombobstwodnval, \rvvcombobstwoupval] & [\rvvcombexptwodnval, \rvvcombexptwoupval]
\\\bottomrule\hline\hline
\end{tabular}
\label{tab:paper_summary}
\end{table}



%

\FloatBarrier
\section{Conclusion}
%

A measurement of the off-shell Higgs boson production in the \ohtollll decay channel is presented. The measurement uses $140\,\text{fb}^{-1}$ of integrated luminosity collected at $\sqrt{s}=13\,\TeV$ during the Run 2 of the LHC by the ATLAS detector. The data are analyzed with a neural simulation-based inference strategy in which neural networks are used to estimate a per-event contribution to the likelihood ratio between different hypotheses. This result is combined with the most recent measurement of the off-shell Higgs boson production in the $H^*\to ZZ\to 2\ell2\nu$ decay channel. The observed (expected) value of the off-shell Higgs boson signal strength is $\muoffcomb^{+\muoffcombobsoneup}_{-\muoffcombobsonedn}$ ($1.00^{+\muoffcombexponeup}_{-\muoffcombexponedn}$) at 68\% CL\@. The evidence for off-shell Higgs boson production has an observed (expected) significance of $\muoffcombobssig\sigma$ ($\muoffcombexpsig\sigma$). The off-shell Higgs boson production measurement in the $H^*\to ZZ$ decay channel is combined with the on-shell Higgs boson production measurement in the same channel to obtain a constraint on the Higgs boson total width. The observed (expected) value of the Higgs boson total width is $\Gamma_H = \widthcombsimple^{+\widthcombobsoneupsimple}_{-\widthcombobsonednsimple} (4.1^{+\widthcombexponeupsimple}_{-\widthcombexponednsimple})\,\MeV$ at 68\% CL, improved relative to the ATLAS previous result of $\Gamma_H = \widthcombhistosimple^{+\widthcombhistoobsoneupsimple}_{-\widthcombhistoobsonednsimple} (4.1^{+\widthcombhistoexponeupsimple}_{-\widthcombhistoexponednsimple})\,\MeV$ using the same dataset.


\clearpage
\section*{Acknowledgements}

%
%

%
%

We thank CERN for the very successful operation of the LHC and its injectors, as well as the support staff at
CERN and at our institutions worldwide without whom ATLAS could not be operated efficiently.

The crucial computing support from all WLCG partners is acknowledged gratefully, in particular from CERN, the ATLAS Tier-1 facilities at TRIUMF/SFU (Canada), NDGF (Denmark, Norway, Sweden), CC-IN2P3 (France), KIT/GridKA (Germany), INFN-CNAF (Italy), NL-T1 (Netherlands), PIC (Spain), RAL (UK) and BNL (USA), the Tier-2 facilities worldwide and large non-WLCG resource providers. Major contributors of computing resources are listed in Ref.~\cite{ATL-SOFT-PUB-2025-001}.

We gratefully acknowledge the support of ANPCyT, Argentina; YerPhI, Armenia; ARC, Australia; BMWFW and FWF, Austria; ANAS, Azerbaijan; CNPq and FAPESP, Brazil; NSERC, NRC and CFI, Canada; CERN; ANID, Chile; CAS, MOST and NSFC, China; Minciencias, Colombia; MEYS CR, Czech Republic; DNRF and DNSRC, Denmark; IN2P3-CNRS and CEA-DRF/IRFU, France; SRNSFG, Georgia; BMBF, HGF and MPG, Germany; GSRI, Greece; RGC and Hong Kong SAR, China; ICHEP and Academy of Sciences and Humanities, Israel; INFN, Italy; MEXT and JSPS, Japan; CNRST, Morocco; NWO, Netherlands; RCN, Norway; MNiSW, Poland; FCT, Portugal; MNE/IFA, Romania; MSTDI, Serbia; MSSR, Slovakia; ARIS and MVZI, Slovenia; DSI/NRF, South Africa; MICIU/AEI, Spain; SRC and Wallenberg Foundation, Sweden; SERI, SNSF and Cantons of Bern and Geneva, Switzerland; NSTC, Taipei; TENMAK, T\"urkiye; STFC/UKRI, United Kingdom; DOE and NSF, United States of America.

Individual groups and members have received support from BCKDF, CANARIE, CRC and DRAC, Canada; CERN-CZ, FORTE and PRIMUS, Czech Republic; COST, ERC, ERDF, Horizon 2020, ICSC-NextGenerationEU and Marie Sk{\l}odowska-Curie Actions, European Union; Investissements d'Avenir Labex, Investissements d'Avenir Idex and ANR, France; DFG and AvH Foundation, Germany; Herakleitos, Thales and Aristeia programmes co-financed by EU-ESF and the Greek NSRF, Greece; BSF-NSF and MINERVA, Israel; NCN and NAWA, Poland; La Caixa Banking Foundation, CERCA Programme Generalitat de Catalunya and PROMETEO and GenT Programmes Generalitat Valenciana, Spain; G\"{o}ran Gustafssons Stiftelse, Sweden; The Royal Society and Leverhulme Trust, United Kingdom.

In addition, individual members wish to acknowledge support from Armenia: Yerevan Physics Institute (FAPERJ); CERN: European Organization for Nuclear Research (CERN DOCT); Chile: Agencia Nacional de Investigaci\'on y Desarrollo (FONDECYT 1230812, FONDECYT 1230987, FONDECYT 1240864); China: Chinese Ministry of Science and Technology (MOST-2023YFA1605700, MOST-2023YFA1609300), National Natural Science Foundation of China (NSFC - 12175119, NSFC 12275265, NSFC-12075060); Czech Republic: Czech Science Foundation (GACR - 24-11373S), Ministry of Education Youth and Sports (ERC-CZ-LL2327, FORTE CZ.02.01.01/00/22\_008/0004632), PRIMUS Research Programme (PRIMUS/21/SCI/017); EU: H2020 European Research Council (ERC - 101002463); European Union: European Research Council (ERC - 948254, ERC 101089007, ERC, BARD, 101116429), Horizon 2020 Framework Programme (MUCCA - CHIST-ERA-19-XAI-00), European Union, Future Artificial Intelligence Research (FAIR-NextGenerationEU PE00000013), Italian Center for High Performance Computing, Big Data and Quantum Computing (ICSC, NextGenerationEU); France: Agence Nationale de la Recherche (ANR-20-CE31-0013, ANR-21-CE31-0013, ANR-21-CE31-0022, ANR-22-EDIR-0002); Germany: Baden-Württemberg Stiftung (BW Stiftung-Postdoc Eliteprogramme), Deutsche Forschungsgemeinschaft (DFG - 469666862, DFG - CR 312/5-2); Italy: Istituto Nazionale di Fisica Nucleare (ICSC, NextGenerationEU), Ministero dell'Università e della Ricerca (NextGenEU PRIN20223N7F8K M4C2.1.1); Japan: Japan Society for the Promotion of Science (JSPS KAKENHI JP22H01227, JSPS KAKENHI JP22H04944, JSPS KAKENHI JP22KK0227, JSPS KAKENHI JP23KK0245); Norway: Research Council of Norway (RCN-314472); Poland: Ministry of Science and Higher Education (IDUB AGH, POB8, D4 no 9722), Polish National Agency for Academic Exchange (PPN/PPO/2020/1/00002/U/00001), Polish National Science Centre (NCN 2021/42/E/ST2/00350, NCN OPUS 2023/51/B/ST2/02507, NCN OPUS nr 2022/47/B/ST2/03059, NCN UMO-2019/34/E/ST2/00393, NCN \& H2020 MSCA 945339, UMO-2020/37/B/ST2/01043, UMO-2021/40/C/ST2/00187, UMO-2022/47/O/ST2/00148, UMO-2023/49/B/ST2/04085, UMO-2023/51/B/ST2/00920); Portugal: Foundation for Science and Technology (FCT); Spain: Generalitat Valenciana (Artemisa, FEDER, IDIFEDER/2018/048), Ministry of Science and Innovation (MCIN \& NextGenEU PCI2022-135018-2, MICIN \& FEDER PID2021-125273NB, RYC2019-028510-I, RYC2020-030254-I, RYC2021-031273-I, RYC2022-038164-I); Sweden: Carl Trygger Foundation (Carl Trygger Foundation CTS 22:2312), Swedish Research Council (Swedish Research Council 2023-04654, VR 2018-00482, VR 2021-03651, VR 2022-03845, VR 2022-04683, VR 2023-03403), Knut and Alice Wallenberg Foundation (KAW 2018.0458, KAW 2019.0447, KAW 2022.0358); Switzerland: Swiss National Science Foundation (SNSF - PCEFP2\_194658); United Kingdom: Leverhulme Trust (Leverhulme Trust RPG-2020-004), Royal Society (NIF-R1-231091); United States of America: U.S. Department of Energy (ECA DE-AC02-76SF00515), Neubauer Family Foundation.

%
%


This work started as part of the ATLAS Google Cloud Platform project~\cite{SOFT-2023-02}. Later stages of the work utilized computing resources at Southern Methodist University’s O’Donnell Data Science and Research Computing Institute and computing resources at the University of Massachusetts Amherst Research Computing.
\clearpage
\appendix

\FloatBarrier

\ifshowappendix
\part*{Appendix}
\addcontentsline{toc}{part}{Appendix}

\section{Neural network structure and training}
\label{sec:nn_parameters}

Several parameters related to the probability-density estimation NNs were optimized to reach the accuracy required to do the analysis presented in this paper. All NNs used in this analysis are multi-layer perceptrons trained with a binary cross-entropy functional. Other loss functionals~\cite{Rizvi:2023mws} were considered, but no significant increase in accuracy was observed.

\Tab{\ref{tab:nn_hyperparameters}} shows a complete list of parameters used in this analysis. The choice of parameters depends strongly on how large the phase space is over which the training is done. In practice, the preselection threshold ($\predisc > -0.85$) was chosen as loose as possible, to include as many events in the signal region as possible while allowing for a successful training.

The number of simulated events available for NN training varies from approximately 240k for the $VVV$ process to 2.4M for the $\qqbar \to ZZ$ process. The full set of samples has approximately 3.1M events. The statistical power of the sample depends on the distribution of weights and, more importantly, on the fraction of negatively-weighted events. The number of events used in each training is a fraction of the total number of events available in each sample: 10\% of the total events are kept for 10-fold cross validation, and only 80\% of the remaining events are selected for each ensemble member.

\begin{table}[tbp]
\centering
\caption{Hyperparameters used for the neural networks using in this work. Ranges are given when different values were used for different $\probx/\probref$ estimates.\label{tab:nn_hyperparameters}}
\begin{tabular}{lc}\hline\hline
\toprule\textbf{Parameter} & \textbf{Value} \\\hline
Number of hidden layers & 5\\
Number of neurons per layer & $1\,000$\\
Activation function & \textit{swish}~\cite{DBLP:journals/corr/abs-1710-05941}\\
Optimizer (initial learning rate) & \textit{NAdam}~\cite{DBLP:journals/corr/KingmaB14,dozat.2016} (0.1)\\
Batch Size & 256 -- $4\,096$\\
Training epochs & 100\\
k-fold cross validation & $k=10$\\
Ensemble members (total) & 100 -- 700\\
Ensemble fraction (sampled without replacement) & 80\%\\
Train-validation split & 90--10\% \\\bottomrule
\hline\hline
\end{tabular}
\end{table}


The most important targets for optimization of the hyperparameters were the \textit{calibration test} done over a large number of different samples, and the \textit{reweighting test} done over a very large number of different observables, including observables not included as input for the NNs. Both tests are described briefly in \Sect{\ref{sec:neural}} and in more detail in Ref.~\cite{atlasML}. Closure of the MLE, described in \Sect{\ref{sec:result}}, is also a strict target for optimization used in this analysis. While calibration layers improve the calibration accuracy, they can spoil the other tests and were not used. Using wide and deep NN ensembles was found sufficient to achieve good calibration. The initial learning rate was reduced by a factor of $1/100$ every 30 epochs. Large batch sizes improve the training when samples have a large number of negatively-weighted events and they help the extrapolation of density ratio estimates to different phase spaces.


\section{Comparison with the histogram-based analysis}
\label{sec:histo_comparison}

This appendix provides a comparison of the results presented in this paper with the results in Ref.~\cite{HIGG-2018-32}, which uses a histogram-based analysis. The results of  Ref.~\cite{HIGG-2018-32} used in this comparison have been updated to use the most recent measurement of integrated luminosity of $140\,\ifb$~\cite{DAPR-2021-01}.

\begin{figure}[tbp]
\centering
\includegraphics[width=0.5\linewidth]{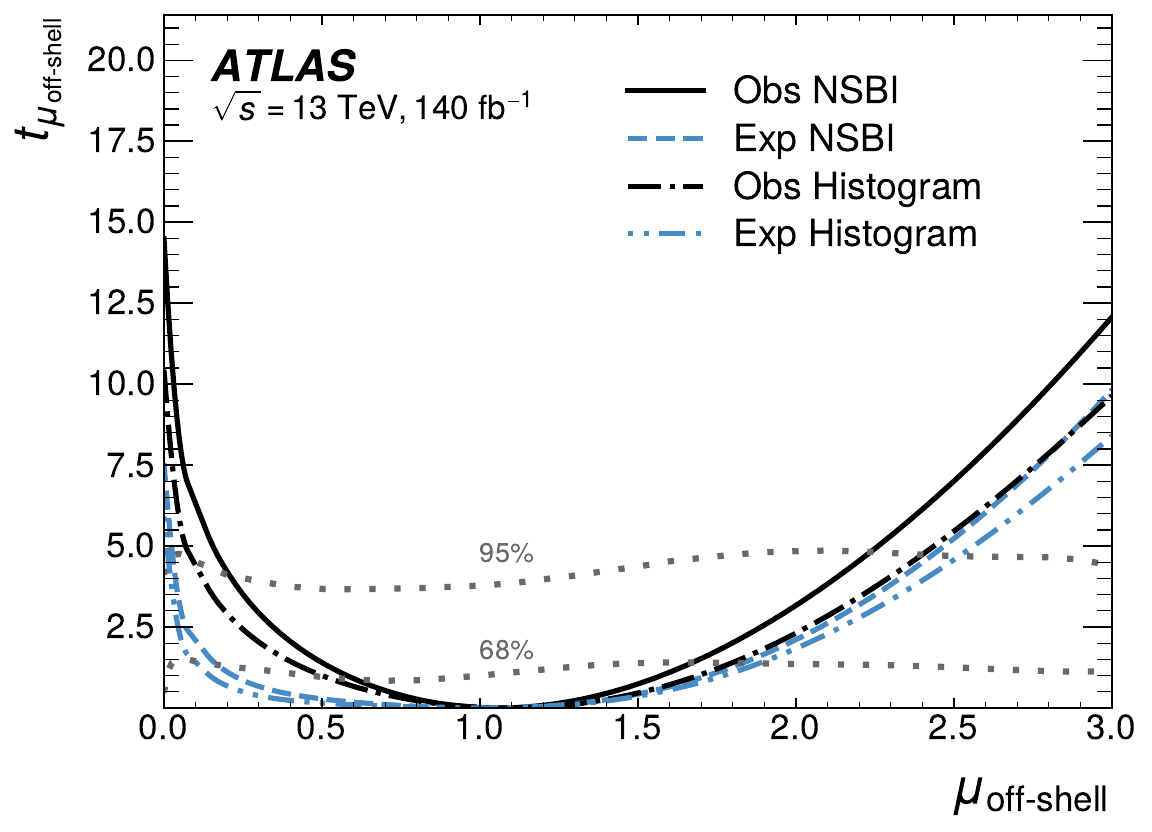}
\caption{Values of the test statistic \tmuoff\ assuming a single parameter of interest \muoff\ obtained with an Asimov dataset (expected, dashed blue) and with data (observed, solid black) when combining the \ohtollll and \ohtollnunu decay channels. The values from the histogram-based analysis~\cite{HIGG-2018-32} are added with dash-dotted lines for comparison. The dotted gray lines show the 68\% and 95\% confidence belt, obtained from the Neyman construction, for the NSBI analysis.}
\label{fig:nll_4l_2l2nu_histo_comparison}
\end{figure}

\Fig{\ref{fig:muoffshell_4l_results_histo}} provides a comparison of the values of the test statistic $t_{\mu_{\text{off-shell}}}$ assuming a single parameter of interest \muoff\ between the NSBI result and the histogram-based hypothesis using only the \ohtollll\ decay channel. \Fig{\ref{fig:nll_4l_2l2nu_histo_comparison}} shows the same comparison after combining the result in the \ohtollnunu\ channel. \Tab{\ref{tab:old_syst_impact}} compares the impact of different groups of systematic uncertainties in the combined result by using the method described in~\cite{HIGG-2018-32}.  The impact of a group of systematic uncertainties is assessed by fixing the associated group of NPs and re-estimating the value of $\mu_{\text{off-shell}}$ for which the value of the test statistic is equal to four, corresponding to 95.5\% CL in the asymptotic approximation. This method highlights the importance of different sources of uncertainties in the signal-dominated region. The importance of an uncertainty is given by its deviation from the last row, which has the uncertainty for the full result. The most important sources of systematic uncertainty are common for both analyses and their impacts on $\muoff$ are similar.

\begin{table}[tbp]
\caption{
The impact of most important systematic uncertainties in the observed upper value of \muoff\ for which $\tmuoff = 4$, obtained by the combined fit. This value corresponds to the two standard deviation upper limit of \muoff\ in the asymptotic approximation. The first column denotes the systematic uncertainty that was excluded from the fit. The last row gives the nominal upper limit, where all uncertainties are included. The importance of an uncertainty its given by its deviation from the last row.
}
\label{tab:old_syst_impact}
\centering
\renewcommand{\arraystretch}{1.2}
\renewcommand{\tabcolsep}{8pt}
\begin{tabular}{lcc}
\hline\hline\toprule
\textbf{Systematic uncertainty fixed} & \multicolumn{2}{c}{$\mu_{\text{off-shell}}$ value at which $t_{\mu_{\text{off-shell}}} = 4$} \\
& \textbf{NSBI analysis} & \textbf{Histogram-based} \\
\hline
All (stat-only) & 1.96 & 2.13 \\\hline
Parton shower uncertainty for $gg\rightarrow ZZ$ (normalization) & 2.07 & 2.26 \\
Parton shower uncertainty for $gg\rightarrow ZZ$ (shape)& 2.12 & 2.29 \\
NLO EW uncertainty for $q\bar{q}\rightarrow ZZ$ & 2.10 & 2.27  \\
NLO QCD uncertainty for $gg\rightarrow ZZ$ & 2.09 & 2.29  \\
Parton shower uncertainty for $q\bar{q}\rightarrow ZZ$ (shape) & 2.12 & 2.29 \\
\hline
Jet energy scale and resolution uncertainty & 2.11 & 2.26  \\
\hline
None (full result)  & 2.12 & 2.30  \\
\bottomrule\hline\hline\noalign{\smallskip}
\end{tabular}
\end{table}


\Fig{\ref{fig:width_hist_4lonly}} shows a comparison between the two analyses for the values of the test statistic for the combination of the off-shell and on-shell Higgs boson production measurements for the \htollll decay channel alone as a function of \kappaH while profiling $\thetaZZ$. \Fig{\ref{fig:width_hist_4l2l2nu}} shows the same comparison after combination with the result in the \ohtollnunu\ decay channel.

\begin{figure}[tbp]
\centering
\subfloat[]{
\includegraphics[width=0.48\linewidth]{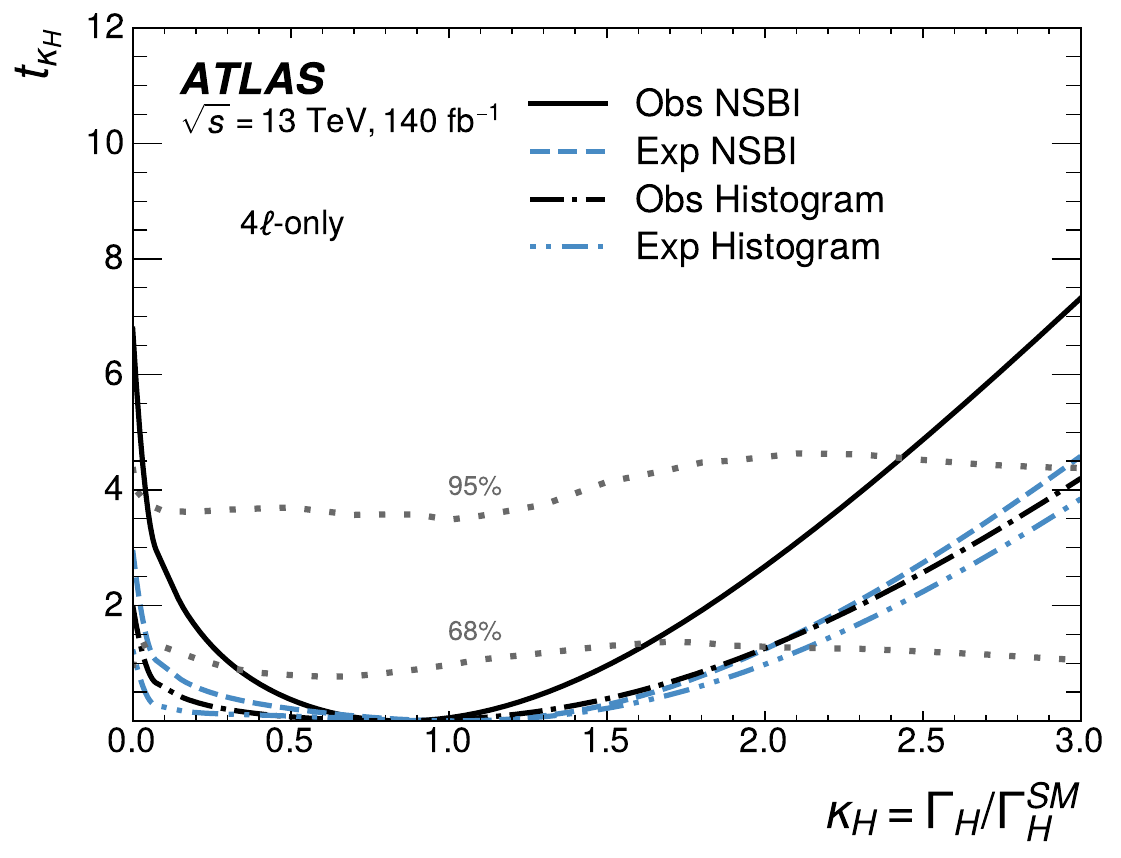}\label{fig:width_hist_4lonly}}\hfill
\subfloat[]{
\includegraphics[width=0.482\linewidth]{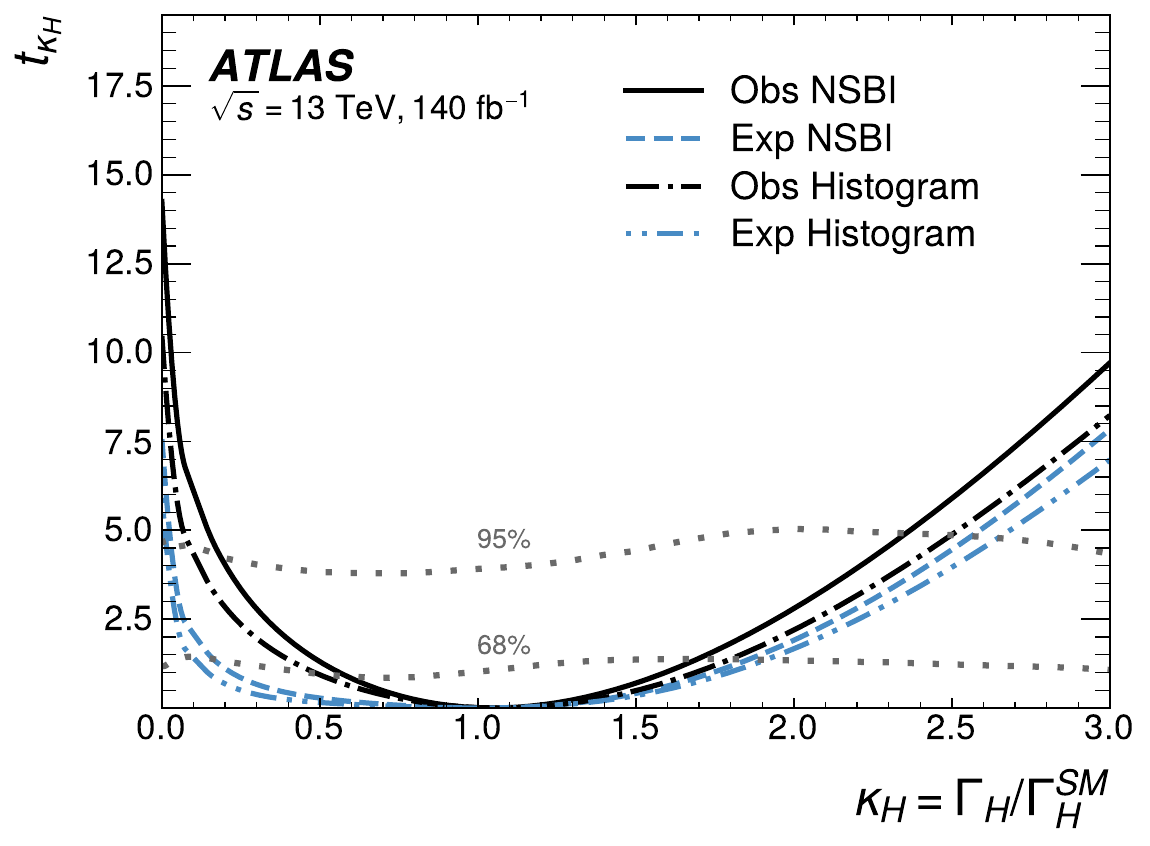}\label{fig:width_hist_4l2l2nu}}
\caption{\protect\subref{fig:width_hist_4lonly} Values of the test statistic $t_{\kappaH}$ as a function of $\kappaH = \Gamma_H/\Gamma_H^\text{SM}$ for the \htollll channel obtained with an Asimov dataset (expected, dashed blue) and with data (observed, solid black). The values from the histogram-based analysis~\cite{HIGG-2018-32} are added with dash-dotted lines for comparison.  The dotted gray lines show the 68\% and 95\% confidence belt, obtained from the Neyman construction, for the NSBI analysis.  \protect\subref{fig:width_hist_4l2l2nu} Same comparison of the values of the test statistic for $\kappaH = \Gamma_H/\Gamma_H^\text{SM}$ after combination with the \ohtollnunu\ channel.}
\end{figure}

\Fig{{\ref{fig:Rgg_results}}} and~\ref{fig:Rvv_results} compares the values of the test statistic for $R_{gg}$ and $R_{vv}$ with the results from the histogram-based analysis. \Tab{\ref{tab:appendix_summary}} summarizes the 68\% and 95\% CL results presented in this appendix, including  $\muoff$ ($4\ell$ only). These are compared with the results of the previous measurement Ref.~\cite{HIGG-2018-32}.

\begin{figure}[tbp]
\centering
\subfloat[]{\includegraphics[width=0.48\linewidth]{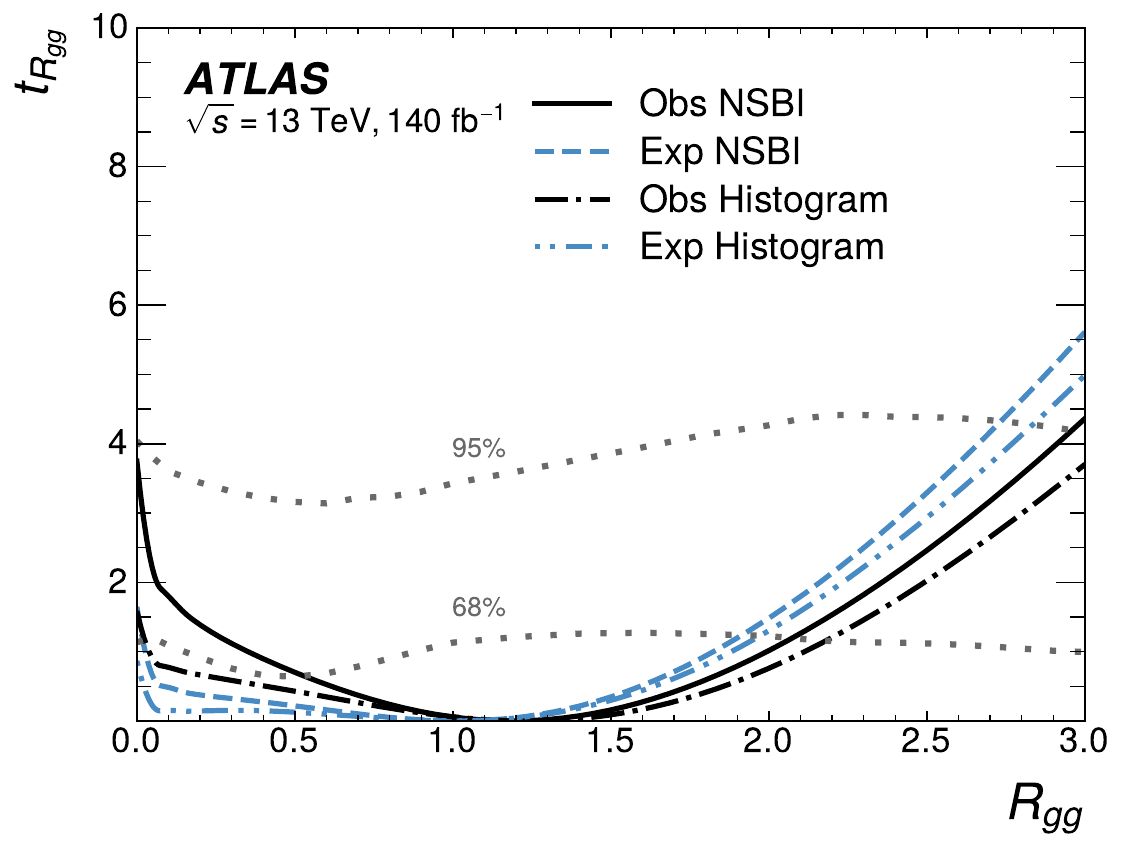}\label{fig:Rgg_results}}\hfill
\subfloat[]{\includegraphics[width=0.482\linewidth]{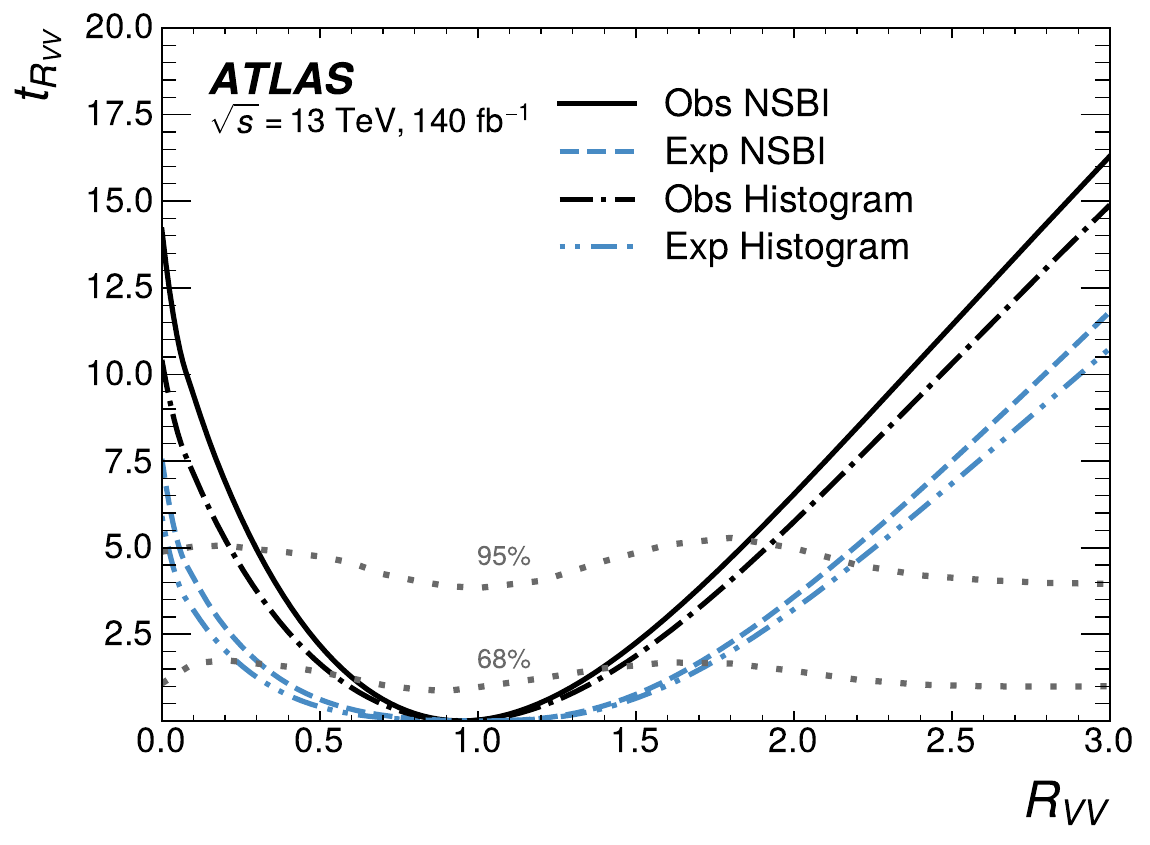}\label{fig:Rvv_results}}\\
\caption{Values of \protect\subref{fig:Rgg_results} the test statistic $t_{R_{gg}}$ as a function of $R_{gg}$, and of \protect\subref{fig:Rvv_results} the test statistic $t_{R_{VV}}$ as a function of $R_{vv}$ for the combined \htollll and \ohtollnunu channels obtained with an Asimov dataset (expected, dashed blue) and with data (observed, solid black). The equivalent graphs from the histogram-based analysis~\cite{HIGG-2018-32} are added with dash-dotted lines for comparison. The dotted gray lines show the 68\% and 95\% confidence belt, obtained from the Neyman construction, for the NSBI analysis.
}
\end{figure}

\begin{table}[tbp]
\centering
\caption{Summary of the results for the NSBI analysis with a comparison to the corresponding results of the histogram analysis of Ref.~\cite{HIGG-2018-32}. The value of the parameter estimates and the 68\% and 95\% confidence intervals are provided.  All results use the full \RunTwo dataset with $140\,\text{fb}^{-1}$ of integrated luminosity.}
\begin{tabular}{cccccc}\hline\hline
\toprule & & \multicolumn{2}{c}{\textbf{68\% CL interval}} & \multicolumn{2}{c}{\textbf{95\% CL interval}}  \\
\textbf{Parameter} & \textbf{Value} & \textbf{Observed}  & \textbf{Expected} & \textbf{Observed} & \textbf{Expected} \\
\hline && \\[-2.0ex]
\multicolumn{6}{c}{\textbf{NSBI analysis}} \\
\hline && \\[-2.0ex]

$\muoff$ ($4\ell$ only) & $\muofffourlnsbi$ & [\muofffourlnsbiobsonednval, \muofffourlnsbiobsoneupval] & [\muofffourlnsbiexponednval, \muofffourlnsbiexponeupval] & [\muofffourlnsbiobstwodnval, \muofffourlnsbiobstwoupval] & < \muofffourlnsbiobstwoupval  \\[1.5ex]
$\muoff$ & $\muoffcomb$ & [\muoffcombobsonednval, \muoffcombobsoneupval] & [\muoffcombexponednval, \muoffcombexponeupval] & [\muoffcombobstwodnval, \muoffcombobstwoupval] & [\muoffcombexptwodnval, \muoffcombexptwoupval] \\[1.5ex]
$\Gamma_H$ [MeV] ($4\ell$ only) & $\widthfourlhisto$ & [\widthfourlobsonednval, \widthfourlobsoneupval] & [\widthfourlexponednval, \widthfourlexponeupval] & [\widthfourlobstwodnval, \widthfourlobstwoupval] & $<\widthfourlexptwoupval$  \\[1.5ex]
$\Gamma_H$ [MeV]                & $\widthcomb$ & [\widthcombobsonednval, \widthcombobsoneupval] & [\widthcombexponednval, \widthcombexponeupval] & [\widthcombobstwodnval, \widthcombobstwoupval] & [\widthcombexptwodnval, \widthcombexptwoupval] \\[1.5ex]
$R_{gg}$                       & $\rggcomb$ & [\rggcombobsonednval, \rggcombobsoneupval] & [\rggcombexponednval, \rggcombexponeupval] & < \rggcombobstwoupval &  < \rggcombexptwoupval  \\[1.5ex]
$R_{VV}$                       & $\rvvcomb$ & [\rvvcombobsonednval, \rvvcombobsoneupval] & [\rvvcombexponednval, \rvvcombexponeupval] & [\rvvcombobstwodnval, \rvvcombobstwoupval] & [\rvvcombexptwodnval, \rvvcombexptwoupval] \\

\hline && \\[-2.0ex]
\multicolumn{6}{c}{\textbf{Histogram-based analysis}} \\
\hline && \\[-2.0ex]

$\muoff$ ($4\ell$ only)         & $\muofffourlhisto$ & [\muofffourlhistoobsonednval, \muofffourlhistoobsoneupval] &  $<\muofffourlhistoexponeupval$ & $< \muofffourlhistoobstwoupval$ &  $<\muofffourlhistoexptwoupval$ \\[1.5ex]
$\muoff$                        & $\muoffcombhisto$ & [\muoffcombhistoobsonednval, \muoffcombhistoobsoneupval] & [\muoffcombhistoexponednval, \muoffcombhistoexponeupval] & [\muoffcombhistoobstwodnval, \muoffcombhistoobstwoupval] & [\muoffcombhistoexptwodnval, \muoffcombhistoexptwoupval] \\[1.5ex]
$\Gamma_H$ [MeV] ($4\ell$ only) & $\widthfourlhisto$ & [\widthfourlhistoobsonednval, \widthfourlhistoobsoneupval] & $<\widthfourlhistoexponeupval$ & $<\widthfourlhistoobstwoupval$ & $< \widthfourlhistoexptwoupval$  \\[1.5ex]
$\Gamma_H$ [MeV]                & $\widthcombhisto$ & [\widthcombhistoobsonednval, \widthcombhistoobsoneupval] & [\widthcombhistoexponednval, \widthcombhistoexponeupval] & [\widthcombhistoobstwodnval, \widthcombhistoobstwoupval] & $< \widthcombhistoexptwoupval$   \\[1.5ex]
$R_{gg}$                       & $\rggcombhisto$ & [\rggcombhistoobsonednval, \rggcombhistoobsoneupval] & $< \rggcombhistoexponeupval$ & $< \rggcombhistoobstwoupval$ & $< \rggcombhistoexptwoupval$ \\[1.5ex]
$R_{VV}$                       & $\rvvcombhisto$ & [\rvvcombhistoobsonednval, \rvvcombhistoobsoneupval] & [\rvvcombhistoexponednval, \rvvcombhistoexponeupval] & [\rvvcombhistoobstwodnval, \rvvcombhistoobstwoupval] & [\rvvcombhistoexptwodnval, \rvvcombhistoexptwoupval]
\\\bottomrule \hline\hline

\end{tabular}%
\label{tab:appendix_summary}
\end{table}



\stepcounter{section}
\section{Construction of optimal observables}
\label{sec:oo_construction}

Despite the high dimensionality of the final-state phase space of processes such as \ggprod and \ewprod, the signal being studied tends to be concentrated in a small region. Histogram-based multivariate analyses can accurately isolate the regions with large statistical significance. In the case of non-linear analyses, like the measurement of off-shell Higgs boson production, these regions will change depending on the value of the parameter of interest.

The explicit estimate of the density ratios $p(x|\mu,\alpha)/p_{\text{ref}}(x)$ allows the creation of optimal observables $p(x|\mu,\widehat{\alpha})/p(x|\mu=1,\widehat{\alpha})$ that depend on the value of the signal strength $\mu$. \Fig{\ref{fig:oo_comparison}} shows this observable for $\mu_{\text{off-shell}}=0.3$ and $1.7$. Reference~\cite{atlasML} shows that the test statistic built from the optimal observable multinomial density converges to the NSBI value in the limit of large number of bins.

\begin{figure}[tbp]
\centering
\subfloat[]{\includegraphics[width=0.46\linewidth]{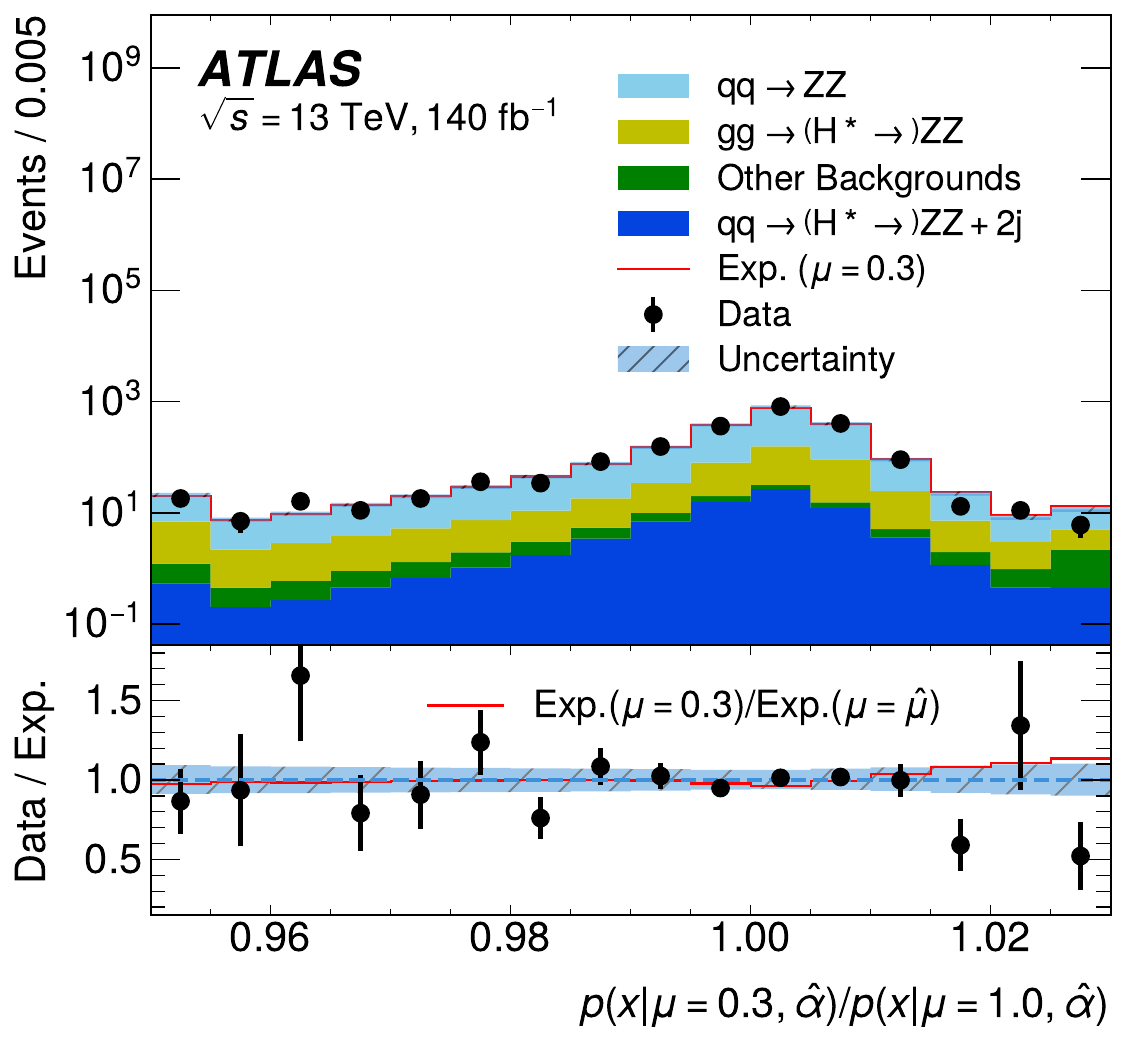}\label{fig:oo_comparison_0p3}}\hfill
\subfloat[]{\includegraphics[width=0.46\linewidth]{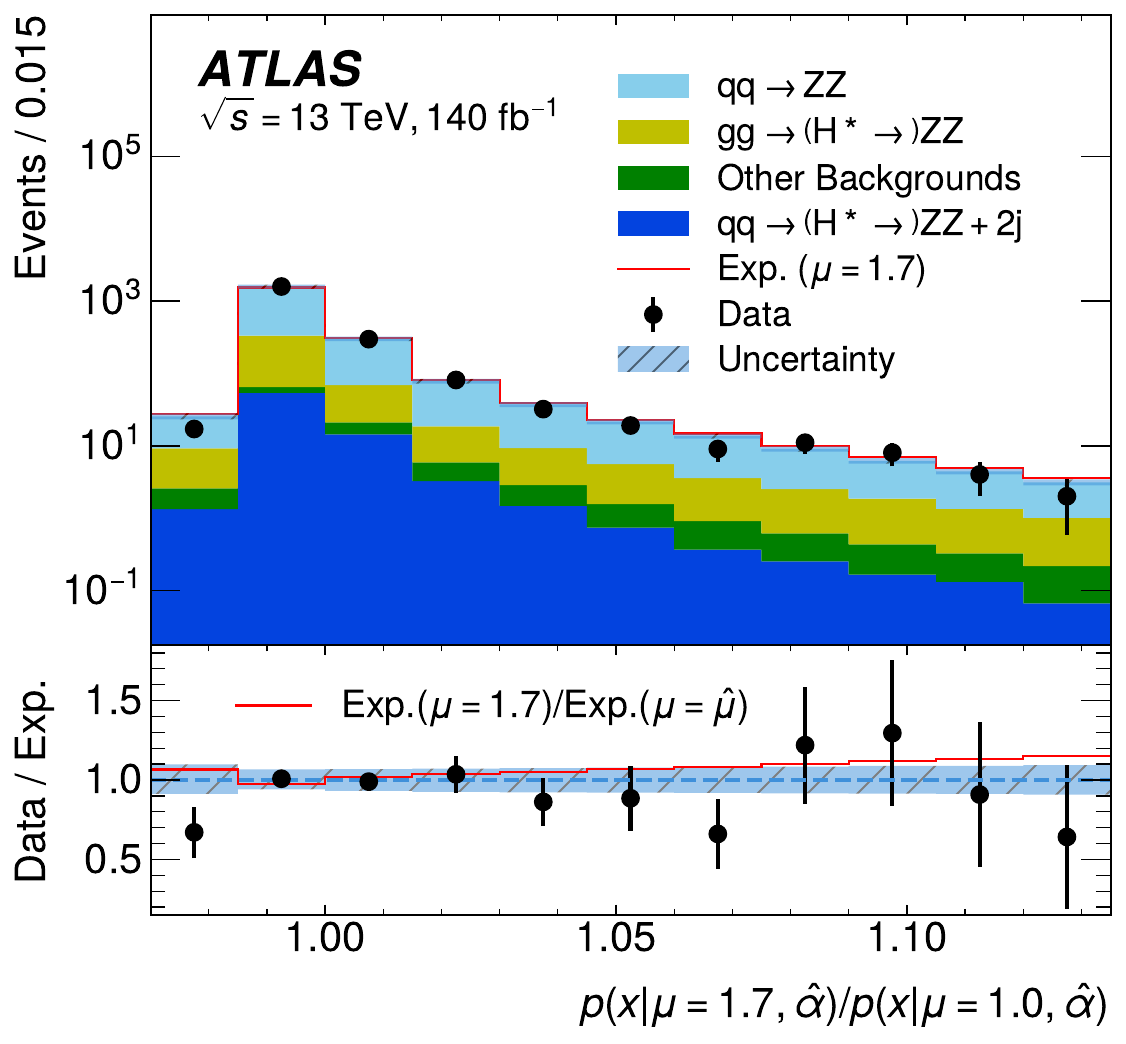}\label{fig:oo_comparison_1p7}}
\caption{Comparison between data and expectation in the SR ($\predisc >-0.85$) of the optimal observables \protect\subref{fig:oo_comparison_0p3} $p(x|\mu=0.3,\hatalpha)/p(x|1.0,\hatalpha)$ and \protect\subref{fig:oo_comparison_1p7} $p(x|\mu=1.7,\hatalpha)/p(x|1.0,\hatalpha)$. The solid red line shows the expected distribution for which the observable is optimal. The lower panels show the ratio of the data to expectation. The background is estimated under the SM hypothesis (post-fit, $\muoff=1$). The hatched area corresponds to the total systematic uncertainty in the expected distributions. The first and last bins contains overflow events.}
\label{fig:oo_comparison}
\end{figure}

The explicit estimate of the density ratios $p(x|\mu,\alpha)/p_{\text{ref}}(x)$ can also be used to study distributions enhancing the weight of those events that contribute the most to differentiate the SM hypothesis and another hypothesis with different value of \muoff. \Figs{\ref{fig:datamc_Dpre_weighted_0p3}}{\ref{fig:datamc_Dpre_weighted_1p7}} show a comparison between the observed and expected distributions of $\predisc $ when events are weighted by $p(x|\mu_{\text{off-shell}}=0.3)/p(x|\mu_{\text{off-shell}}=1)$ and $p(x|\mu_{\text{off-shell}}=1.7)/p(x|\mu_{\text{off-shell}}=1)$, respectively.

\begin{figure}[!htbp]
\centering
\subfloat[]{\includegraphics[width=0.46\linewidth]{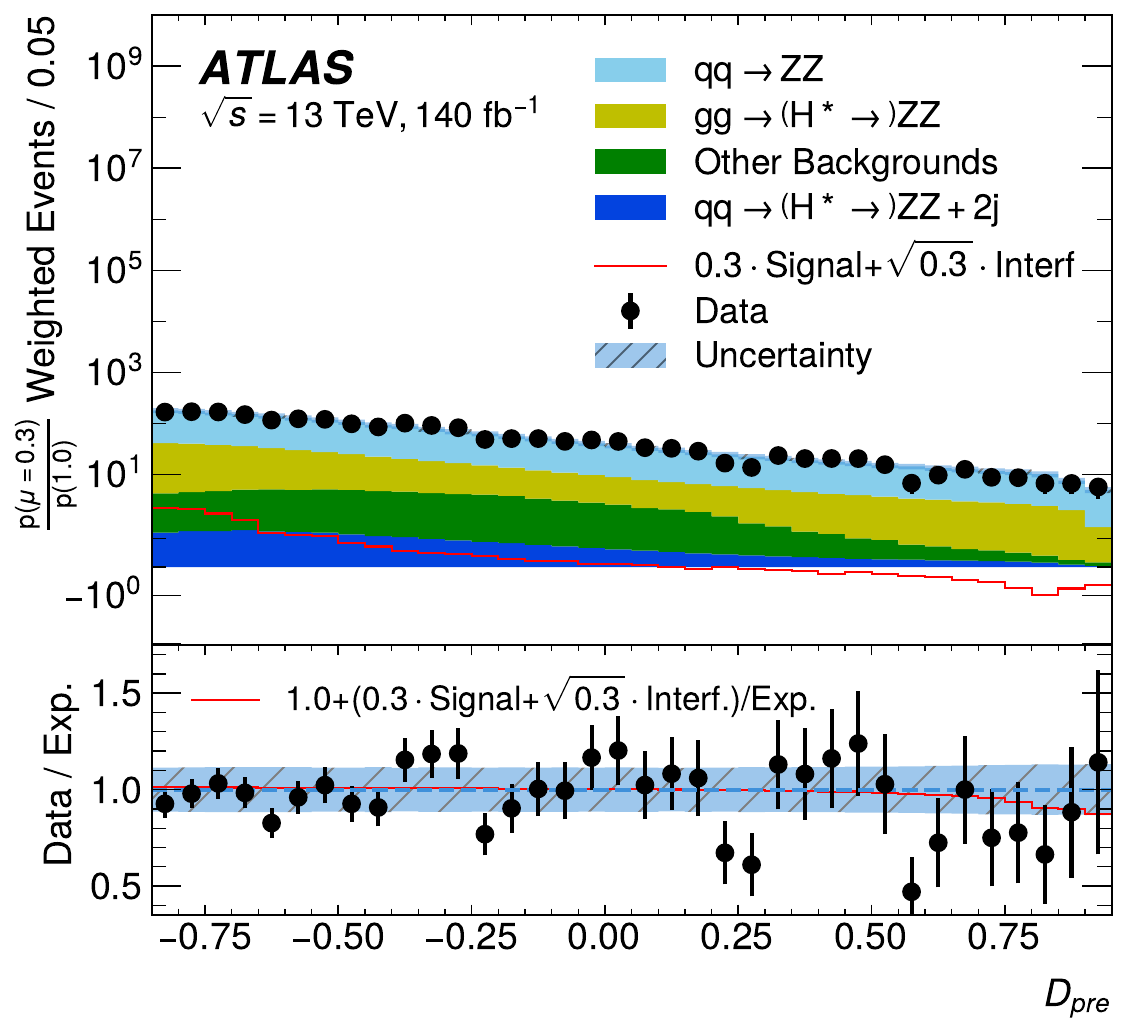}\label{fig:datamc_Dpre_weighted_0p3}}\hfill
\subfloat[]{\includegraphics[width=0.46\linewidth]{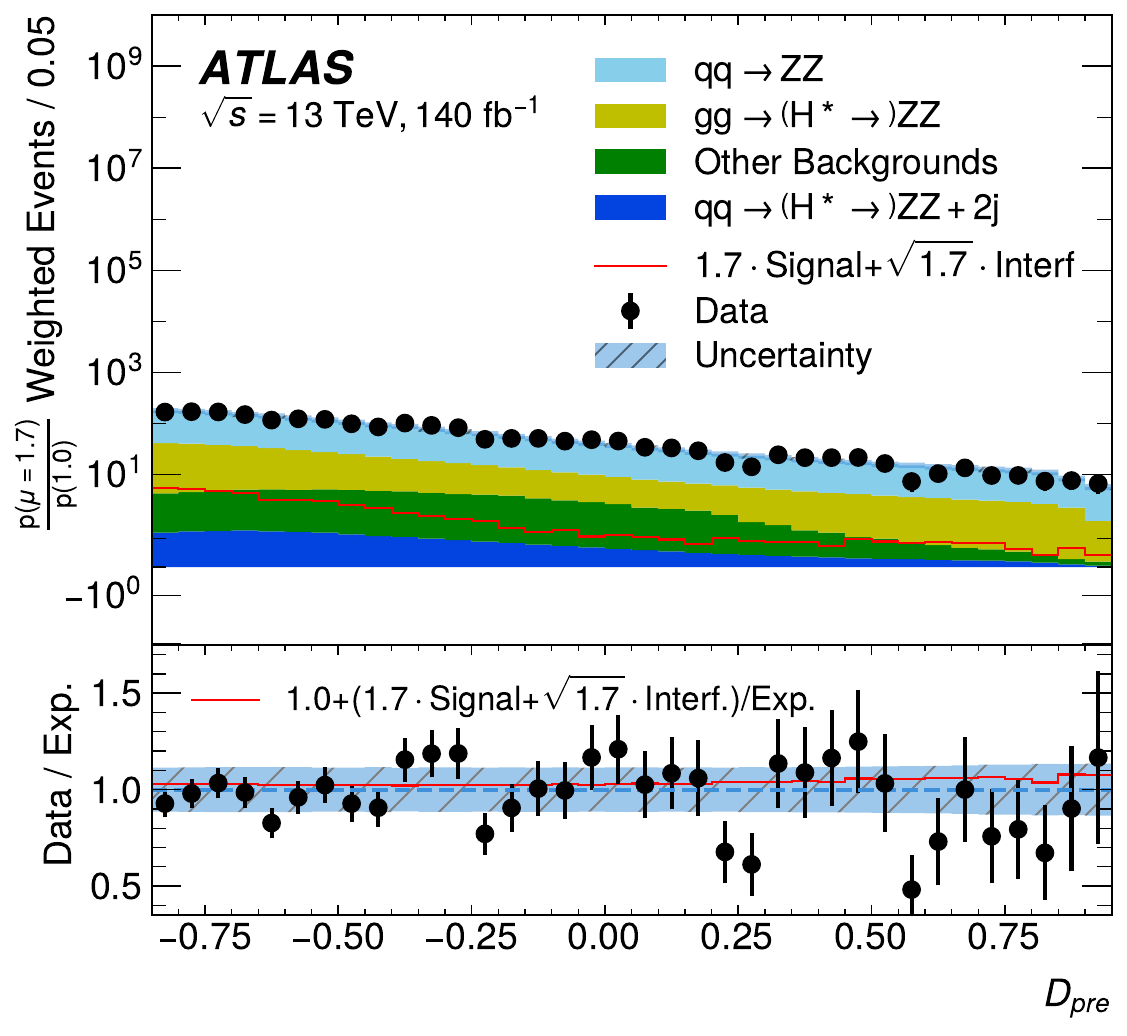}\label{fig:datamc_Dpre_weighted_1p7}}
\caption{Comparison between data and expectation in the SR ($\predisc >-0.85$) of the preselection NN\@. In \protect\subref{fig:datamc_Dpre_weighted_0p3}, events are weighted by $p(x|\mu=0.3)/p(x|\mu=1)$ to highlight the low-\muoff\ region dominated by the interference component. In \protect\subref{fig:datamc_Dpre_weighted_1p7}, events are weighted by $p(x|\mu=1.7)/p(x|\mu=1)$ to highlight the high-\muoff\ region dominated by the signal component. The solid red line shows the sum of the signal and interference components estimated with the $\muoff$ hypothesis used to reweight the distributions. The lower panels show the ratio of the data to expectation. The background is estimated under the SM hypothesis (post-fit, $\muoff=1$). The hatched area corresponds to the total systematic uncertainty in the expected distributions. The first and last bins contains overflow events.}
\end{figure}


\else
\addcontentsline{toc}{part}{Appendix}

\phantomsection
\refstepcounter{section}
\label{sec:nn_parameters}
\addtocontents{toc}{\protect\contentsline{section}{\protect\numberline{A} Neural network structure and training}{}{appendix.A}%
}
\refstepcounter{section}
\phantomsection
\label{sec:histo_comparison}
\addtocontents{toc}{\protect\contentsline{section}{\protect\numberline{B} Comparison with the histogram-based analysis}{}{appendix.B}%
}
\refstepcounter{section}
\phantomsection
\label{sec:oo_construction}
\addtocontents{toc}{\protect\contentsline{section}{\protect\numberline{C} Construction of optimal observables}{}{appendix.C}%
}
\fi

\clearpage
\printbibliography
\clearpage
 
\begin{flushleft}
\hypersetup{urlcolor=black}
{\Large The ATLAS Collaboration}

\bigskip

\AtlasOrcid[0000-0002-6665-4934]{G.~Aad}$^\textrm{\scriptsize 104}$,
\AtlasOrcid[0000-0001-7616-1554]{E.~Aakvaag}$^\textrm{\scriptsize 17}$,
\AtlasOrcid[0000-0002-5888-2734]{B.~Abbott}$^\textrm{\scriptsize 123}$,
\AtlasOrcid[0000-0002-0287-5869]{S.~Abdelhameed}$^\textrm{\scriptsize 119a}$,
\AtlasOrcid[0000-0002-1002-1652]{K.~Abeling}$^\textrm{\scriptsize 56}$,
\AtlasOrcid[0000-0001-5763-2760]{N.J.~Abicht}$^\textrm{\scriptsize 50}$,
\AtlasOrcid[0000-0002-8496-9294]{S.H.~Abidi}$^\textrm{\scriptsize 30}$,
\AtlasOrcid[0009-0003-6578-220X]{M.~Aboelela}$^\textrm{\scriptsize 46}$,
\AtlasOrcid[0000-0002-9987-2292]{A.~Aboulhorma}$^\textrm{\scriptsize 36e}$,
\AtlasOrcid[0000-0001-5329-6640]{H.~Abramowicz}$^\textrm{\scriptsize 155}$
% ...
\bigskip
\\

% ...
$^{17}$Department for Physics and Technology, University of Bergen, Bergen; Norway.\\
% ...
$^{30}$Physics Department, Brookhaven National Laboratory, Upton NY; United States of America.\\
% ...
$^{36}$$^{(a)}$Facult\'e des Sciences Ain Chock, Universit\'e Hassan II de Casablanca;$^{(b)}$Facult\'{e} des Sciences, Universit\'{e} Ibn-Tofail, K\'{e}nitra;$^{(c)}$Facult\'e des Sciences Semlalia, Universit\'e Cadi Ayyad, LPHEA-Marrakech;$^{(d)}$LPMR, Facult\'e des Sciences, Universit\'e Mohamed Premier, Oujda;$^{(e)}$Facult\'e des sciences, Universit\'e Mohammed V, Rabat;$^{(f)}$Institute of Applied Physics, Mohammed VI Polytechnic University, Ben Guerir; Morocco.\\
% ...
$^{46}$Physics Department, Southern Methodist University, Dallas TX; United States of America.\\
% ...
$^{50}$Fakult\"{a}t Physik , Technische Universit{\"a}t Dortmund, Dortmund; Germany.\\
% ...
$^{56}$II. Physikalisches Institut, Georg-August-Universit\"{a}t G\"ottingen, G\"ottingen; Germany.\\
% ...
$^{104}$CPPM, Aix-Marseille Universit\'e, CNRS/IN2P3, Marseille; France.\\
% ...
$^{119}$$^{(a)}$New York University Abu Dhabi, Abu Dhabi;$^{(b)}$United Arab Emirates University, Al Ain; United Arab Emirates.\\
% ...
$^{123}$Homer L. Dodge Department of Physics and Astronomy, University of Oklahoma, Norman OK; United States of America.\\
% ...
$^{155}$Raymond and Beverly Sackler School of Physics and Astronomy, Tel Aviv University, Tel Aviv; Israel.\\
% ...
$^{a}$ Also Affiliated with an institute covered by a cooperation agreement with CERN.\\
$^{b}$ Also at An-Najah National University, Nablus; Palestine.\\
$^{c}$ Also at Borough of Manhattan Community College, City University of New York, New York NY; United States of America.\\
$^{d}$ Also at Center for High Energy Physics, Peking University; China.\\
$^{e}$ Also at Center for Interdisciplinary Research and Innovation (CIRI-AUTH), Thessaloniki; Greece.\\
$^{f}$ Also at CERN, Geneva; Switzerland.\\
% ...

\end{flushleft}


%

%
%

%

%
%
%
%

\end{document}